\documentclass[longauth]{aa}

\usepackage{graphicx}
\usepackage{natbib}
\usepackage{scalerel}
\usepackage{csquotes}
\usepackage{threeparttable}

\usepackage[table]{xcolor}

\usepackage{txfonts}
\usepackage[pdfencoding=auto,psdextra]{hyperref}
\hypersetup{
    colorlinks=true,
    linkcolor=blue,
    filecolor=magenta,      
    urlcolor=blue,
    citecolor=blue
}
\makeatletter
\renewcommand*\aa@pageof{, page \thepage{} of \pageref*{LastPage}}
\makeatother

\usepackage[utf8]{inputenc}

\usepackage[switch, modulo]{lineno}

\usepackage{euclid}
\def\photoz{photo-\textit{z}}

\def\Msol{{\rm M}_\odot}

\def\lephare{\texttt{LePHARE}}
\def\eazy{\texttt{EAZY}}
\def\tractor{\texttt{The Tractor}}
\def\farmer{\texttt{The Farmer}}

\def\chOne{[3.6$\mu$m]}
\def\chTwo{[4.5$\mu$m]}

\def\Ms{$\mathcal{M}$\xspace}

\defcitealias{Weaver2023SMF}{W23}
\defcitealias{Davidzon2017}{D17}
\defcitealias{Grazian2015}{G15}
\defcitealias{Weibel2024}{W24}
\defcitealias{Zalesky2024}{EC-Z24}

\begin{document}
%
%

\title{\Euclid\/ preparation. Cosmic Dawn Survey:\\evolution of the galaxy stellar mass function across $0.2 < z \leq 6.5$ measured over 10 square degrees}

   
\newcommand{\orcid}[1]{} 
\author{Euclid Collaboration: L.~Zalesky\orcid{0000-0001-5680-2326}\thanks{\email{zalesky@hawaii.edu}}\inst{\ref{aff1}}
\and J.~R.~Weaver\orcid{0000-0003-1614-196X}\inst{\ref{aff2}}
\and C.~J.~R.~McPartland\orcid{0000-0003-0639-025X}\inst{\ref{aff3},\ref{aff4}}
\and G.~Murphree\orcid{0009-0007-7266-8914}\inst{\ref{aff1}}
\and I.~Valdes\orcid{0009-0002-8551-9372}\inst{\ref{aff1}}
\and C.~K.~Jespersen\orcid{0000-0002-8896-6496}\inst{\ref{aff5}}
\and S.~Taamoli\orcid{0000-0003-0749-4667}\inst{\ref{aff6}}
\and N.~Chartab\orcid{0000-0003-3691-937X}\inst{\ref{aff7}}
\and N.~Allen\orcid{0000-0001-9610-7950}\inst{\ref{aff8},\ref{aff4}}
\and S.~W.~J.~Barrow\orcid{0009-0001-9835-3650}\inst{\ref{aff8}}
\and D.~B.~Sanders\orcid{0000-0002-1233-9998}\inst{\ref{aff1}}
\and S.~Toft\orcid{0000-0003-3631-7176}\inst{\ref{aff8},\ref{aff4}}
\and B.~Mobasher\orcid{0000-0001-5846-4404}\inst{\ref{aff6}}
\and I.~Szapudi\orcid{0000-0003-2274-0301}\inst{\ref{aff1}}
\and B.~Altieri\orcid{0000-0003-3936-0284}\inst{\ref{aff9}}
\and A.~Amara\inst{\ref{aff10}}
\and S.~Andreon\orcid{0000-0002-2041-8784}\inst{\ref{aff11}}
\and N.~Auricchio\orcid{0000-0003-4444-8651}\inst{\ref{aff12}}
\and C.~Baccigalupi\orcid{0000-0002-8211-1630}\inst{\ref{aff13},\ref{aff14},\ref{aff15},\ref{aff16}}
\and M.~Baldi\orcid{0000-0003-4145-1943}\inst{\ref{aff17},\ref{aff12},\ref{aff18}}
\and S.~Bardelli\orcid{0000-0002-8900-0298}\inst{\ref{aff12}}
\and P.~Battaglia\orcid{0000-0002-7337-5909}\inst{\ref{aff12}}
\and A.~Biviano\orcid{0000-0002-0857-0732}\inst{\ref{aff14},\ref{aff13}}
\and D.~Bonino\orcid{0000-0002-3336-9977}\inst{\ref{aff19}}
\and E.~Branchini\orcid{0000-0002-0808-6908}\inst{\ref{aff20},\ref{aff21},\ref{aff11}}
\and M.~Brescia\orcid{0000-0001-9506-5680}\inst{\ref{aff22},\ref{aff23}}
\and J.~Brinchmann\orcid{0000-0003-4359-8797}\inst{\ref{aff24},\ref{aff25}}
\and A.~Caillat\inst{\ref{aff26}}
\and S.~Camera\orcid{0000-0003-3399-3574}\inst{\ref{aff27},\ref{aff28},\ref{aff19}}
\and G.~Ca\~nas-Herrera\orcid{0000-0003-2796-2149}\inst{\ref{aff29},\ref{aff30}}
\and V.~Capobianco\orcid{0000-0002-3309-7692}\inst{\ref{aff19}}
\and C.~Carbone\orcid{0000-0003-0125-3563}\inst{\ref{aff31}}
\and J.~Carretero\orcid{0000-0002-3130-0204}\inst{\ref{aff32},\ref{aff33}}
\and S.~Casas\orcid{0000-0002-4751-5138}\inst{\ref{aff34}}
\and F.~J.~Castander\orcid{0000-0001-7316-4573}\inst{\ref{aff35},\ref{aff36}}
\and M.~Castellano\orcid{0000-0001-9875-8263}\inst{\ref{aff37}}
\and G.~Castignani\orcid{0000-0001-6831-0687}\inst{\ref{aff12}}
\and S.~Cavuoti\orcid{0000-0002-3787-4196}\inst{\ref{aff23},\ref{aff38}}
\and K.~C.~Chambers\orcid{0000-0001-6965-7789}\inst{\ref{aff1}}
\and A.~Cimatti\inst{\ref{aff39}}
\and C.~Colodro-Conde\inst{\ref{aff40}}
\and G.~Congedo\orcid{0000-0003-2508-0046}\inst{\ref{aff41}}
\and C.~J.~Conselice\orcid{0000-0003-1949-7638}\inst{\ref{aff42}}
\and L.~Conversi\orcid{0000-0002-6710-8476}\inst{\ref{aff43},\ref{aff9}}
\and Y.~Copin\orcid{0000-0002-5317-7518}\inst{\ref{aff44}}
\and F.~Courbin\orcid{0000-0003-0758-6510}\inst{\ref{aff45},\ref{aff46}}
\and H.~M.~Courtois\orcid{0000-0003-0509-1776}\inst{\ref{aff47}}
\and A.~Da~Silva\orcid{0000-0002-6385-1609}\inst{\ref{aff48},\ref{aff49}}
\and H.~Degaudenzi\orcid{0000-0002-5887-6799}\inst{\ref{aff50}}
\and G.~De~Lucia\orcid{0000-0002-6220-9104}\inst{\ref{aff14}}
\and A.~M.~Di~Giorgio\orcid{0000-0002-4767-2360}\inst{\ref{aff51}}
\and H.~Dole\orcid{0000-0002-9767-3839}\inst{\ref{aff52}}
\and F.~Dubath\orcid{0000-0002-6533-2810}\inst{\ref{aff50}}
\and C.~A.~J.~Duncan\orcid{0009-0003-3573-0791}\inst{\ref{aff42}}
\and X.~Dupac\inst{\ref{aff9}}
\and S.~Dusini\orcid{0000-0002-1128-0664}\inst{\ref{aff53}}
\and A.~Ealet\orcid{0000-0003-3070-014X}\inst{\ref{aff44}}
\and S.~Escoffier\orcid{0000-0002-2847-7498}\inst{\ref{aff54}}
\and M.~Farina\orcid{0000-0002-3089-7846}\inst{\ref{aff51}}
\and R.~Farinelli\inst{\ref{aff12}}
\and S.~Farrens\orcid{0000-0002-9594-9387}\inst{\ref{aff55}}
\and F.~Faustini\orcid{0000-0001-6274-5145}\inst{\ref{aff56},\ref{aff37}}
\and S.~Ferriol\inst{\ref{aff44}}
\and F.~Finelli\orcid{0000-0002-6694-3269}\inst{\ref{aff12},\ref{aff57}}
\and P.~Fosalba\orcid{0000-0002-1510-5214}\inst{\ref{aff36},\ref{aff35}}
\and S.~Fotopoulou\orcid{0000-0002-9686-254X}\inst{\ref{aff58}}
\and M.~Frailis\orcid{0000-0002-7400-2135}\inst{\ref{aff14}}
\and E.~Franceschi\orcid{0000-0002-0585-6591}\inst{\ref{aff12}}
\and M.~Fumana\orcid{0000-0001-6787-5950}\inst{\ref{aff31}}
\and K.~George\orcid{0000-0002-1734-8455}\inst{\ref{aff59}}
\and B.~Gillis\orcid{0000-0002-4478-1270}\inst{\ref{aff41}}
\and C.~Giocoli\orcid{0000-0002-9590-7961}\inst{\ref{aff12},\ref{aff18}}
\and J.~Gracia-Carpio\inst{\ref{aff60}}
\and A.~Grazian\orcid{0000-0002-5688-0663}\inst{\ref{aff61}}
\and F.~Grupp\inst{\ref{aff60},\ref{aff59}}
\and S.~Gwyn\orcid{0000-0001-8221-8406}\inst{\ref{aff62}}
\and S.~V.~H.~Haugan\orcid{0000-0001-9648-7260}\inst{\ref{aff63}}
\and W.~Holmes\inst{\ref{aff64}}
\and I.~Hook\orcid{0000-0002-2960-978X}\inst{\ref{aff65}}
\and F.~Hormuth\inst{\ref{aff66}}
\and A.~Hornstrup\orcid{0000-0002-3363-0936}\inst{\ref{aff67},\ref{aff3}}
\and K.~Jahnke\orcid{0000-0003-3804-2137}\inst{\ref{aff68}}
\and M.~Jhabvala\inst{\ref{aff69}}
\and B.~Joachimi\orcid{0000-0001-7494-1303}\inst{\ref{aff70}}
\and E.~Keih\"anen\orcid{0000-0003-1804-7715}\inst{\ref{aff71}}
\and S.~Kermiche\orcid{0000-0002-0302-5735}\inst{\ref{aff54}}
\and A.~Kiessling\orcid{0000-0002-2590-1273}\inst{\ref{aff64}}
\and B.~Kubik\orcid{0009-0006-5823-4880}\inst{\ref{aff44}}
\and K.~Kuijken\orcid{0000-0002-3827-0175}\inst{\ref{aff72}}
\and M.~K\"ummel\orcid{0000-0003-2791-2117}\inst{\ref{aff59}}
\and M.~Kunz\orcid{0000-0002-3052-7394}\inst{\ref{aff73}}
\and H.~Kurki-Suonio\orcid{0000-0002-4618-3063}\inst{\ref{aff74},\ref{aff75}}
\and A.~M.~C.~Le~Brun\orcid{0000-0002-0936-4594}\inst{\ref{aff76}}
\and D.~Le~Mignant\orcid{0000-0002-5339-5515}\inst{\ref{aff26}}
\and S.~Ligori\orcid{0000-0003-4172-4606}\inst{\ref{aff19}}
\and P.~B.~Lilje\orcid{0000-0003-4324-7794}\inst{\ref{aff63}}
\and V.~Lindholm\orcid{0000-0003-2317-5471}\inst{\ref{aff74},\ref{aff75}}
\and I.~Lloro\orcid{0000-0001-5966-1434}\inst{\ref{aff77}}
\and G.~Mainetti\orcid{0000-0003-2384-2377}\inst{\ref{aff78}}
\and D.~Maino\inst{\ref{aff79},\ref{aff31},\ref{aff80}}
\and E.~Maiorano\orcid{0000-0003-2593-4355}\inst{\ref{aff12}}
\and O.~Mansutti\orcid{0000-0001-5758-4658}\inst{\ref{aff14}}
\and O.~Marggraf\orcid{0000-0001-7242-3852}\inst{\ref{aff81}}
\and K.~Markovic\orcid{0000-0001-6764-073X}\inst{\ref{aff64}}
\and M.~Martinelli\orcid{0000-0002-6943-7732}\inst{\ref{aff37},\ref{aff82}}
\and N.~Martinet\orcid{0000-0003-2786-7790}\inst{\ref{aff26}}
\and F.~Marulli\orcid{0000-0002-8850-0303}\inst{\ref{aff83},\ref{aff12},\ref{aff18}}
\and R.~Massey\orcid{0000-0002-6085-3780}\inst{\ref{aff84}}
\and S.~Maurogordato\inst{\ref{aff85}}
\and H.~J.~McCracken\orcid{0000-0002-9489-7765}\inst{\ref{aff86}}
\and E.~Medinaceli\orcid{0000-0002-4040-7783}\inst{\ref{aff12}}
\and S.~Mei\orcid{0000-0002-2849-559X}\inst{\ref{aff87},\ref{aff88}}
\and Y.~Mellier\inst{\ref{aff89},\ref{aff86}}
\and M.~Meneghetti\orcid{0000-0003-1225-7084}\inst{\ref{aff12},\ref{aff18}}
\and E.~Merlin\orcid{0000-0001-6870-8900}\inst{\ref{aff37}}
\and G.~Meylan\inst{\ref{aff90}}
\and A.~Mora\orcid{0000-0002-1922-8529}\inst{\ref{aff91}}
\and M.~Moresco\orcid{0000-0002-7616-7136}\inst{\ref{aff83},\ref{aff12}}
\and L.~Moscardini\orcid{0000-0002-3473-6716}\inst{\ref{aff83},\ref{aff12},\ref{aff18}}
\and R.~Nakajima\orcid{0009-0009-1213-7040}\inst{\ref{aff81}}
\and C.~Neissner\orcid{0000-0001-8524-4968}\inst{\ref{aff92},\ref{aff33}}
\and S.-M.~Niemi\inst{\ref{aff29}}
\and J.~W.~Nightingale\orcid{0000-0002-8987-7401}\inst{\ref{aff93}}
\and C.~Padilla\orcid{0000-0001-7951-0166}\inst{\ref{aff92}}
\and S.~Paltani\orcid{0000-0002-8108-9179}\inst{\ref{aff50}}
\and F.~Pasian\orcid{0000-0002-4869-3227}\inst{\ref{aff14}}
\and K.~Pedersen\inst{\ref{aff94}}
\and V.~Pettorino\inst{\ref{aff29}}
\and G.~Polenta\orcid{0000-0003-4067-9196}\inst{\ref{aff56}}
\and M.~Poncet\inst{\ref{aff95}}
\and L.~A.~Popa\inst{\ref{aff96}}
\and L.~Pozzetti\orcid{0000-0001-7085-0412}\inst{\ref{aff12}}
\and F.~Raison\orcid{0000-0002-7819-6918}\inst{\ref{aff60}}
\and R.~Rebolo\inst{\ref{aff40},\ref{aff97},\ref{aff98}}
\and A.~Renzi\orcid{0000-0001-9856-1970}\inst{\ref{aff99},\ref{aff53}}
\and J.~Rhodes\orcid{0000-0002-4485-8549}\inst{\ref{aff64}}
\and G.~Riccio\inst{\ref{aff23}}
\and E.~Romelli\orcid{0000-0003-3069-9222}\inst{\ref{aff14}}
\and M.~Roncarelli\orcid{0000-0001-9587-7822}\inst{\ref{aff12}}
\and R.~Saglia\orcid{0000-0003-0378-7032}\inst{\ref{aff59},\ref{aff60}}
\and Z.~Sakr\orcid{0000-0002-4823-3757}\inst{\ref{aff100},\ref{aff101},\ref{aff102}}
\and D.~Sapone\orcid{0000-0001-7089-4503}\inst{\ref{aff103}}
\and B.~Sartoris\orcid{0000-0003-1337-5269}\inst{\ref{aff59},\ref{aff14}}
\and J.~A.~Schewtschenko\orcid{0000-0002-4913-6393}\inst{\ref{aff41}}
\and M.~Schirmer\orcid{0000-0003-2568-9994}\inst{\ref{aff68}}
\and P.~Schneider\orcid{0000-0001-8561-2679}\inst{\ref{aff81}}
\and T.~Schrabback\orcid{0000-0002-6987-7834}\inst{\ref{aff104}}
\and A.~Secroun\orcid{0000-0003-0505-3710}\inst{\ref{aff54}}
\and E.~Sefusatti\orcid{0000-0003-0473-1567}\inst{\ref{aff14},\ref{aff13},\ref{aff15}}
\and G.~Seidel\orcid{0000-0003-2907-353X}\inst{\ref{aff68}}
\and S.~Serrano\orcid{0000-0002-0211-2861}\inst{\ref{aff36},\ref{aff105},\ref{aff35}}
\and P.~Simon\inst{\ref{aff81}}
\and C.~Sirignano\orcid{0000-0002-0995-7146}\inst{\ref{aff99},\ref{aff53}}
\and G.~Sirri\orcid{0000-0003-2626-2853}\inst{\ref{aff18}}
\and L.~Stanco\orcid{0000-0002-9706-5104}\inst{\ref{aff53}}
\and J.-L.~Starck\orcid{0000-0003-2177-7794}\inst{\ref{aff55}}
\and J.~Steinwagner\orcid{0000-0001-7443-1047}\inst{\ref{aff60}}
\and P.~Tallada-Cresp\'{i}\orcid{0000-0002-1336-8328}\inst{\ref{aff32},\ref{aff33}}
\and D.~Tavagnacco\orcid{0000-0001-7475-9894}\inst{\ref{aff14}}
\and A.~N.~Taylor\inst{\ref{aff41}}
\and H.~I.~Teplitz\orcid{0000-0002-7064-5424}\inst{\ref{aff106}}
\and I.~Tereno\inst{\ref{aff48},\ref{aff107}}
\and R.~Toledo-Moreo\orcid{0000-0002-2997-4859}\inst{\ref{aff108}}
\and F.~Torradeflot\orcid{0000-0003-1160-1517}\inst{\ref{aff33},\ref{aff32}}
\and A.~Tsyganov\inst{\ref{aff109}}
\and I.~Tutusaus\orcid{0000-0002-3199-0399}\inst{\ref{aff101}}
\and L.~Valenziano\orcid{0000-0002-1170-0104}\inst{\ref{aff12},\ref{aff57}}
\and J.~Valiviita\orcid{0000-0001-6225-3693}\inst{\ref{aff74},\ref{aff75}}
\and T.~Vassallo\orcid{0000-0001-6512-6358}\inst{\ref{aff59},\ref{aff14}}
\and G.~Verdoes~Kleijn\orcid{0000-0001-5803-2580}\inst{\ref{aff110}}
\and A.~Veropalumbo\orcid{0000-0003-2387-1194}\inst{\ref{aff11},\ref{aff21},\ref{aff20}}
\and Y.~Wang\orcid{0000-0002-4749-2984}\inst{\ref{aff106}}
\and J.~Weller\orcid{0000-0002-8282-2010}\inst{\ref{aff59},\ref{aff60}}
\and A.~Zacchei\orcid{0000-0003-0396-1192}\inst{\ref{aff14},\ref{aff13}}
\and G.~Zamorani\orcid{0000-0002-2318-301X}\inst{\ref{aff12}}
\and E.~Zucca\orcid{0000-0002-5845-8132}\inst{\ref{aff12}}
\and M.~Bolzonella\orcid{0000-0003-3278-4607}\inst{\ref{aff12}}
\and E.~Bozzo\orcid{0000-0002-8201-1525}\inst{\ref{aff50}}
\and C.~Burigana\orcid{0000-0002-3005-5796}\inst{\ref{aff111},\ref{aff57}}
\and M.~Calabrese\orcid{0000-0002-2637-2422}\inst{\ref{aff112},\ref{aff31}}
\and D.~Di~Ferdinando\inst{\ref{aff18}}
\and J.~A.~Escartin~Vigo\inst{\ref{aff60}}
\and L.~Gabarra\orcid{0000-0002-8486-8856}\inst{\ref{aff113}}
\and S.~Matthew\orcid{0000-0001-8448-1697}\inst{\ref{aff41}}
\and N.~Mauri\orcid{0000-0001-8196-1548}\inst{\ref{aff39},\ref{aff18}}
\and A.~Pezzotta\orcid{0000-0003-0726-2268}\inst{\ref{aff60}}
\and M.~P\"ontinen\orcid{0000-0001-5442-2530}\inst{\ref{aff74}}
\and C.~Porciani\orcid{0000-0002-7797-2508}\inst{\ref{aff81}}
\and V.~Scottez\inst{\ref{aff89},\ref{aff114}}
\and M.~Tenti\orcid{0000-0002-4254-5901}\inst{\ref{aff18}}
\and M.~Viel\orcid{0000-0002-2642-5707}\inst{\ref{aff13},\ref{aff14},\ref{aff16},\ref{aff15},\ref{aff115}}
\and M.~Wiesmann\orcid{0009-0000-8199-5860}\inst{\ref{aff63}}
\and Y.~Akrami\orcid{0000-0002-2407-7956}\inst{\ref{aff116},\ref{aff117}}
\and V.~Allevato\orcid{0000-0001-7232-5152}\inst{\ref{aff23}}
\and I.~T.~Andika\orcid{0000-0001-6102-9526}\inst{\ref{aff118},\ref{aff119}}
\and S.~Anselmi\orcid{0000-0002-3579-9583}\inst{\ref{aff53},\ref{aff99},\ref{aff76}}
\and M.~Archidiacono\orcid{0000-0003-4952-9012}\inst{\ref{aff79},\ref{aff80}}
\and F.~Atrio-Barandela\orcid{0000-0002-2130-2513}\inst{\ref{aff120}}
\and M.~Ballardini\orcid{0000-0003-4481-3559}\inst{\ref{aff121},\ref{aff12},\ref{aff122}}
\and D.~Bertacca\orcid{0000-0002-2490-7139}\inst{\ref{aff99},\ref{aff61},\ref{aff53}}
\and M.~Bethermin\orcid{0000-0002-3915-2015}\inst{\ref{aff123}}
\and A.~Blanchard\orcid{0000-0001-8555-9003}\inst{\ref{aff101}}
\and L.~Blot\orcid{0000-0002-9622-7167}\inst{\ref{aff124},\ref{aff76}}
\and S.~Borgani\orcid{0000-0001-6151-6439}\inst{\ref{aff125},\ref{aff13},\ref{aff14},\ref{aff15},\ref{aff115}}
\and M.~L.~Brown\orcid{0000-0002-0370-8077}\inst{\ref{aff42}}
\and S.~Bruton\orcid{0000-0002-6503-5218}\inst{\ref{aff126}}
\and R.~Cabanac\orcid{0000-0001-6679-2600}\inst{\ref{aff101}}
\and A.~Calabro\orcid{0000-0003-2536-1614}\inst{\ref{aff37}}
\and B.~Camacho~Quevedo\orcid{0000-0002-8789-4232}\inst{\ref{aff36},\ref{aff35}}
\and A.~Cappi\inst{\ref{aff12},\ref{aff85}}
\and F.~Caro\inst{\ref{aff37}}
\and C.~S.~Carvalho\inst{\ref{aff107}}
\and T.~Castro\orcid{0000-0002-6292-3228}\inst{\ref{aff14},\ref{aff15},\ref{aff13},\ref{aff115}}
\and R.~Chary\orcid{0000-0001-7583-0621}\inst{\ref{aff106},\ref{aff127}}
\and F.~Cogato\orcid{0000-0003-4632-6113}\inst{\ref{aff83},\ref{aff12}}
\and T.~Contini\orcid{0000-0003-0275-938X}\inst{\ref{aff101}}
\and A.~R.~Cooray\orcid{0000-0002-3892-0190}\inst{\ref{aff128}}
\and O.~Cucciati\orcid{0000-0002-9336-7551}\inst{\ref{aff12}}
\and S.~Davini\orcid{0000-0003-3269-1718}\inst{\ref{aff21}}
\and F.~De~Paolis\orcid{0000-0001-6460-7563}\inst{\ref{aff129},\ref{aff130},\ref{aff131}}
\and G.~Desprez\orcid{0000-0001-8325-1742}\inst{\ref{aff110}}
\and A.~D\'iaz-S\'anchez\orcid{0000-0003-0748-4768}\inst{\ref{aff132}}
\and S.~Di~Domizio\orcid{0000-0003-2863-5895}\inst{\ref{aff20},\ref{aff21}}
\and J.~M.~Diego\orcid{0000-0001-9065-3926}\inst{\ref{aff133}}
\and A.~G.~Ferrari\orcid{0009-0005-5266-4110}\inst{\ref{aff18}}
\and A.~Finoguenov\orcid{0000-0002-4606-5403}\inst{\ref{aff74}}
\and K.~Ganga\orcid{0000-0001-8159-8208}\inst{\ref{aff87}}
\and J.~Garc\'ia-Bellido\orcid{0000-0002-9370-8360}\inst{\ref{aff116}}
\and T.~Gasparetto\orcid{0000-0002-7913-4866}\inst{\ref{aff14}}
\and E.~Gaztanaga\orcid{0000-0001-9632-0815}\inst{\ref{aff35},\ref{aff36},\ref{aff134}}
\and F.~Giacomini\orcid{0000-0002-3129-2814}\inst{\ref{aff18}}
\and F.~Gianotti\orcid{0000-0003-4666-119X}\inst{\ref{aff12}}
\and G.~Gozaliasl\orcid{0000-0002-0236-919X}\inst{\ref{aff135},\ref{aff74}}
\and A.~Gregorio\orcid{0000-0003-4028-8785}\inst{\ref{aff125},\ref{aff14},\ref{aff15}}
\and M.~Guidi\orcid{0000-0001-9408-1101}\inst{\ref{aff17},\ref{aff12}}
\and C.~M.~Gutierrez\orcid{0000-0001-7854-783X}\inst{\ref{aff136}}
\and A.~Hall\orcid{0000-0002-3139-8651}\inst{\ref{aff41}}
\and W.~G.~Hartley\inst{\ref{aff50}}
\and S.~Hemmati\orcid{0000-0003-2226-5395}\inst{\ref{aff7}}
\and H.~Hildebrandt\orcid{0000-0002-9814-3338}\inst{\ref{aff137}}
\and J.~Hjorth\orcid{0000-0002-4571-2306}\inst{\ref{aff94}}
\and M.~Huertas-Company\orcid{0000-0002-1416-8483}\inst{\ref{aff40},\ref{aff138},\ref{aff139},\ref{aff140}}
\and O.~Ilbert\orcid{0000-0002-7303-4397}\inst{\ref{aff26}}
\and J.~J.~E.~Kajava\orcid{0000-0002-3010-8333}\inst{\ref{aff141},\ref{aff142}}
\and Y.~Kang\orcid{0009-0000-8588-7250}\inst{\ref{aff50}}
\and V.~Kansal\orcid{0000-0002-4008-6078}\inst{\ref{aff143},\ref{aff144}}
\and D.~Karagiannis\orcid{0000-0002-4927-0816}\inst{\ref{aff121},\ref{aff145}}
\and C.~C.~Kirkpatrick\inst{\ref{aff71}}
\and S.~Kruk\orcid{0000-0001-8010-8879}\inst{\ref{aff9}}
\and M.~Lattanzi\orcid{0000-0003-1059-2532}\inst{\ref{aff122}}
\and J.~Le~Graet\orcid{0000-0001-6523-7971}\inst{\ref{aff54}}
\and L.~Legrand\orcid{0000-0003-0610-5252}\inst{\ref{aff146},\ref{aff147}}
\and M.~Lembo\orcid{0000-0002-5271-5070}\inst{\ref{aff121},\ref{aff122}}
\and G.~Leroy\orcid{0009-0004-2523-4425}\inst{\ref{aff148},\ref{aff84}}
\and J.~Lesgourgues\orcid{0000-0001-7627-353X}\inst{\ref{aff34}}
\and T.~I.~Liaudat\orcid{0000-0002-9104-314X}\inst{\ref{aff149}}
\and A.~Loureiro\orcid{0000-0002-4371-0876}\inst{\ref{aff150},\ref{aff151}}
\and J.~Macias-Perez\orcid{0000-0002-5385-2763}\inst{\ref{aff152}}
\and G.~Maggio\orcid{0000-0003-4020-4836}\inst{\ref{aff14}}
\and M.~Magliocchetti\orcid{0000-0001-9158-4838}\inst{\ref{aff51}}
\and C.~Mancini\orcid{0000-0002-4297-0561}\inst{\ref{aff31}}
\and F.~Mannucci\orcid{0000-0002-4803-2381}\inst{\ref{aff153}}
\and R.~Maoli\orcid{0000-0002-6065-3025}\inst{\ref{aff154},\ref{aff37}}
\and J.~Mart\'{i}n-Fleitas\orcid{0000-0002-8594-569X}\inst{\ref{aff91}}
\and C.~J.~A.~P.~Martins\orcid{0000-0002-4886-9261}\inst{\ref{aff155},\ref{aff24}}
\and L.~Maurin\orcid{0000-0002-8406-0857}\inst{\ref{aff52}}
\and R.~B.~Metcalf\orcid{0000-0003-3167-2574}\inst{\ref{aff83},\ref{aff12}}
\and M.~Miluzio\inst{\ref{aff9},\ref{aff156}}
\and P.~Monaco\orcid{0000-0003-2083-7564}\inst{\ref{aff125},\ref{aff14},\ref{aff15},\ref{aff13}}
\and C.~Moretti\orcid{0000-0003-3314-8936}\inst{\ref{aff16},\ref{aff115},\ref{aff14},\ref{aff13},\ref{aff15}}
\and G.~Morgante\inst{\ref{aff12}}
\and C.~Murray\inst{\ref{aff87}}
\and K.~Naidoo\orcid{0000-0002-9182-1802}\inst{\ref{aff134}}
\and P.~Natoli\orcid{0000-0003-0126-9100}\inst{\ref{aff121},\ref{aff122}}
\and A.~Navarro-Alsina\orcid{0000-0002-3173-2592}\inst{\ref{aff81}}
\and S.~Nesseris\orcid{0000-0002-0567-0324}\inst{\ref{aff116}}
\and K.~Paterson\orcid{0000-0001-8340-3486}\inst{\ref{aff68}}
\and L.~Patrizii\inst{\ref{aff18}}
\and A.~Pisani\orcid{0000-0002-6146-4437}\inst{\ref{aff54},\ref{aff5}}
\and D.~Potter\orcid{0000-0002-0757-5195}\inst{\ref{aff157}}
\and I.~Risso\orcid{0000-0003-2525-7761}\inst{\ref{aff158}}
\and P.-F.~Rocci\inst{\ref{aff52}}
\and M.~Sahl\'en\orcid{0000-0003-0973-4804}\inst{\ref{aff159}}
\and E.~Sarpa\orcid{0000-0002-1256-655X}\inst{\ref{aff16},\ref{aff115},\ref{aff15}}
\and J.~Schaye\orcid{0000-0002-0668-5560}\inst{\ref{aff72}}
\and A.~Schneider\orcid{0000-0001-7055-8104}\inst{\ref{aff157}}
\and M.~Schultheis\inst{\ref{aff85}}
\and D.~Sciotti\orcid{0009-0008-4519-2620}\inst{\ref{aff37},\ref{aff82}}
\and E.~Sellentin\inst{\ref{aff160},\ref{aff72}}
\and M.~Sereno\orcid{0000-0003-0302-0325}\inst{\ref{aff12},\ref{aff18}}
\and F.~Shankar\orcid{0000-0001-8973-5051}\inst{\ref{aff161}}
\and L.~C.~Smith\orcid{0000-0002-3259-2771}\inst{\ref{aff162}}
\and S.~A.~Stanford\orcid{0000-0003-0122-0841}\inst{\ref{aff163}}
\and K.~Tanidis\inst{\ref{aff113}}
\and C.~Tao\orcid{0000-0001-7961-8177}\inst{\ref{aff54}}
\and G.~Testera\inst{\ref{aff21}}
\and R.~Teyssier\orcid{0000-0001-7689-0933}\inst{\ref{aff5}}
\and S.~Tosi\orcid{0000-0002-7275-9193}\inst{\ref{aff20},\ref{aff158}}
\and A.~Troja\orcid{0000-0003-0239-4595}\inst{\ref{aff99},\ref{aff53}}
\and M.~Tucci\inst{\ref{aff50}}
\and C.~Valieri\inst{\ref{aff18}}
\and A.~Venhola\orcid{0000-0001-6071-4564}\inst{\ref{aff164}}
\and D.~Vergani\orcid{0000-0003-0898-2216}\inst{\ref{aff12}}
\and G.~Verza\orcid{0000-0002-1886-8348}\inst{\ref{aff165}}
\and P.~Vielzeuf\orcid{0000-0003-2035-9339}\inst{\ref{aff54}}
\and N.~A.~Walton\orcid{0000-0003-3983-8778}\inst{\ref{aff162}}}
										   
\institute{Institute for Astronomy, University of Hawaii, 2680 Woodlawn Drive, Honolulu, HI 96822, USA\label{aff1}
\and
Department of Astronomy, University of Massachusetts, Amherst, MA 01003, USA\label{aff2}
\and
Cosmic Dawn Center (DAWN), Denmark\label{aff3}
\and
Niels Bohr Institute, University of Copenhagen, Jagtvej 128, 2200 Copenhagen, Denmark\label{aff4}
\and
Department of Astrophysical Sciences, Peyton Hall, Princeton University, Princeton, NJ 08544, USA\label{aff5}
\and
Physics and Astronomy Department, University of California, 900 University Ave., Riverside, CA 92521, USA\label{aff6}
\and
Caltech/IPAC, 1200 E. California Blvd., Pasadena, CA 91125, USA\label{aff7}
\and
Cosmic Dawn Center (DAWN)\label{aff8}
\and
ESAC/ESA, Camino Bajo del Castillo, s/n., Urb. Villafranca del Castillo, 28692 Villanueva de la Ca\~nada, Madrid, Spain\label{aff9}
\and
School of Mathematics and Physics, University of Surrey, Guildford, Surrey, GU2 7XH, UK\label{aff10}
\and
INAF-Osservatorio Astronomico di Brera, Via Brera 28, 20122 Milano, Italy\label{aff11}
\and
INAF-Osservatorio di Astrofisica e Scienza dello Spazio di Bologna, Via Piero Gobetti 93/3, 40129 Bologna, Italy\label{aff12}
\and
IFPU, Institute for Fundamental Physics of the Universe, via Beirut 2, 34151 Trieste, Italy\label{aff13}
\and
INAF-Osservatorio Astronomico di Trieste, Via G. B. Tiepolo 11, 34143 Trieste, Italy\label{aff14}
\and
INFN, Sezione di Trieste, Via Valerio 2, 34127 Trieste TS, Italy\label{aff15}
\and
SISSA, International School for Advanced Studies, Via Bonomea 265, 34136 Trieste TS, Italy\label{aff16}
\and
Dipartimento di Fisica e Astronomia, Universit\`a di Bologna, Via Gobetti 93/2, 40129 Bologna, Italy\label{aff17}
\and
INFN-Sezione di Bologna, Viale Berti Pichat 6/2, 40127 Bologna, Italy\label{aff18}
\and
INAF-Osservatorio Astrofisico di Torino, Via Osservatorio 20, 10025 Pino Torinese (TO), Italy\label{aff19}
\and
Dipartimento di Fisica, Universit\`a di Genova, Via Dodecaneso 33, 16146, Genova, Italy\label{aff20}
\and
INFN-Sezione di Genova, Via Dodecaneso 33, 16146, Genova, Italy\label{aff21}
\and
Department of Physics "E. Pancini", University Federico II, Via Cinthia 6, 80126, Napoli, Italy\label{aff22}
\and
INAF-Osservatorio Astronomico di Capodimonte, Via Moiariello 16, 80131 Napoli, Italy\label{aff23}
\and
Instituto de Astrof\'isica e Ci\^encias do Espa\c{c}o, Universidade do Porto, CAUP, Rua das Estrelas, PT4150-762 Porto, Portugal\label{aff24}
\and
Faculdade de Ci\^encias da Universidade do Porto, Rua do Campo de Alegre, 4150-007 Porto, Portugal\label{aff25}
\and
Aix-Marseille Universit\'e, CNRS, CNES, LAM, Marseille, France\label{aff26}
\and
Dipartimento di Fisica, Universit\`a degli Studi di Torino, Via P. Giuria 1, 10125 Torino, Italy\label{aff27}
\and
INFN-Sezione di Torino, Via P. Giuria 1, 10125 Torino, Italy\label{aff28}
\and
European Space Agency/ESTEC, Keplerlaan 1, 2201 AZ Noordwijk, The Netherlands\label{aff29}
\and
Institute Lorentz, Leiden University, Niels Bohrweg 2, 2333 CA Leiden, The Netherlands\label{aff30}
\and
INAF-IASF Milano, Via Alfonso Corti 12, 20133 Milano, Italy\label{aff31}
\and
Centro de Investigaciones Energ\'eticas, Medioambientales y Tecnol\'ogicas (CIEMAT), Avenida Complutense 40, 28040 Madrid, Spain\label{aff32}
\and
Port d'Informaci\'{o} Cient\'{i}fica, Campus UAB, C. Albareda s/n, 08193 Bellaterra (Barcelona), Spain\label{aff33}
\and
Institute for Theoretical Particle Physics and Cosmology (TTK), RWTH Aachen University, 52056 Aachen, Germany\label{aff34}
\and
Institute of Space Sciences (ICE, CSIC), Campus UAB, Carrer de Can Magrans, s/n, 08193 Barcelona, Spain\label{aff35}
\and
Institut d'Estudis Espacials de Catalunya (IEEC),  Edifici RDIT, Campus UPC, 08860 Castelldefels, Barcelona, Spain\label{aff36}
\and
INAF-Osservatorio Astronomico di Roma, Via Frascati 33, 00078 Monteporzio Catone, Italy\label{aff37}
\and
INFN section of Naples, Via Cinthia 6, 80126, Napoli, Italy\label{aff38}
\and
Dipartimento di Fisica e Astronomia "Augusto Righi" - Alma Mater Studiorum Universit\`a di Bologna, Viale Berti Pichat 6/2, 40127 Bologna, Italy\label{aff39}
\and
Instituto de Astrof\'{\i}sica de Canarias, V\'{\i}a L\'actea, 38205 La Laguna, Tenerife, Spain\label{aff40}
\and
Institute for Astronomy, University of Edinburgh, Royal Observatory, Blackford Hill, Edinburgh EH9 3HJ, UK\label{aff41}
\and
Jodrell Bank Centre for Astrophysics, Department of Physics and Astronomy, University of Manchester, Oxford Road, Manchester M13 9PL, UK\label{aff42}
\and
European Space Agency/ESRIN, Largo Galileo Galilei 1, 00044 Frascati, Roma, Italy\label{aff43}
\and
Universit\'e Claude Bernard Lyon 1, CNRS/IN2P3, IP2I Lyon, UMR 5822, Villeurbanne, F-69100, France\label{aff44}
\and
Institut de Ci\`{e}ncies del Cosmos (ICCUB), Universitat de Barcelona (IEEC-UB), Mart\'{i} i Franqu\`{e}s 1, 08028 Barcelona, Spain\label{aff45}
\and
Instituci\'o Catalana de Recerca i Estudis Avan\c{c}ats (ICREA), Passeig de Llu\'{\i}s Companys 23, 08010 Barcelona, Spain\label{aff46}
\and
UCB Lyon 1, CNRS/IN2P3, IUF, IP2I Lyon, 4 rue Enrico Fermi, 69622 Villeurbanne, France\label{aff47}
\and
Departamento de F\'isica, Faculdade de Ci\^encias, Universidade de Lisboa, Edif\'icio C8, Campo Grande, PT1749-016 Lisboa, Portugal\label{aff48}
\and
Instituto de Astrof\'isica e Ci\^encias do Espa\c{c}o, Faculdade de Ci\^encias, Universidade de Lisboa, Campo Grande, 1749-016 Lisboa, Portugal\label{aff49}
\and
Department of Astronomy, University of Geneva, ch. d'Ecogia 16, 1290 Versoix, Switzerland\label{aff50}
\and
INAF-Istituto di Astrofisica e Planetologia Spaziali, via del Fosso del Cavaliere, 100, 00100 Roma, Italy\label{aff51}
\and
Universit\'e Paris-Saclay, CNRS, Institut d'astrophysique spatiale, 91405, Orsay, France\label{aff52}
\and
INFN-Padova, Via Marzolo 8, 35131 Padova, Italy\label{aff53}
\and
Aix-Marseille Universit\'e, CNRS/IN2P3, CPPM, Marseille, France\label{aff54}
\and
Universit\'e Paris-Saclay, Universit\'e Paris Cit\'e, CEA, CNRS, AIM, 91191, Gif-sur-Yvette, France\label{aff55}
\and
Space Science Data Center, Italian Space Agency, via del Politecnico snc, 00133 Roma, Italy\label{aff56}
\and
INFN-Bologna, Via Irnerio 46, 40126 Bologna, Italy\label{aff57}
\and
School of Physics, HH Wills Physics Laboratory, University of Bristol, Tyndall Avenue, Bristol, BS8 1TL, UK\label{aff58}
\and
Universit\"ats-Sternwarte M\"unchen, Fakult\"at f\"ur Physik, Ludwig-Maximilians-Universit\"at M\"unchen, Scheinerstrasse 1, 81679 M\"unchen, Germany\label{aff59}
\and
Max Planck Institute for Extraterrestrial Physics, Giessenbachstr. 1, 85748 Garching, Germany\label{aff60}
\and
INAF-Osservatorio Astronomico di Padova, Via dell'Osservatorio 5, 35122 Padova, Italy\label{aff61}
\and
NRC Herzberg, 5071 West Saanich Rd, Victoria, BC V9E 2E7, Canada\label{aff62}
\and
Institute of Theoretical Astrophysics, University of Oslo, P.O. Box 1029 Blindern, 0315 Oslo, Norway\label{aff63}
\and
Jet Propulsion Laboratory, California Institute of Technology, 4800 Oak Grove Drive, Pasadena, CA, 91109, USA\label{aff64}
\and
Department of Physics, Lancaster University, Lancaster, LA1 4YB, UK\label{aff65}
\and
Felix Hormuth Engineering, Goethestr. 17, 69181 Leimen, Germany\label{aff66}
\and
Technical University of Denmark, Elektrovej 327, 2800 Kgs. Lyngby, Denmark\label{aff67}
\and
Max-Planck-Institut f\"ur Astronomie, K\"onigstuhl 17, 69117 Heidelberg, Germany\label{aff68}
\and
NASA Goddard Space Flight Center, Greenbelt, MD 20771, USA\label{aff69}
\and
Department of Physics and Astronomy, University College London, Gower Street, London WC1E 6BT, UK\label{aff70}
\and
Department of Physics and Helsinki Institute of Physics, Gustaf H\"allstr\"omin katu 2, 00014 University of Helsinki, Finland\label{aff71}
\and
Leiden Observatory, Leiden University, Einsteinweg 55, 2333 CC Leiden, The Netherlands\label{aff72}
\and
Universit\'e de Gen\`eve, D\'epartement de Physique Th\'eorique and Centre for Astroparticle Physics, 24 quai Ernest-Ansermet, CH-1211 Gen\`eve 4, Switzerland\label{aff73}
\and
Department of Physics, P.O. Box 64, 00014 University of Helsinki, Finland\label{aff74}
\and
Helsinki Institute of Physics, Gustaf H{\"a}llstr{\"o}min katu 2, University of Helsinki, Helsinki, Finland\label{aff75}
\and
Laboratoire Univers et Th\'eorie, Observatoire de Paris, Universit\'e PSL, Universit\'e Paris Cit\'e, CNRS, 92190 Meudon, France\label{aff76}
\and
SKA Observatory, Jodrell Bank, Lower Withington, Macclesfield, Cheshire SK11 9FT, UK\label{aff77}
\and
Centre de Calcul de l'IN2P3/CNRS, 21 avenue Pierre de Coubertin 69627 Villeurbanne Cedex, France\label{aff78}
\and
Dipartimento di Fisica "Aldo Pontremoli", Universit\`a degli Studi di Milano, Via Celoria 16, 20133 Milano, Italy\label{aff79}
\and
INFN-Sezione di Milano, Via Celoria 16, 20133 Milano, Italy\label{aff80}
\and
Universit\"at Bonn, Argelander-Institut f\"ur Astronomie, Auf dem H\"ugel 71, 53121 Bonn, Germany\label{aff81}
\and
INFN-Sezione di Roma, Piazzale Aldo Moro, 2 - c/o Dipartimento di Fisica, Edificio G. Marconi, 00185 Roma, Italy\label{aff82}
\and
Dipartimento di Fisica e Astronomia "Augusto Righi" - Alma Mater Studiorum Universit\`a di Bologna, via Piero Gobetti 93/2, 40129 Bologna, Italy\label{aff83}
\and
Department of Physics, Institute for Computational Cosmology, Durham University, South Road, Durham, DH1 3LE, UK\label{aff84}
\and
Universit\'e C\^{o}te d'Azur, Observatoire de la C\^{o}te d'Azur, CNRS, Laboratoire Lagrange, Bd de l'Observatoire, CS 34229, 06304 Nice cedex 4, France\label{aff85}
\and
Institut d'Astrophysique de Paris, UMR 7095, CNRS, and Sorbonne Universit\'e, 98 bis boulevard Arago, 75014 Paris, France\label{aff86}
\and
Universit\'e Paris Cit\'e, CNRS, Astroparticule et Cosmologie, 75013 Paris, France\label{aff87}
\and
CNRS-UCB International Research Laboratory, Centre Pierre Bin\'etruy, IRL2007, CPB-IN2P3, Berkeley, USA\label{aff88}
\and
Institut d'Astrophysique de Paris, 98bis Boulevard Arago, 75014, Paris, France\label{aff89}
\and
Institute of Physics, Laboratory of Astrophysics, Ecole Polytechnique F\'ed\'erale de Lausanne (EPFL), Observatoire de Sauverny, 1290 Versoix, Switzerland\label{aff90}
\and
Aurora Technology for European Space Agency (ESA), Camino bajo del Castillo, s/n, Urbanizacion Villafranca del Castillo, Villanueva de la Ca\~nada, 28692 Madrid, Spain\label{aff91}
\and
Institut de F\'{i}sica d'Altes Energies (IFAE), The Barcelona Institute of Science and Technology, Campus UAB, 08193 Bellaterra (Barcelona), Spain\label{aff92}
\and
School of Mathematics, Statistics and Physics, Newcastle University, Herschel Building, Newcastle-upon-Tyne, NE1 7RU, UK\label{aff93}
\and
DARK, Niels Bohr Institute, University of Copenhagen, Jagtvej 155, 2200 Copenhagen, Denmark\label{aff94}
\and
Centre National d'Etudes Spatiales -- Centre spatial de Toulouse, 18 avenue Edouard Belin, 31401 Toulouse Cedex 9, France\label{aff95}
\and
Institute of Space Science, Str. Atomistilor, nr. 409 M\u{a}gurele, Ilfov, 077125, Romania\label{aff96}
\and
Consejo Superior de Investigaciones Cientificas, Calle Serrano 117, 28006 Madrid, Spain\label{aff97}
\and
Universidad de La Laguna, Departamento de Astrof\'{\i}sica, 38206 La Laguna, Tenerife, Spain\label{aff98}
\and
Dipartimento di Fisica e Astronomia "G. Galilei", Universit\`a di Padova, Via Marzolo 8, 35131 Padova, Italy\label{aff99}
\and
Institut f\"ur Theoretische Physik, University of Heidelberg, Philosophenweg 16, 69120 Heidelberg, Germany\label{aff100}
\and
Institut de Recherche en Astrophysique et Plan\'etologie (IRAP), Universit\'e de Toulouse, CNRS, UPS, CNES, 14 Av. Edouard Belin, 31400 Toulouse, France\label{aff101}
\and
Universit\'e St Joseph; Faculty of Sciences, Beirut, Lebanon\label{aff102}
\and
Departamento de F\'isica, FCFM, Universidad de Chile, Blanco Encalada 2008, Santiago, Chile\label{aff103}
\and
Universit\"at Innsbruck, Institut f\"ur Astro- und Teilchenphysik, Technikerstr. 25/8, 6020 Innsbruck, Austria\label{aff104}
\and
Satlantis, University Science Park, Sede Bld 48940, Leioa-Bilbao, Spain\label{aff105}
\and
Infrared Processing and Analysis Center, California Institute of Technology, Pasadena, CA 91125, USA\label{aff106}
\and
Instituto de Astrof\'isica e Ci\^encias do Espa\c{c}o, Faculdade de Ci\^encias, Universidade de Lisboa, Tapada da Ajuda, 1349-018 Lisboa, Portugal\label{aff107}
\and
Universidad Polit\'ecnica de Cartagena, Departamento de Electr\'onica y Tecnolog\'ia de Computadoras,  Plaza del Hospital 1, 30202 Cartagena, Spain\label{aff108}
\and
Centre for Information Technology, University of Groningen, P.O. Box 11044, 9700 CA Groningen, The Netherlands\label{aff109}
\and
Kapteyn Astronomical Institute, University of Groningen, PO Box 800, 9700 AV Groningen, The Netherlands\label{aff110}
\and
INAF, Istituto di Radioastronomia, Via Piero Gobetti 101, 40129 Bologna, Italy\label{aff111}
\and
Astronomical Observatory of the Autonomous Region of the Aosta Valley (OAVdA), Loc. Lignan 39, I-11020, Nus (Aosta Valley), Italy\label{aff112}
\and
Department of Physics, Oxford University, Keble Road, Oxford OX1 3RH, UK\label{aff113}
\and
ICL, Junia, Universit\'e Catholique de Lille, LITL, 59000 Lille, France\label{aff114}
\and
ICSC - Centro Nazionale di Ricerca in High Performance Computing, Big Data e Quantum Computing, Via Magnanelli 2, Bologna, Italy\label{aff115}
\and
Instituto de F\'isica Te\'orica UAM-CSIC, Campus de Cantoblanco, 28049 Madrid, Spain\label{aff116}
\and
CERCA/ISO, Department of Physics, Case Western Reserve University, 10900 Euclid Avenue, Cleveland, OH 44106, USA\label{aff117}
\and
Technical University of Munich, TUM School of Natural Sciences, Physics Department, James-Franck-Str.~1, 85748 Garching, Germany\label{aff118}
\and
Max-Planck-Institut f\"ur Astrophysik, Karl-Schwarzschild-Str.~1, 85748 Garching, Germany\label{aff119}
\and
Departamento de F{\'\i}sica Fundamental. Universidad de Salamanca. Plaza de la Merced s/n. 37008 Salamanca, Spain\label{aff120}
\and
Dipartimento di Fisica e Scienze della Terra, Universit\`a degli Studi di Ferrara, Via Giuseppe Saragat 1, 44122 Ferrara, Italy\label{aff121}
\and
Istituto Nazionale di Fisica Nucleare, Sezione di Ferrara, Via Giuseppe Saragat 1, 44122 Ferrara, Italy\label{aff122}
\and
Universit\'e de Strasbourg, CNRS, Observatoire astronomique de Strasbourg, UMR 7550, 67000 Strasbourg, France\label{aff123}
\and
Center for Data-Driven Discovery, Kavli IPMU (WPI), UTIAS, The University of Tokyo, Kashiwa, Chiba 277-8583, Japan\label{aff124}
\and
Dipartimento di Fisica - Sezione di Astronomia, Universit\`a di Trieste, Via Tiepolo 11, 34131 Trieste, Italy\label{aff125}
\and
California Institute of Technology, 1200 E California Blvd, Pasadena, CA 91125, USA\label{aff126}
\and
University of California, Los Angeles, CA 90095-1562, USA\label{aff127}
\and
Department of Physics \& Astronomy, University of California Irvine, Irvine CA 92697, USA\label{aff128}
\and
Department of Mathematics and Physics E. De Giorgi, University of Salento, Via per Arnesano, CP-I93, 73100, Lecce, Italy\label{aff129}
\and
INFN, Sezione di Lecce, Via per Arnesano, CP-193, 73100, Lecce, Italy\label{aff130}
\and
INAF-Sezione di Lecce, c/o Dipartimento Matematica e Fisica, Via per Arnesano, 73100, Lecce, Italy\label{aff131}
\and
Departamento F\'isica Aplicada, Universidad Polit\'ecnica de Cartagena, Campus Muralla del Mar, 30202 Cartagena, Murcia, Spain\label{aff132}
\and
Instituto de F\'isica de Cantabria, Edificio Juan Jord\'a, Avenida de los Castros, 39005 Santander, Spain\label{aff133}
\and
Institute of Cosmology and Gravitation, University of Portsmouth, Portsmouth PO1 3FX, UK\label{aff134}
\and
Department of Computer Science, Aalto University, PO Box 15400, Espoo, FI-00 076, Finland\label{aff135}
\and
Instituto de Astrof\'\i sica de Canarias, c/ Via Lactea s/n, La Laguna 38200, Spain. Departamento de Astrof\'\i sica de la Universidad de La Laguna, Avda. Francisco Sanchez, La Laguna, 38200, Spain\label{aff136}
\and
Ruhr University Bochum, Faculty of Physics and Astronomy, Astronomical Institute (AIRUB), German Centre for Cosmological Lensing (GCCL), 44780 Bochum, Germany\label{aff137}
\and
Instituto de Astrof\'isica de Canarias (IAC); Departamento de Astrof\'isica, Universidad de La Laguna (ULL), 38200, La Laguna, Tenerife, Spain\label{aff138}
\and
Universit\'e PSL, Observatoire de Paris, Sorbonne Universit\'e, CNRS, LERMA, 75014, Paris, France\label{aff139}
\and
Universit\'e Paris-Cit\'e, 5 Rue Thomas Mann, 75013, Paris, France\label{aff140}
\and
Department of Physics and Astronomy, Vesilinnantie 5, 20014 University of Turku, Finland\label{aff141}
\and
Serco for European Space Agency (ESA), Camino bajo del Castillo, s/n, Urbanizacion Villafranca del Castillo, Villanueva de la Ca\~nada, 28692 Madrid, Spain\label{aff142}
\and
ARC Centre of Excellence for Dark Matter Particle Physics, Melbourne, Australia\label{aff143}
\and
Centre for Astrophysics \& Supercomputing, Swinburne University of Technology,  Hawthorn, Victoria 3122, Australia\label{aff144}
\and
Department of Physics and Astronomy, University of the Western Cape, Bellville, Cape Town, 7535, South Africa\label{aff145}
\and
DAMTP, Centre for Mathematical Sciences, Wilberforce Road, Cambridge CB3 0WA, UK\label{aff146}
\and
Kavli Institute for Cosmology Cambridge, Madingley Road, Cambridge, CB3 0HA, UK\label{aff147}
\and
Department of Physics, Centre for Extragalactic Astronomy, Durham University, South Road, Durham, DH1 3LE, UK\label{aff148}
\and
IRFU, CEA, Universit\'e Paris-Saclay 91191 Gif-sur-Yvette Cedex, France\label{aff149}
\and
Oskar Klein Centre for Cosmoparticle Physics, Department of Physics, Stockholm University, Stockholm, SE-106 91, Sweden\label{aff150}
\and
Astrophysics Group, Blackett Laboratory, Imperial College London, London SW7 2AZ, UK\label{aff151}
\and
Univ. Grenoble Alpes, CNRS, Grenoble INP, LPSC-IN2P3, 53, Avenue des Martyrs, 38000, Grenoble, France\label{aff152}
\and
INAF-Osservatorio Astrofisico di Arcetri, Largo E. Fermi 5, 50125, Firenze, Italy\label{aff153}
\and
Dipartimento di Fisica, Sapienza Universit\`a di Roma, Piazzale Aldo Moro 2, 00185 Roma, Italy\label{aff154}
\and
Centro de Astrof\'{\i}sica da Universidade do Porto, Rua das Estrelas, 4150-762 Porto, Portugal\label{aff155}
\and
HE Space for European Space Agency (ESA), Camino bajo del Castillo, s/n, Urbanizacion Villafranca del Castillo, Villanueva de la Ca\~nada, 28692 Madrid, Spain\label{aff156}
\and
Department of Astrophysics, University of Zurich, Winterthurerstrasse 190, 8057 Zurich, Switzerland\label{aff157}
\and
INAF-Osservatorio Astronomico di Brera, Via Brera 28, 20122 Milano, Italy, and INFN-Sezione di Genova, Via Dodecaneso 33, 16146, Genova, Italy\label{aff158}
\and
Theoretical astrophysics, Department of Physics and Astronomy, Uppsala University, Box 515, 751 20 Uppsala, Sweden\label{aff159}
\and
Mathematical Institute, University of Leiden, Einsteinweg 55, 2333 CA Leiden, The Netherlands\label{aff160}
\and
School of Physics \& Astronomy, University of Southampton, Highfield Campus, Southampton SO17 1BJ, UK\label{aff161}
\and
Institute of Astronomy, University of Cambridge, Madingley Road, Cambridge CB3 0HA, UK\label{aff162}
\and
Department of Physics and Astronomy, University of California, Davis, CA 95616, USA\label{aff163}
\and
Space physics and astronomy research unit, University of Oulu, Pentti Kaiteran katu 1, FI-90014 Oulu, Finland\label{aff164}
\and
Center for Computational Astrophysics, Flatiron Institute, 162 5th Avenue, 10010, New York, NY, USA\label{aff165}}        

%
%
\abstract{
The Cosmic Dawn Survey Pre-launch (PL) catalogues cover an effective 10.13 deg$^{2}$ area with uniform deep \textit{Spitzer}/IRAC data ($m\sim25$ mag, 5$\sigma$), the largest area covered to these depths at infrared wavelengths. These data are used to gain new insight into the growth of stellar mass across cosmic history by characterising the evolution of the galaxy stellar mass function through $0.2 < z \leq 6.5$. The total volume (0.62 Gpc$^{3}$) represents an order of magnitude increase compared to previous works that have explored $z > 3$ and significantly reduces cosmic variance, thus yielding strong constraints on the abundance of galaxies above the characteristic stellar mass (\Ms{}$^{\star}$) across this ten billion year time period. The evolution of the galaxy stellar mass function is generally consistent with results from the literature but now provide firm estimates of number density where  only upper limits were previously available. 
Contrasting the galaxy stellar mass function with the dark matter halo mass function suggests that massive galaxies ($\mathcal{M} \gtrsim10^{11}$ M$_{\odot}$) at $z > 3.5$ required integrated star-formation efficiencies of $\mathcal{M}/(\mathcal{M}_{\rm h}f_{\rm b}) \gtrsim$ 0.25--0.5, in excess of the commonly-held view of \enquote{universal peak efficiency} from studies on the stellar-to-halo mass relation. Such increased efficiencies imply an evolving peak in the stellar-to-halo mass relation at $z > 3.5$ which can be maintained if feedback mechanisms from active galactic nuclei and stellar processes are ineffective at early times. In addition, a significant fraction of the most massive quiescent galaxies are observed to be in place already by $z\sim 2.5$--3. The apparent lack in change of their number density by $z\sim 0.2$ is consistent with relatively little mass growth from mergers. Utilising the unique volume, evidence for an environmental dependence of the galaxy stellar mass function is found all the way through $z\sim 3.5$ for the first time, though a more careful characterisation of the density field is ultimately required for confirmation.
    }
%
%
%
%
\titlerunning{\Euclid preparation: DAWN galaxy stellar mass function}
\authorrunning{Euclid Collaboration: L. Zalesky et al.}
   
   \maketitle
%
%
%
%
   
\section{\label{sc:Intro}Introduction}

The galaxy stellar mass function (SMF) quantifies the number of galaxies per unit co-moving volume as a function of their stellar mass. Despite its apparent simplicity, the evolution of the galaxy SMF over cosmic time provides a basic framework to understand the growth of galaxies. At any point in time, the stellar mass of a galaxy is defined by its star-formation and merger history up until that moment. Whether a galaxy can efficiently convert gas into stars is impacted by the competing actions of gas inflow and accretion \citep{Dekel2009,Tacconi2020}, internal gas dynamics and heat exchange \citep{Scoville2012}, merger scenarios \citep{Conselice2014,Pearson2019}, as well as energy feedback from stellar processes \citep{Hopkins2012,Agertz2013} and active galactic nuclei \citep{Fabian2012,Beckman2017}, each of which change with time \citep{Madau2014}. Star formation is further correlated with local environment \citep{Gomez2003,Kauffmann2004,taamoli2024} and properties of the host dark matter halo \citep{Behroozi2013,Schaye2015,Wechsler2018}. Thus, measuring the evolution of the galaxy SMF from one epoch to another provides powerful insight into the general processes that govern stellar mass growth because the time integral of these processes over that interval entirely determines the changes in the shape of the galaxy SMF that are observed. 

The most massive galaxies provide strong constraints for theories of mass assembly. In the local Universe, the abundance of massive galaxies revealed the importance of feedback from AGN in shaping the galaxy SMF as well as the galaxy luminosity function with respect to the dark matter halo mass function \citep{Silk1998,Bower2006}. A consensus has emerged that at low redshift, the growth of stellar mass beyond some threshold is coincident with the cessation of star formation, a phenomenon dubbed \enquote{mass quenching} by \cite{Peng2010}. This is also supported by the relationship between star formation and stellar mass, i.e., the \enquote{main sequence} \citep{Brinchmann2004, Daddi2007, Elbaz2007, Noeske2007, Salim2007, Speagle2014, Whitaker2012, Popesso2023}. In the early Universe, massive galaxies challenge formation models because there is little elapsed time for them to form \citep{Steinhardt2016,Behroozi_Silk2018,Boylan2023}. These systems have required reconsideration of feedback \citep{Dekel2023,Li2023,Silk2024} and the assumption of a \enquote{universal} stellar initial mass function \citep{Chary2008,Riaz2021,Steinhardt2022}. At all redshifts, massive galaxies further provide an important link to the shape and growth of large-scale structure, as the most massive galaxies are embedded within the most massive dark matter halos that anchor the cosmic web \citep{Springel2005,Kolchin2009,Metuki2015,Chen2023}. Massive dark matter halos, over time, grow into dense galaxy environments and clusters, and their abundance and spatial distribution provides further constraints for cosmological models \citep{Bahcall1993,Zitrin2012,Hung2021}

Galaxy stellar mass is generally held to be the most robust intrinsic quantity that can be inferred from broadband photometry, and inference methods generally agree to within $\sim$0.15\,dex when the photometry measures rest-frame optical emission (\citealt{Mobasher2015,Pacifici2023}; see also \citealt{Conroy2013} for a review). 
As such, not only does measuring the evolution of the galaxy SMF promote an understanding of the processes of stellar mass growth, but its measurement is pragmatically accessible for large numbers of galaxies through photometric surveys. During the past two decades, photometric surveys of galaxies have matured alongside techniques used to accurately identify and characterise galaxies at different redshifts (\citealt{Weaver2021} and citations therein). Owing to these advancements, the build up and cessation of growth in galaxies since $z \sim 2$ has been well studied, including the most massive systems (see \citealt{Schreiber2020} for a recent review). The galaxy SMF of high-redshift ($z\gtrsim 2$) galaxies have mostly been studied through deep space-based surveys enabled primarily by the \textit{Hubble} Space Telescope (HST) and often in conjunction with \Spitzer \citep{Stark2009,Marchesini2009,Santini2012,Gonzalez2011,Duncan2014,Grazian2015,Davidzon2017,Kikuchihara2020,Stefanon2021,Adams2021, Weaver2023SMF}.  Recent observations from the \textit{James Webb} Space Telescope (JWST) have led to mid-infrared, mass-selected samples enabling new constraints on the evolution of the galaxy SMF to $z\sim 9$ and higher \citep{Harvey2024,Wang2024,Weibel2024,Shuntov2024}. 

Historically, space-based surveys have been deep but limited by the small fields of view offered by space telescopes. While such surveys have proved extraordinarily successful in characterising the abundance and growth of low- and intermediate-mass galaxies \citep{Furtak2021}, they are generally unable to constrain the growth of stellar mass within the most massive galaxies, i.e., beyond the characteristic mass \Ms{}$^{\star}$ (or \enquote{knee} of the Schechter function). The number density of massive galaxies rapidly declines with increasing stellar mass above the characteristic mass, making them intrinsically rare \citep{Weaver2023SMF}. Further, the galaxy bias is always greater for massive galaxies, implying that the uncertainty in their abundance (i.e., their \enquote{cosmic variance}) is well above what would be ordinarily expected from pure Poisson noise \citep{Moster2010,Jesperson2024}. Consequently, a statistically significant characterisation of the evolution of the most massive galaxies in the early Universe is missing.

The Euclid Wide Survey (EWS; \citealt{Scaramella22}) will probe enormous cosmic volumes ($>$\num{14000} deg$^{2}$) with high-resolution imaging in the optical and near-infrared wavelengths ($m < 24$ in the near-IR at $5\sigma$ for point sources; \citealt{Laureijs11}). The EWS is thus expected to provide photometry for over a billion galaxies and spectroscopic redshifts for several tens of millions \citep{Mellier2024}, thereby sampling cosmologically representative structures and including many massive galaxies. At these depths and in the absence of deep mid-IR imaging, the EWS will be particularly suited to address galaxy evolution at low redshift ($z \leq 2$). At higher redshifts, deep mid-infrared imaging is required to accurately measure stellar masses, as the emission of K- and M-class stars is progressively redshifted to longer wavelengths. For an $i$-band selected catalogue, \cite{Chartab2023} showed that the \textit{Spitzer}/IRAC \chOne{} and \chTwo{} bands contain the most information (compared to UV-NIR bands) related to galaxy stellar mass over all galaxies, implying its ubiquitous value across redshift.

Approximately 20\% of the \Euclid mission time will be devoted to observing six Euclid Deep and Auxiliary Fields (EDFs and EAFs, respectively) to acquire deeper imaging ($m < 26$ in the near-IR at $5\sigma$ for point sources), measure spectroscopic redshifts across a multitude of dispersion angles, and otherwise perform calibration operations \citep{Scaramella22,Mellier2024}. The EDFs are well positioned to be observed repeatedly throughout the \Euclid mission and include Euclid Deep Field North (20 deg$^{2}$), Euclid Deep Field Fornax (10 deg$^{2}$), and Euclid Deep Field South (23 deg$^{2}$). A 2.5 deg$^{2}$ region in EDF-N, referred to as the \enquote{self-calibration} field, will be observed to even greater depths ($m < 27.7$ in the near-IR at $5\sigma$ for point sources). The EAFs include four regions of the sky with significant archival observations from other facilities: AEGIS \citep{Davis2007}, GOODS-N \citep{Giavalisco2004}, COSMOS \citep{Scoville2007}, and XMM-LSS \citep{Clerc2014}. Altogether, the EDFs and EAFs comprise 59 deg$^{2}$. 


All previously acquired \textit{Spitzer}/IRAC data over the EDFs and EAFs were uniformly processed as part of the Cosmic Dawn Survey of the Euclid Deep and Auxiliary Fields (DAWN; \citealt{Moneti2022,McPartland2024}). The DAWN survey further provides depth-matched UV/optical imaging to the deep \textit{Spitzer}/IRAC across the entire 59 deg$^{2}$ of the combined EDFs and EAFs. One of the primary goals of the DAWN survey is thus to self-consistently measure photometry from the UV to mid-IR from these data, and thereby produce source catalogues optimised for high-redshift science. In a recent work, \cite{Zalesky2024}, hereafter referred to as \citetalias{Zalesky2024}, provided the first public release of pre-launch data from the DAWN survey (\enquote{DAWN PL}), which includes multiwavelength photometry and galaxy properties measured over Euclid Deep Field North (EDF-N) and Euclid Deep Field Fornax (EDF-F), collectively spanning over 16 deg$^{2}$. Consisting entirely of pre-launch data, the DAWN PL catalogues do not include \Euclid photometry. Nonetheless, they include exceptionally deep UV/optical photometry paired with deep \textit{Spitzer}/IRAC. Consequently, the DAWN PL catalogues currently provide the widest survey area mapped by \textit{Spitzer}/IRAC to depths of $m\sim25$ mag (5$\sigma$), despite lacking \Euclid photometry (i.e., being \enquote{pre-launch}). 

In this work, the contents of the DAWN PL catalogues are used to measure the evolution of the galaxy SMF across $0.2 < z \leq 6.5$, a significant majority (10.2 billion years) of cosmic history. The volume sampled across this redshift interval is 0.62 Gpc$^{3}$, an order of magnitude increase over \cite{Weaver2023SMF} and a factor of twenty increase over \cite{Shuntov2024}, the only other works to self-consistently (i.e., from a single dataset) measure the stellar mass function across this redshift interval. Such a volume drastically reduces uncertainty due to cosmic variance, while also providing diverse environments of high and low density from which to identify galaxies. A key objective of this work is to further  exploit the significant volume of DAWN PL to investigate the abundance of the most massive galaxies and their growth, as a population, over time. Eventually, the DAWN survey will provide deep multiwavelength photometry over a combined area of 59 deg$^{2}$ and include deep \Euclid near-IR photometry, which will prompt a reanalysis of the galaxy SMF in these fields. Therefore, this work also serves to benchmark the improvement that will inevitably be obtained due to the contribution of \Euclid (i.e., \enquote{post-launch}).

At the time of writing, only a few wide areas of the sky have been covered to depths beyond $m = 24$ mag (5$\sigma$) in the mid-infrared, thanks to \textit{Spitzer}/IRAC \citep{Ashby2018,Moneti2022}. The deepest fields with areas greater than $10$ deg$^{2}$ mapped to $m \sim 25$ are the EDF-N and EDF-F \citep{McPartland2024}. In addition, no other fields will benefit from the combination of deep mid-infrared and depth-matched UV/optical imaging until ten years after the Legacy Survey of Space and Time with the Rubin Observatory \citep{Ivezic2019}. With the decommissioning of \textit{Spitzer}, JWST is now the only facility currently capable of reaching similar depths in the mid-IR. However, due to the small field of view of JWST (NIRCam: 0.003 deg$^{2}$, MIRI 0.00065 deg$^{2}$), it is not clear when there will ever be larger (or additional, similarly large) areas with such deep mid-IR imaging.

This paper is organised as follows. Section~\ref{sc:data} provides a brief description of the DAWN PL catalogues and the measurements used to construct the galaxy SMF at each epoch. In Sect.~\ref{sc:sample} the selection criteria used to identify a reliable galaxy sample and to separate star forming galaxies from quiescent (at $z \leq 3$) are detailed, as well as the methods for determining sources of uncertainty associated with the sample. Section~\ref{sc:formalism} introduces the formalisms used for inferring the intrinsic galaxy SMF from the observed galaxy SMF and includes a description of the Schechter function and treatment of Eddington bias. The results of both the observed and intrinsic galaxy SMFs are presented in Sect.~\ref{sc:results} and compared with the literature. Section~\ref{sc:discussion} discusses the results in view of broader considerations of galaxy evolution, including the connection to dark matter and local environment. Finally, the work is summarised in Sect.~\ref{sc:summary}. 

This work assumes a standard $\Lambda$CDM cosmology with $H_0=70$\,km\,s$^{-1}$\,Mpc$^{-1}$, $\Omega_{\rm m}=0.3$, and $\Omega_{\Lambda}=0.7$ throughout, such that the dimensionless Hubble parameter $h_{70} \equiv H_{0}/(70\,\mathrm{km}\,\mathrm{s}^{-1}\,\mathrm{Mpc}^{-1})=1$. Galaxy stellar masses scale as the square of the luminosity distance (i.e., $d_{\rm L}^{2}$) when derived from SED fitting, and therefore a factor of $h_{70}^{-1}$ is retained implicitly for all relevant measurements \citep{Croton2013}. Estimates of stellar mass (hereafter, $\mathcal{M}$) assume a \citet{Chabrier2003} initial mass function (IMF). All magnitudes are expressed in the AB system \citep{Oke1974}, for which a flux $f_\nu$ in $\mu$Jy ($10^{-29}$~erg~s$^{-1}$~cm$^{-2}$~Hz$^{-1}$) corresponds to AB$_\nu=23.9-2.5\,\logten(f_\nu/\mu{\rm Jy})$.



\section{\label{sc:data}Data: DAWN survey PL catalogues}

The data utilised in this analysis are the Cosmic Dawn Survey PL catalogues, hereafter referred to as \enquote{DAWN PL}. DAWN PL provides multiwavelength photometry from the ultraviolet (UV) to mid-infrared wavelengths with derived galaxy properties across EDF-N and EDF-F. Ultraviolet (UV) and optical coverage is primarily provided by the Hawaii Twenty Square Degree Survey (H20).  H20 utilises the MegaCam instrument \citep{Boulade2003} on the Canada-France-Hawaii Telescope (CFHT) to obtain UV imaging in the $u$ band and the Hyper Suprime-Cam instrument \citep{Miyazaki2018} on the Subaru telescope to obtain optical imaging in the $griz$ bands. The DAWN survey PL catalogues were created also utilising archival Subaru HSC data in EDF-N from HEROES \citep{Taylor2023} and \textit{AKARI} \citep{Oi2021} along with privately shared CFHT MegaCam data from the Deep \Euclid U-band Survey \citep[DEUS; designed similarly to][]{Sawicki_2019}. Mid-infrared coverage over EDF-N and EDF-F is provided by the DAWN survey \textit{Spitzer}/IRAC data \citep{Moneti2022}, where the primary contribution is from the \textit{Spitzer} Legacy Survey \cite[SLS;][]{Capak2016} which obtained deep imaging in two channels, \chOne{} and \chTwo{} over the entirety of EDF-N and EDF-F. This work was conducted before the launch of \Euclid, making it a `pre-launch' SMF. Future work will extend this study from \Euclid-selected samples with greater mass completeness and higher redshifts.

Measuring photometry from images reaching great depths but with a wide range of resolution is challenging. To this end, the DAWN survey PL catalogues utilise the model-based photometry method introduced in the creation of the most recent COSMOS catalogue \enquote{COSMOS2020,} \citep{Weaver2021} called \farmer{}~\citep{WeaverFarmer}. \farmer{} is built around \tractor{}~\citep{Lang2016} to self-consistently measure total flux and flux uncertainties from images of varying point spread functions and is well suited for handling crowded fields of deep imaging. DAWN PL photometric bands include CFHT $u$, HSC $griz$, and \textit{Spitzer}/IRAC \chOne{} and \chTwo{}. In EDF-N, photometry from archival HSC $y$-band imaging is also measured.  \citetalias{Zalesky2024} provides a full description of DAWN PL. Following COSMOS2020, photometric redshifts (\photoz{}s) and galaxy properties are computed using two independent codes, \eazy{}~\citep{Brammer2008} and \lephare{}~\citep{Arnouts2002,Ilbert2006}. The inferred galaxy properties are internally validated and further supported by an in-depth comparison with galaxies from COSMOS2020 in \cite{Zalesky2024}. At the outset, the properties inferred from SED fitting appear robust. 

The DAWN PL catalogue of EDF-N spans a total area of 16.87 deg$^{2}$, with 9.37 deg$^{2}$ in the center reaching final survey depths (see Fig.~1 of  \citetalias{Zalesky2024}). Meanwhile, the DAWN survey PL catalogue of EDF-F contains 2.85 deg$^{2}$ of the deepest presently available data, with 1.77 deg$^{2}$ reaching final survey depths in all but the HSC $z$ band, which is shallower by 0.5 mag. However, the HSC $i$-band imaging used in the DAWN survey PL catalogue of EDF-F is 0.3 mag deeper than in EDF-N. Thus, being selected from an HSC $r+i+z$ stack, the two catalogues reach approximately equal depth. Notably, the full-depth regions of each catalogue are fully covered by \textit{Spitzer}/IRAC \chOne{} and \chTwo{} to a depth of $\sim$25 mag ($5\sigma$). Altogether, DAWN PL represents the only present galaxy catalogues reaching 5$\sigma$ depths of $\sim$27 mag in optical bands and $\sim$25 mag in \chOne{} and \chTwo{} spanning a combined area of greater than 10 deg$^{2}$.

\section{\label{sc:sample}Characterisation of galaxies and sample uncertainties}

\subsection{\label{subsec:selection}Galaxy sample}

All sources from the DAWN survey PL catalogues are detected from a composite stack of the HSC $r+i+z$ images, the deepest and reddest bands currently available\footnote{Future releases from the DAWN survey will be detected from \Euclid near-IR imaging.}. Specifically, a \texttt{CHI-MEAN} co-added image is created using \texttt{SWARP}~\citep{Szalay1999,Bertin2002,Bertin2010} and objects are detected using \texttt{SEP}~\citep{Barbary2016}. Although the use of HSC $r$, $i$, and $z$ effectively establishes an optical selection function, the significant depths of the individual images, further improved by their combination, yields a substantial number of high-redshift galaxies as demonstrated in Sect.~\ref{subsec:m_completeness},  Sect.~\ref{sc:results}, and Appendix~\ref{app:validation}. In addition, the observed bandpasses provide a reliable identification of quiescent galaxies, at least through $z = 1.6$, after which the rest-frame Balmer break exits the observed HSC $z$ bandpass (see Sect.~\ref{subsec:SF_Q}). Intrinsically bright quiescent galaxies may be also detected from the deep optical imaging above $z = 1.6$, similar to high-redshift galaxies, i.e., because their relatively fainter emission from the blue side of the Balmer break is detected. Importantly, all galaxy stellar masses are properly constrained by \textit{Spitzer}/IRAC. Yet, the lack of near-infrared coverage (soon to be provided by \Euclid) prevents the inclusion of optically-weak samples such as quiescent and/or dusty galaxies at higher redshifts and/or lower stellar masses. These sources of incompleteness are explored in Sect.~\ref{subsec:SF_Q}.

As described above and in  \citetalias{Zalesky2024}, the DAWN survey PL catalogues span areas of 16.87 deg$^{2}$ for EDF-N and 2.85 deg$^{2}$ for EDF-F but do not have uniform coverage. To minimise systematic uncertainties and biases, sources are only considered from regions with the most reliable and homogeneous photometry. Each DAWN survey PL catalogue include a \enquote{full-depth} region, defined by a flower petal pattern of seven HSC pointings in EDF-N spanning 9.37 deg$^{2}$, and a single HSC pointing in EDF-F spanning 1.77 deg$^{2}$. After accounting for areas masked by stars and other artefacts, the effective areas are 8.42 deg$^{2}$ in EDF-N and 1.71 deg$^{2}$ in EDF-F, where EDF-N is more heavily affected by stars given its low Galactic latitude. Several works have characterised the galaxy SMF at low redshift  utilising larger areas \citep{Weigel2016,Capozzi2017,Kawinwanichakij2020}. However, at $z > 2$, the combined area of 10.13 deg$^{2}$ of the present work surpasses the next closest field by area with deep \textit{Spitzer}/IRAC coverage, i.e., COSMOS2020 \citep{Weaver2021,Weaver2023SMF}, by an order of magnitude. As such, the impact due to Poisson uncertainty is immediately improved by at least a factor of three in comparison. Moreover, the comoving volume probed by the DAWN survey PL full-depth regions at $0 < z < 5$ is nearly 0.5 Gpc$^{3}$, again an order of magnitude larger the COSMOS2020, providing the largest variety in cosmic structure and environment accounted for in consideration of the galaxy SMF at $z > 2$. 

The DAWN survey PL catalogues include \photoz{} measurements from both \eazy{} and \lephare{}. In comparison with spectroscopic samples,  \citetalias{Zalesky2024} demonstrated that \lephare{} achieved a smaller outlier fraction and spread than \eazy{} for bright galaxies with HSC $i < 24$, but the codes performed similarly for faint galaxies with HSC $i > 25$. However, the large gap in wavelength coverage between HSC $z$ (or the shallow HSC $y$ in EDF-N) and \textit{Spitzer}/IRAC \chOne{} in the DAWN survey PL catalogues poses a problem for deriving rest-frame properties using \eazy{} due to the flexible nature of the code. More specifically, \eazy{} fits a combination of templates to each galaxy without any physical priors, therefore unphysical rest-frame colours and stellar masses may arise due to this wavelength-space gap. By contrast, \lephare{} utilises only single templates, and a judicious choice of allowed templates mitigates the possibility of unphysical quantities arising. In light of this distinction, the present analysis uses \photoz{}s and stellar masses measured by \lephare{}. Utilising the output from \lephare{} also enables a direct and straightforward comparison with both \cite{Davidzon2017}, hereafter referred to as \citetalias{Davidzon2017}, and \cite{Weaver2023SMF}, hereafter referred to as \citetalias{Weaver2023SMF}, the most recent characterisations of the galaxy SMF from the COSMOS field, where each work also used \lephare{}. Further, as described by  \citetalias{Zalesky2024}, the configuration of \lephare{} used in creating the DAWN survey PL catalogues, including the choice of galaxy and star templates, closely follows the prescription used in the creation of the COSMOS2020 catalogue \citep{Weaver2021} and subsequently the analysis of the galaxy SMF in \citetalias{Weaver2023SMF}. 

Staying consistent with \citetalias{Davidzon2017} and \citetalias{Weaver2023SMF}, the redshift of each galaxy is defined to be the median of the redshift probability distribution resulting from SED fitting (column name = `\texttt{lp\textunderscore zPDF}'). After fixing the redshift of each galaxy to the appropriate \photoz{}, the stellar mass of each galaxy is measured by determining the median of the posterior distribution for stellar mass obtained after marginalising over the other free parameters varied by \lephare{} (column name = `\texttt{lp\textunderscore mass\textunderscore med}'). Hereafter, the stellar mass is denoted \Ms{}. As a consistency check, the inferred \Ms{} is compared with the best-fit stellar mass (i.e., corresponding to the single template with the minimum $\chi^{2}$), and the median difference is $< 0.01$\,dex with a 1$\sigma$ scatter of 0.09\,dex, similar to both \citetalias{Davidzon2017} and \citetalias{Weaver2023SMF}.

\begin{figure*}
    \centering
    \includegraphics[width=\textwidth]{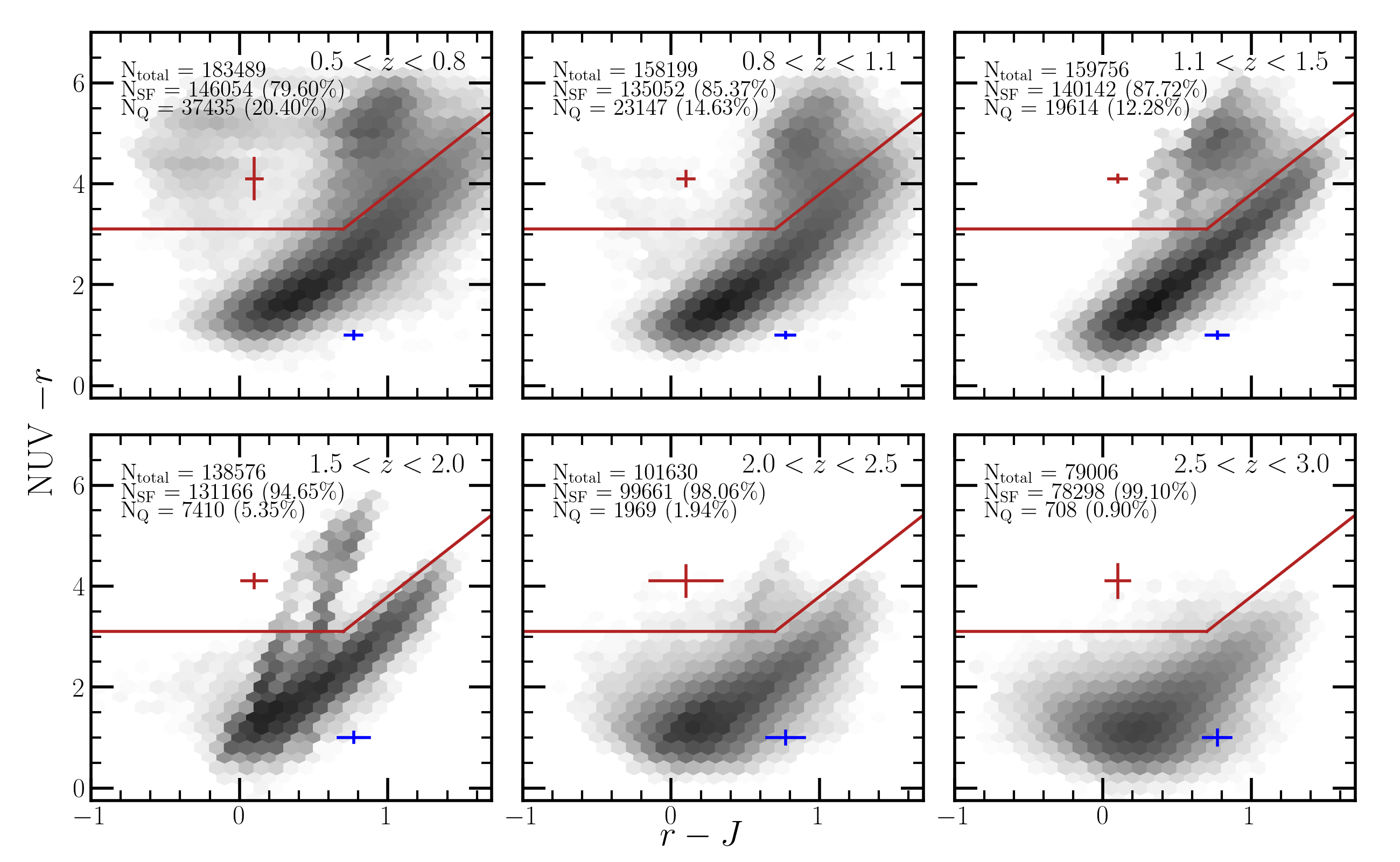}
    \caption{Galaxies are identified as either star-forming or quiescent based on their ${\rm NUV}-r$ and $r-J$ rest-frame colours at $z \leq 3$. Rest-frame colours are measured using \lephare{} following \cite{Ilbert2005}. Representative uncertainties associated with the rest-frame colours of the star-forming (blue) and quiescent (maroon) subsamples are plotted. At $z > 1.5$, the Balmer break is redshifted between two observed bands (HSC $z$ and IRAC \chOne{}), and so the inferred rest-frame colours become increasingly model-dependent. Shading corresponds to logarithmic density.}
    \label{fig:NUVrJ}
\end{figure*}

Stars that are brighter than 17 mag in the \textit{Gaia} $G$ band, according to \textit{Gaia} DR3 \citep{Gaia2022}, are masked in the DAWN survey PL catalogues. Fainter stars are removed through SED fitting.  As alluded to, \lephare{} is capable of fitting star templates to photometric data in addition to galaxy templates. In this work, the same set of stellar templates is utilised as in \cite{Weaver2021} and \citetalias{Weaver2023SMF}. The contribution from \textit{Spitzer}/IRAC provides a strong constraint on the likelihood of stellar contamination, and stars are effectively removed by requiring galaxy candidates to have a smaller $\chi^{2}$ from the best-fit galaxy template compared to the best-fit stellar template. However, the lack of NIR coverage at $z > 4$ poses a challenge for distinguishing some high-$z$ galaxy candidates from brown dwarf stars; see Sect.~\ref{subsubsec:interlopers} for further discussion.

The DAWN survey PL catalogues (EDF-N and EDF-F) includes \num{5195940} sources (\num{4336651} and \num{859289}). Restricting the selection of objects to the full-depth region (column name = `\texttt{FULL\textunderscore DEPTH\textunderscore DR1}') provides \num{3274786} sources (\num{2697776} and \num{577010}). Seeking to include only those sources with the most secure \photoz{}s and \Ms{} estimates, and to further ensure that a reliable distinction between stellar interlopers can be made, every source is further required to have a signal-to-noise ratio (S/N) of at least 3 in each of the HSC $r$, $i$, and $z$ bands as well as in \textit{Spitzer}/IRAC \chOne{} and \chTwo{}. However, in an effort to include intrinsically redder sources at high redshift, the S/N requirement in the HSC $r$ band is dropped for sources at $z = 3.5$ and above. Following \citetalias{Weaver2023SMF}, galaxies that have significantly uncertain redshifts are not included, requiring the same criterion that 68\% of the redshift probability distribution is contained within the interval $z_{\rm phot} \pm 0.5$. Inspection showed that the majority of these objects fall below the limiting stellar mass (see Sect.~\ref{subsec:m_completeness}). Finally, for each redshift bin (see Table~\ref{tab:volumes}), the 95th percentile of the distribution of best-fit SED reduced $\chi^{2}$ is calculated (e.g. 10 at $z\approx$\,1--4, $\sim40$ at $z\gtrsim5$), and those above are removed. This approach yields the same $\chi^{2}$ cut at intermediate redshifts (approximately $1 < z \leq 3$) as \citetalias{Weaver2023SMF}, which used a uniform $\chi^{2} < 10$ cut, but consistently removes the same fraction of objects from each redshift bin. This choice is further discussed in Sect.~\ref{subsec:massive}, but in short, the $\chi^{2} < 10$ of \citetalias{Weaver2023SMF} primarily affects only the highest redshift bin. The final sample considered in this work includes \num{2091740} galaxies total (\num{1758410} and \num{333330}). Note that EDF-F alone provides a sample of galaxies approximately equal to that of \citetalias{Weaver2023SMF}. 

\subsection{\label{subsec:SF_Q}Star-forming vs. quiescent classification}

At any given redshift, the total galaxy SMF calculated over the survey is a sum of individual SMFs of galaxies of different types, for example, star-forming and quiescent (e.g., \citealt{Peng2010}). Thus, the total galaxy SMF is more fully understood by examining its constituent components. Primarily motivated by ease of comparison, this work adopts the classification of star-forming and quiescent galaxies set forth by \cite{Ilbert2013} and used by both \citetalias{Davidzon2017} and \citetalias{Weaver2023SMF}. In short, galaxies are classified as either star-forming or quiescent according to their position in the ${\rm NUV} - r$, $r-J$ diagram. Quiescent galaxies are defined as those satisfying
\begin{equation} \label{eq:NUVrJ}
    M_{\rm NUV} - M_{r} > 3(M_{r} - M_{J}) + 1 {~\mathrm{and}~} M_{\rm NUV} - M_{r} > 3.1.
\end{equation}
The requirement of Eq.~(\ref{eq:NUVrJ}) separates red and blue galaxies, like the classic $UVJ$ requirement of \cite{Williams2009}, but primarily through the difference in $M_{\rm NUV} - M_{r}$ which is comparatively more sensitive to recent star formation \citep{Arnouts2007,Martin2007,Ilbert2013}. Dusty star-forming galaxies, though apparently red in $M_{\rm NUV} - M_{r}$, are distinguished from galaxies with genuinely old stellar populations by the difference in $M_{r} - M_{J}$, where increasing dust attenuation advances galaxies in a direction parallel to the slope of the bounding region (see, e.g., Figure 1 of \citealt{Leja2019}). 

\begin{figure*}
    \centering
    \includegraphics[width=\textwidth]{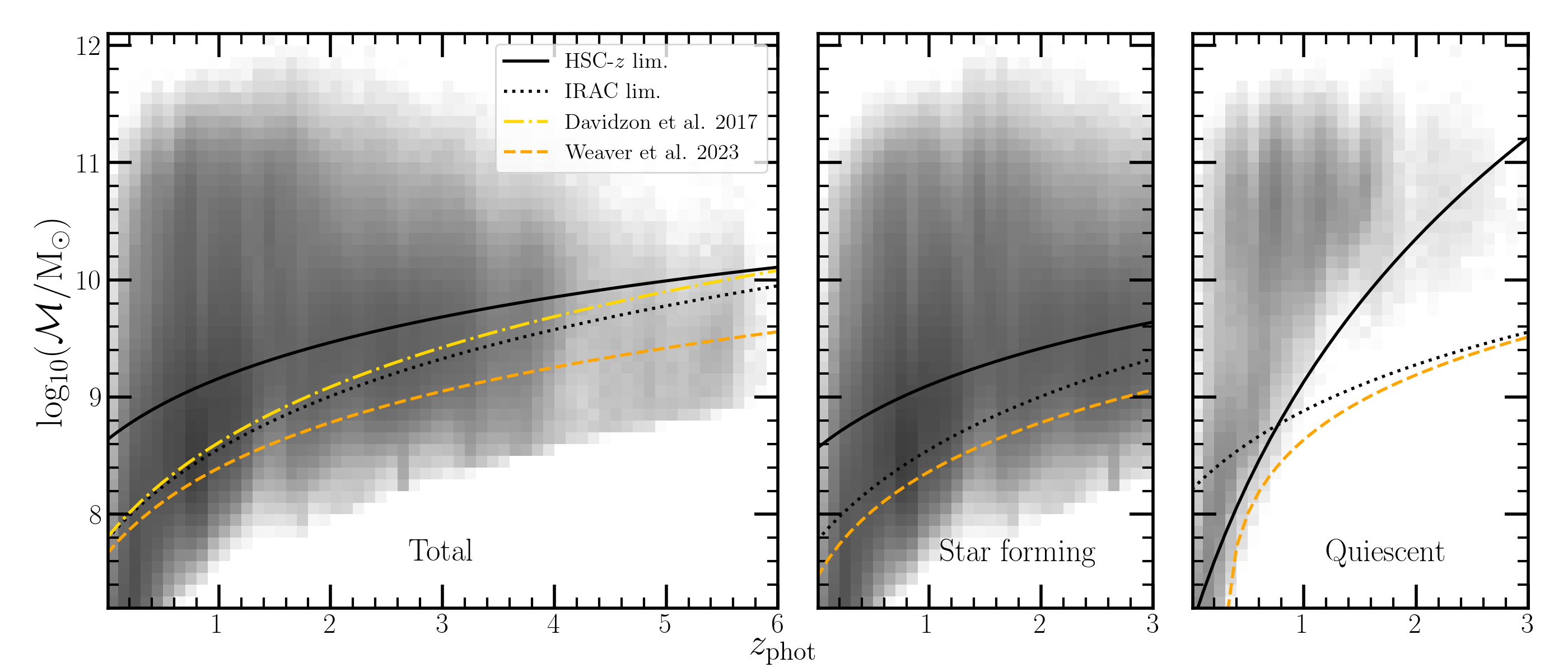}
    \caption{Galaxy stellar mass distribution as a function of redshift. The limiting stellar mass of the total (left), star-forming (middle), and quiescent (right) samples are determined as a function of redshift following \cite{Pozzetti2010} and are shown by the solid black curves. Two estimates are computed based on the limiting magnitude of HSC $z$ (solid) and the IRAC \chOne{} limiting magnitude (dotted), respectively. The more conservative estimate using HSC $z$ is used in the remainder of this work and are presented as Eqs.~(\ref{eq:tot_mcomp})--(\ref{eq:q_mcomp}) for the total, star-forming, and quiescent samples. For comparison, the stellar mass limits of \citetalias{Davidzon2017} and \citetalias{Weaver2023SMF} are also shown. Shading corresponds to logarithmic density.}
    \label{fig:mass_lim}
\end{figure*}

Absolute magnitudes are best constrained when the wavelengths of an observed filter overlaps directly to the corresponding rest-frame wavelengths of the desired absolute magnitude, thus minimising the $k$-correction \citep{Hogg2002}. Without such direct overlap, an extrapolation must be made from the best-fit SED model. In this work, absolute magnitudes are calculated with \lephare{} according to the method described by \cite{Ilbert2005}, where the absolute magnitude in a given filter $\lambda_{\rm abs}$ is related to observed flux in the observed-frame filter nearest to  $\lambda_{\rm abs}(1+z)$ in order to minimise the dependence of the $k$-correction on the assumed galaxy template. However, when the distance between the nearest observed-frame filter and the rest-frame filter is large, the predicted absolute magnitude becomes more reliant on the best-fit SED model. This is the same approach used by \citetalias{Davidzon2017} and \citetalias{Weaver2023SMF}.

The NUV$rJ$ selection applied to DAWN PL is shown in Fig.~\ref{fig:NUVrJ}. Only galaxies above the respective mass completeness limits (see Sect.~\ref{subsec:m_completeness}) are shown. For each population (i.e., star-forming and quiescent), the median photometric error on the rest-frame colour is illustrated by the coloured cross. Here, the median photometric error corresponds to the photometric uncertainties in the observed-frame filters nearest to each of the NUV$rJ$ for the given redshift bin, added in quadrature. As noted by \citetalias{Weaver2023SMF}, the median photometric uncertainty is more representative of the faint galaxies that dominate in abundance compared to the brighter, more massive systems. Note that the uncertainty of the best-fit SED is not propagated to the photometric uncertainty displayed in Fig.~\ref{fig:NUVrJ}.  

The decrease in the number of quiescent galaxies with increasing redshift apparent from Fig.~\ref{fig:NUVrJ} is a consequence of both well-understood aspects of galaxy evolution \citep{Ilbert2013} and observational effects.  Considering the latter, the selection function described in Sect~\ref{subsec:selection} determines the ability to detect quiescent galaxies from the DAWN survey PL images. After redshift $z\sim 1.6$, the Balmer breaks drops out of the HSC $z$ band, and consequently galaxies above $z\sim1.6$ are detected on the basis of increasingly blue rest-frame light. 

These results broadly agree with \citetalias{Weaver2023SMF}. Namely, the fraction of massive quiescent galaxies is similar, up to $z\sim$\,2--3. However, the DAWN PL catalog is optically-selected leading to a dearth of detected quiescent galaxies at $z>3$. While expected, it limits the exploration of quiescent (and generally red) objects to $z<3$. Consequently, the total sample at $z>3$ contains only blue, star-forming galaxies without significant dust attenuation. Future work incorporating \Euclid's near-infrared bands will complete this picture.

\subsection{\label{subsec:m_completeness}Galaxy stellar mass limit}

Measuring the galaxy SMF requires identifying the minimum stellar mass at each redshift above which galaxies are detected. \cite{Pozzetti2010} presented a method that is often used \citep{Ilbert2010,Ilbert2013,Muzzin2013,Tomczak2017,Davidzon2017,Stefanon2021,Weaver2023SMF} to empirically measure the stellar mass limit of a survey based on the measured stellar masses of detected galaxies and the limiting flux of the survey. The method consists of converting the flux limit of a given survey to a stellar mass limit by first inferring a mass-to-light ratio, applying a transformation to the measured stellar masses given the difference between their measured flux and the limiting flux, and using the rescaled stellar masses to describe the completeness limit.  Here, the implementation of the \cite{Pozzetti2010} method by \citetalias{Weaver2023SMF} is followed, considering only the 99\% best-fit objects by $\chi^2$ from Sect.~\ref{subsec:selection}. Subsequently, the galaxies with the 30\% lowest stellar masses are selected as being representative of those near to the stellar mass limit. The stellar masses of this subsample are rescaled following \citet{Pozzetti2010}:
\begin{equation} \label{eq:resc}
    \logten\left(\mathcal{M}_{\text {resc }}/{\rm M}_{\odot}\right)=\logten(\mathcal{M}/{\rm M}_{\odot})+0.4\left(m_{z}-26.9\right),
\end{equation}
where $m_{z}$ is the apparent magnitude in the HSC $z$ band and 26.9 is the approximate $3\sigma$ limiting magnitude in the combined HSC $r+i+z$ images \citep{Zalesky2024}. Galaxies below $\logten(\mathcal{M}/{\rm M}_{\odot}) < 8$ are not expected to be detectable and are therefore not considered in calculating the mass completeness. The galaxies with rescaled stellar masses are binned according to their redshift, with a spacing of $\Delta z = 0.2$. In a further effort to be conservative, the 95th percentile of the rescaled masses ($\mathcal{M}_{\text {resc}}$) is used to identify the stellar mass limit in each bin, in contrast to the 90th percentile used by \citetalias{Weaver2023SMF}. Finally, an analytical function is fit to the binned mass completeness limits of the form $\mathcal{M}_{z} = A(1+z)^{B}$, following \citetalias{Davidzon2017} rather than the second-order expansion in $(1+z)$ used by \citetalias{Weaver2023SMF}, as the former appears to provide a better description of at low $z$. 

The stellar mass limit is computed for the total sample, star-forming sample, and quiescent sample independently according to the method above. The results are as follows
\begin{equation} \label{eq:tot_mcomp}
    {\mathrm{Total:}~} \mathcal{M}_{\rm lim}/\Msol = 4.31\times10^{8}(1+z)^{1.74},
\end{equation}
\begin{equation} \label{eq:sf_mcomp}
    {\mathrm{Star-forming:}~} \mathcal{M}_{\rm lim}/\Msol = 3.65\times10^{8}(1+z)^{1.79},
\end{equation}
\begin{equation} \label{eq:q_mcomp}
    {\mathrm{Quiescent:}~} \mathcal{M}_{\rm lim}/\Msol = 1.13\times10^{7}(1+z)^{6.91}.
\end{equation}

As can be inferred from Eqs.~(\ref{eq:tot_mcomp},~\ref{eq:sf_mcomp}), the stellar mass limit of star-forming galaxies is similar to that of the total sample, differing only at the highest and lowest redshifts. Meanwhile, the stellar mass limit of quiescent galaxies is a significantly stronger function of redshift. The results are depicted in Fig.~\ref{fig:mass_lim} over a two-dimensional histogram of redshift and stellar mass, \Ms{}. Because there are so few quiescent galaxies detected at $z > 2$, the low-redshift galaxies have the dominant weight in the fit of the $\mathcal{M}_{z} = A(1+z)^{B}$ function. In practice, quiescent galaxies below the total stellar mass limit (left panel of Fig.~\ref{fig:mass_lim}) are not considered, thereby ensuring consistency across the samples. The stellar mass limits of \citetalias{Davidzon2017} and \citetalias{Weaver2023SMF} are also shown. By comparison, the stellar mass limits of DAWN PL, computed according to the HSC $z$ band, are shallower. This is not surprising, given that both \citetalias{Davidzon2017} and \citetalias{Weaver2023SMF} utilise a detection image incorporating NIR data. Consequently, DAWN PL is not ideal for describing the galaxy SMF to the lowest stellar masses. Instead, deeper programs (e.g., \citealt{Furtak2021}) are preferable to study the low-mass end. Nonetheless, the stellar mass limit reached by DAWN PL is sufficient to study massive galaxies $\mathcal{M} > 10^{10}$ M$_{\odot}$ at all redshifts. 

It may be reasonable to use the \chOne{} or \chTwo{} apparent magnitude and the corresponding limiting magnitude in the rescaling equation (Eq.~\ref{eq:resc}) to determine the stellar mass limits of DAWN PL. Recall that a 3$\sigma$ detection in the \textit{Spitzer}/IRAC \chOne{} and \chTwo{} bands is required (Sect.~\ref{subsec:selection}) for a source to be considered. The result of using \chOne{} to derive the limiting stellar mass limits is included in Fig.~\ref{fig:mass_lim}. This calculation provides stellar mass limits that are notably deeper than those obtained when using HSC $z$, and similar to that which is obtained by \citetalias{Davidzon2017}. However, to avoid any pitfalls associated with extrapolating from the selection function in the HSC $r+i+z$ bands to a stellar mass limit derived with \chOne{}, the conservative option of the HSC $z$-derived stellar mass limits is used. 

It should be understood that the method of \cite{Pozzetti2010} provides an estimate of the limiting stellar mass above which a galaxy is detected given the selection function. Consequently, the stellar mass limits quantified by Eqs.~(\ref{eq:tot_mcomp})--(\ref{eq:q_mcomp}) may be overly optimistic for galaxies that are difficult to detect, given the selection function defined by the HSC $r+i+z$ bands. Such galaxies primarily include intrinsically red objects, such as significantly dust-attenuated star-forming galaxies and quiescent systems at $z >1.5$. 

For now, it is assumed that the selection function and the significant depth of the detection image is sufficient to identify a substantial fraction of galaxies above the stellar mass limit to $z\sim 6$. To support this hypothesis, consider that the 3$\sigma$ limiting magnitude in the UltraVISTA \citep{McCracken2012} near-IR $K_{s}$ band of the COSMOS2020 catalogue is $\sim 25$ (and shallower for the COSMOS2015 catalogue used by \citetalias{Davidzon2017}, between 24--24.7 mag depending on the area; see \citealt{Laigle2016}). Selecting galaxies brighter than $K_{s} \leq 25$ that also satisfy the selection criteria for the mass function specified by \citetalias{Weaver2023SMF} indicates that galaxies from $3 < z \leq 6$ span a range in colour $z - K_{s}$ where the 90th percentile is approximately 2.5\,mag at $z\sim3$, dropping to $\sim 2$\,mag at $z\sim4$, and continuing to fall $< 1$ by $z\sim 6$. Accordingly, a significant majority of red galaxies according to $z - K_{s}$ should be detected by the combined depth of the HSC $r+i+z$ image that reaches at least 26.9 mag in the HSC $z$. Further, the galaxy SMFs presented in Sect.~\ref{sc:results} appear to support this hypothesis, at least to a similar completeness achieved by \citetalias{Davidzon2017} to $z\sim5$. It is possible that the stellar mass limit is underestimated in the highest-redshift bins considered in this work at $z > 4.5$. However, the significant volume of DAWN PL provides a unique opportunity to characterize the abundance of massive galaxies at this epoch, and so the redshift bins above $z > 4.5$ are included, despite the possibility of incompleteness.

\subsection{\label{subsec:errors}Uncertainty and bias estimation}

\begin{figure*}
    \centering
    \includegraphics[width=\textwidth]{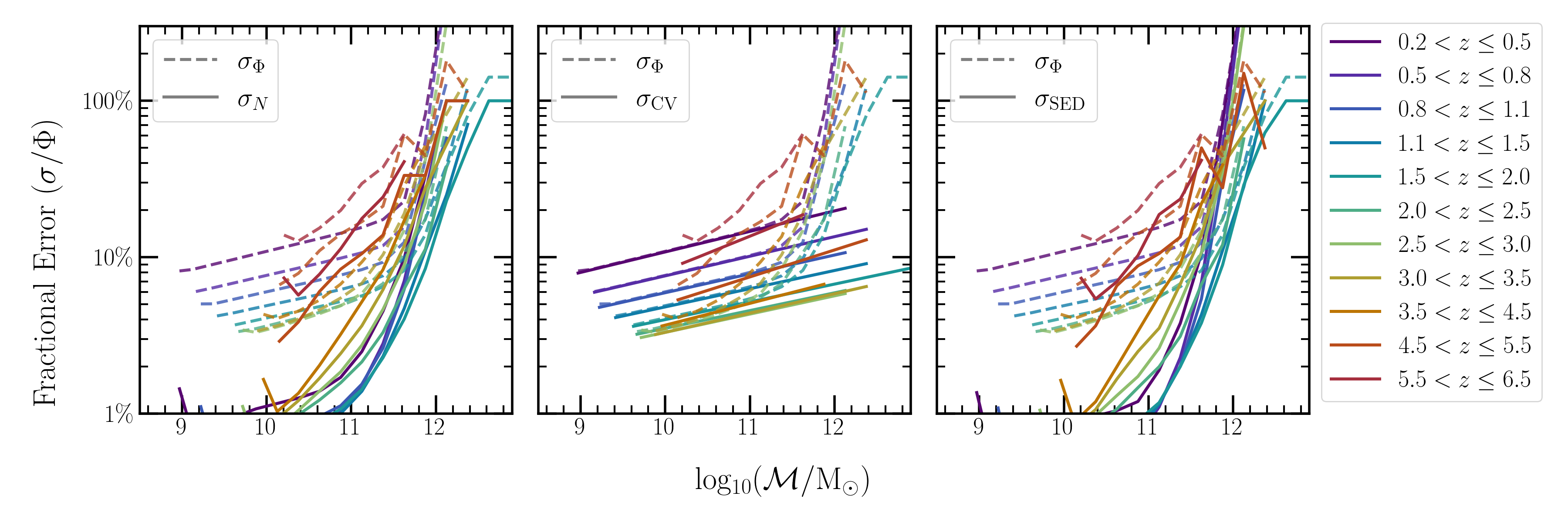}
    \caption{The total uncertainty of the number density in each \Ms{} bin, $\sigma_{\Phi}$, has three components: the Poisson uncertainty $\sigma_{N}$ (left), cosmic variance $\sigma_{\rm CV}$ (middle), and uncertainty arising from SED fitting $\sigma_{\rm SED}$ (right). The total uncertainty, shown by the dashed lines in each panel, is the quadrature sum of the three components, shown by the solid lines in each panel.}
    \label{fig:sigma}
\end{figure*}

The total uncertainty of the observed galaxy SMF has at least three general components. As for any statistical characterisation that involves counting $N$ samples drawn from a parent population, the galaxy SMF is affected by Poisson uncertainty ($\sigma_{N}$; Sect.~\ref{subsubsec:poisson}). An additional source of uncertainty that must be accounted for is the cosmological fluctuation of galaxy properties (in particular stellar mass) on the physical scales observed by the survey, an effect commonly referred to as \enquote{cosmic variance} ($\sigma_{\mathrm{CV}}$; Sect.~\ref{subsubsec:cvar}). Lastly, the galaxy SMF is impacted by the choices and assumptions made during the spectral energy distribution modelling insofar that they impact the photometric redshifts and galaxy stellar masses ($\sigma_{\mathrm{SED}}$; Sect.~\ref{subsubsec:sed}). The total uncertainty is the quadrature added sum of the three main components, i.e., $\sigma_{\mathrm{\Phi}} = (\sigma_{N}^{2} + \sigma_{\mathrm{cv}}^{2} + \sigma_{\mathrm{SED}}^{2})^{1/2}$. 

\subsubsection{\label{subsubsec:poisson}Poisson}

Poisson uncertainty is ubiquitous for all measurements seeking to estimate the abundance of a population by drawing independent random samples. For a given bin, the Poisson uncertainty is simply the square root of the number of galaxies, $N_{\mathrm{bin}}$, in the bin per volume per mass bin width, i.e., $\sigma_{N,{\mathrm{bin}}} = \sqrt{N_{\mathrm{bin}}}\,/\,V\,/\,\delta\mathcal{M}$. In this work, the Poisson uncertainty of each bin is estimated separately for the total, star-forming, and quiescent samples, in each \Ms{} bin. As previously noted, the increase in area of DAWN PL compared to the next most similar field, COSMOS, is approximately an order of magnitude. As such, the Poisson uncertainty is improved by a factor of $\sim$3 compared to \citetalias{Weaver2023SMF}, enabling a statistically significant characterisation of galaxies with $\mathcal{M} \gtrsim 10^{11}$ M$_{\odot}$ at all redshifts. The resulting Poisson uncertainty for each redshift interval and mass bin is presented in the left panel of Fig.~\ref{fig:sigma}. 

For a given volume, the Poisson upper limit may also be determined by considering the $N_{\mathrm{bin}}$ = 0 case. Following \cite{Gehrels1986}, the 1$\sigma$ upper limit for an $N_{\mathrm{bin}}$ = 0 event in a volume $V$ is 1.841/$V$, while the 3$\sigma$ upper limit is 6.61/$V$ (see also \citealt{Ebeling2004}). The 1$\sigma$ upper limit is illustrated in many of the figures shown in Sect.~\ref{sc:results} by a downward pointing arrow.

\subsubsection{\label{subsubsec:cvar}Cosmic variance}

In one aspect, the Poisson uncertainty described above (Sect.~\ref{subsubsec:poisson}) represents the minimum error on the number of galaxies expected in a given bin. However, the Poisson uncertainty assumes that galaxies properties, including the stellar masses and their spatial distributions, are independent. In reality, galaxy properties are known to be spatially correlated  \citep{Kauffmann2004, Wu_Jespersen_env_2023, Wu2024_env}, and moreover, environments differ on cosmological scales, in part due to large-scale fluctuations in the cosmic density field that have grown since the Big Bang \citep{Springel2005}. From observations, it has been shown the galaxies cluster more strongly than their dark matter halos, and from simulations, it is predicted that such galaxy \enquote{bias} increases across redshift \citep{Moster2010}. Importantly, more massive systems are more biased than their lower mass counterparts. Since it is impossible to know a priori where a given field is sitting in the global large-scale structure fluctuation, and smaller areas are naturally affected by both long- and short-wavelength fluctuations, galaxy surveys with small areas are more strongly affected by field-to-field variance, referred to as \enquote{cosmic variance}, $\sigma_{\mathrm{CV}}$, than larger surveys of equal depth. See \cite{Moster2010,Moster2011} for a review.

The estimation of cosmic variance used in this work differs from \citetalias{Weaver2023SMF}, which relied on cosmic variance calculations provided by \cite{Moster2011} that were extrapolated by \cite{Steinhardt2021} to higher redshifts and greater stellar masses. Instead, the method of \cite{Jesperson2024} is followed, wherein the authors demonstrated a method to more robustly derive cosmic variance estimates for massive galaxies at a range of redshifts using the \texttt{UniverseMachine} simulations \citep{Behroozi2019}. A brief summary follows, although the reader is referred to \cite{Jesperson2024} for details. Each survey field of DAWN PL, i.e., EDF-N and EDF-F, is treated independently, as they are highly separated and thus not in causal contact. Number counts of galaxies are sampled from a survey volume corresponding to the areas of the two fields, 8.42 deg$^{2}$ for EDF-N and 1.71 deg$^{2}$ for EDF-F. In addition, the same redshift and mass bins defined by this work (see Table~\ref{tab:volumes} and Sect.~\ref{sc:results}) are imposed on the selection. The variation in the number counts that is in excess of the Poisson uncertainty and due to clustering of galaxies is modelled by a power law in stellar mass with a redshift-dependent normalisation and slope. The calibration of the final model of cosmic variance also includes error terms identified by \cite{Jesperson2024}, which correct the cosmic variance estimates for the impact of the skew of the underlying distribution of number counts in single bins.\footnote{This is important since the highly skewed number count distributions in the limit of high cosmic variance bias the estimates of the cosmic variance, artificially inflating them.} The total cosmic variance $\sigma_{\mathrm{CV, tot}}$ across the combination of EDF-N and EDF-F, is given by
\begin{equation}
    \sigma_{\mathrm{CV,~TOT}} = [(\sigma_{\mathrm{CV,~N}})^{-2} + (\sigma_{\mathrm{CV,~F}})^{-2}]^{-1/2},
\end{equation}
where $\sigma_{\mathrm{CV,~N}}$ is the cosmic variance corresponding to the volume of EDF-N and $\sigma_{\mathrm{CV,~F}}$ is the cosmic variance corresponding to the volume of EDF-F.

The $\sigma_{\mathrm{CV}}$ estimates for the combination of EDF-N is presented in the centre panel of Fig.~\ref{fig:sigma}. Given the significant volume of DAWN PL and the improvement of the method of \cite{Jesperson2024} over \cite{Moster2011}, the impact due to cosmic variance is significantly smaller in comparison to \citetalias{Davidzon2017} and \citetalias{Weaver2023SMF}. For example, the uncertainty due to cosmic variance is at least a factor of 5 times greater in \citetalias{Weaver2023SMF} for galaxies of $\mathcal{M} \sim 10^{10.5}$ M$_{\odot}$ at $z\sim5$.

\subsubsection{\label{subsubsec:sed}SED fitting}

As described above, galaxy properties are measured using \lephare{} closely following the procedure of \citetalias{Weaver2023SMF} (as well as \citealt{Ilbert2013} and \citetalias{Davidzon2017}, though to a less degree). SED modelling assumptions (e.g. parametric star-formation histories)  are the same as detailed in \citet{Ilbert2013}. \citetalias{Zalesky2024} validated the \photoz{} measurements for DAWN PL by comparing with \num{3300} high-quality spectroscopic redshifts matched from GOODS-S \citep{Garilli2021,Kodra2023} that sample $0 < z \leq 5.5$ (see Figs. 8 and 9 of \citetalias{Zalesky2024}). In addition, the authors modified the photometric errors of COSMOS2020 catalogue to match those of the DAWN survey PL catalogue and used \lephare{} to re-measure \photoz{} and \Ms{} utilising only the filters available to DAWN PL. The results demonstrated that more than 80\% of all galaxies had consistent redshifts and stellar masses. The majority of the those that are not consistent are too faint to satisfy the selection criteria used in the present work (see Fig. 11, and Appendix C of  \citetalias{Zalesky2024}). As such, the \photoz{} and \Ms{} measurements provided by DAWN PL are believed to be robust, especially in view of the requirements applied in Sect.~\ref{subsec:selection}. Nonetheless, the impact of the uncertainties associated with modelling the SEDs must be accounted for and propagated through to the resulting uncertainty on the galaxy SMF.

Under the configuration applied across \cite{Ilbert2013}, \citetalias{Davidzon2017}, and \citetalias{Weaver2023SMF}, as well as here, \photoz{} and \Ms{} estimates are obtained separately, utilising a different template set for each calculation (see \citealt{Weaver2021} and  \citetalias{Zalesky2024} for details). More specifically, the physical properties of galaxies are inferred assuming a fixed redshift, $z \equiv z_{\mathrm{phot}}$. At the time of writing, \lephare{} does not support allowing the redshift to vary using one template set while measuring physical properties using another template set. Consequently, the uncertainties on \Ms{} provided in the DAWN survey PL catalogues do not include the covariance between \photoz{} and \Ms{}. Facing a similar situation, \citetalias{Weaver2023SMF} relied on the extreme \photoz{} precision achieved by the COSMOS2020 catalogue and suggested that their $\sigma_{\mathrm{SED}}$ measurements should be considered lower limits. By contrast, \citetalias{Davidzon2017} performed a Markov Chain Monte Carlo (MCMC) simulation of their entire catalogue by varying the photometric measurements within their uncertainties measuring a new \photoz{} and \Ms{} one thousand times. Similar approaches have been used elsewhere (\citetalias{Grazian2015}). However, \citetalias{Davidzon2017} used a sample of galaxies that is more than an order of magnitude smaller than the DAWN PL sample, and applying their approach to DAWN PL is not computationally tractable at present.\footnote{Utilising a computing facility with 200 total cores and over $500\,{\rm Gb}$ of RAM, performing an MCMC simulation for 100 DAWN survey PL catalogues would require 1--2 years.} 

A zeroth-order correction to the impact of the stellar mass uncertainties on the measured galaxy SMF, addressing the covariance between redshift and mass, is obtained as follows. Recall that galaxies are required to have 68\% of their redshift probability distribution contained within the interval $z_{\rm phot} \pm 0.5$. In addition, \citetalias{Zalesky2024} demonstrated with a high-quality sample of \num{3300} spectroscopic redshifts matched from GOODS-S \citep{Garilli2021,Kodra2023}, that 68\% of all galaxies with spectroscopic redshifts had \photoz{} measurements within $1\sigma$ of their spectroscopic value. As such, \photoz{} probability distributions for galaxies used here are narrow and appear well calibrated. Therefore, instead of determining new redshifts for each galaxy by re-running \lephare{}, \num{1000} redshifts for each galaxy are drawn from the respective redshift PDFs. To zeroth order, the stellar mass inferred for a galaxy at $z_{1}+\delta z$ compared to its inferred stellar mass at redshift $z_{1}$ will differ by a re-scaling proportional to $\delta z$, similar to their absolute magnitudes. This is because at small $\delta z$, the inferred galaxy type will not change and therefore the inferred mass-to-light ratio will be approximately constant. Assuming the $k$-correction is small, the difference in absolute magnitudes (d$M$) between two galaxies with the same apparent magnitude but at redshifts $z_{1}$ and $z_{2} = z_{1}+\delta z$ is
\begin{equation} \label{eq:abs_mag_dif}
    {\mathrm{d}}M = 2.5\logten\left(\frac{1+z_{1}}{1+z_{2}}\right) - 5\logten\left(\frac{d_{\rm L}(z_{1})}{d_{\rm L}(z_{2})}\right),
\end{equation}
where $d_{\rm L}(z)$ is the luminosity distance to a galaxy at redshift $z$. 

At each newly drawn redshift, the galaxy stellar mass, \Ms{}, is scaled according to Eq.~(\ref{eq:abs_mag_dif}). It is acknowledged that a range of mass-to-light ratios are allowed for a particular galaxy given its photometric uncertainties, and so an additional perturbation to the rescaling factor is applied in proportion to the relative flux errors. Finally, a final perturbation that is proportional to the difference in mass that would be obtained from a random draw from the \Ms{} probability distribution function from \lephare{} is applied. Note that because $\delta z$ is small, the original stellar mass PDF should still be approximately representative, given that it is largely driven by photometric uncertainties \citep{Ilbert2013,Davidzon2017,Weaver2023SMF}. However, it too must be scaled according to Eq.~(\ref{eq:abs_mag_dif}) to account for the newly assigned redshift. The final result of this exercise is \num{1000} independent realisations of galaxy \photoz{} and \Ms{} estimations, where each \Ms{} value has been adjusted to the newly assumed redshift while accounting for a variety of possible mass-to-light ratios and further scattered due to shape of probability distribution function of stellar mass. This includes 2.3 trillion total measurements.

The simulated \photoz{} and \Ms{} measurements are used to construct \num{1000} realisations of the galaxy SMF, using the same redshift and stellar mass bins as the primary analysis (see Sect.~\ref{sc:results}). The uncertainty due to SED fitting, $\sigma_{\mathrm{SED}}$, is then quantified for each redshift bin by measuring the $1\sigma$ variation in the number density at every \Ms{} bin relative to the median number density. The result is shown in the right panel of Fig.~\ref{fig:sigma}. The scale of $\sigma_{\mathrm{SED}}$ follows a similar trend compared to $\sigma_{N}$, reflecting the abundance of low-mass galaxies that have a number density that does not change appreciably due to SED fitting. By contrast, massive galaxies with $\mathcal{M} >10^{11.5}$ M$_{\odot}$ have uncertainties in their abundance of order unity. 

Although effort has been taken to account for the covariance between stellar mass and redshift, the estimated $\sigma_{\mathrm{SED}}$ are expected to be valid in the regime of small variations in the assumed redshift. The analysis has not accounted for catastrophic outliers in redshifts, or the choice of templates used by \lephare{}, for example, given that systematic errors are not easily combined with random errors. Improving the computational time required to run current SED-fitting codes is imperative for future datasets even larger than DAWN PL.

\subsubsection{\label{subsubsec:validation} Validation}

Several requirements have been established above in order to obtain a sample of galaxies with only the most robust estimates of their properties. With trustworthy estimates of galaxy properties in place, it is possible to characterise the evolution of the galaxy SMF across cosmic time. Interpreting the observed galaxy SMF is strengthened by an understanding of the success of these efforts, or lack thereof, and any resulting systematics. A stronger understanding of these systematics is also vital to infer the intrinsic galaxy SMF with confidence (Sect.~\ref{subsec:modeling}). To this end, Appendix~\ref{app:validation} provides a detailed discussion of a series of validation tests that have been performed on the DAWN PL data. The validation tests and their findings are summarised below. 

\begin{itemize}

    \item Re-fitting COSMOS2020 with only DAWN PL filters: COSMOS2020 \citep{Weaver2021} includes photometry in all of the filters used by DAWN PL, but with deeper imaging.  \citetalias{Zalesky2024} demonstrated a test wherein \photoz{}s and stellar masses were recomputed for all of COSMOS2020, utilising only the filters available to DAWN PL and with flux uncertainties scaled to match DAWN PL. Galaxies are further selected from a detection imaging comprising the same noise properties as that of DAWN PL. Using measurements entirely from the re-fitting, the COSMOS2020 galaxy SMF is measured and compared with \citetalias{Weaver2023SMF}, finding excellent agreement with one primary exception. Galaxies at $z \sim$ 1.1--1.5 are found to have degenerate template solutions, which given the decreasing sample size with redshift, leads to an underestimation of low-mass galaxies at these redshifts and an overestimation of massive galaxies at higher redshifts $1.5 < z \leq 2.5$ (see App.~\ref{subapp:refit_cosmos} and Sect.~\ref{subsec:modeling}). There is no significant bias detected above $z = 2.5$ except for what is predicted by the difference in selection functions. The agreement between \citetalias{Weaver2023SMF} with the SMF derived using only the DAWN PL filters lends confidence to the results presented here. 
    \item Validating galaxy properties through machine learning: Following \cite{Chartab2023}, a random forest regressor model is trained on COSMOS2020 restricted to the DAWN PL bands to predict galaxy properties from the DAWN PL photometry. The performance of both the resulting \photoz{} and stellar mass estimates are consistent with those obtained from SED fitting using \lephare{} with mild improvement from test galaxies in COSMOS2020. The galaxy SMF may be measured using the galaxy properties predicted from the DAWN PL catalogues, and the result improves the characterisation of galaxies at $1.1 < z \leq 2.5$. However, the conservative nature of the random forest regressor prohibits predictions of galaxy properties outside its training data, and therefore its ability to accurately determine the properties of galaxies with extreme values may be limited. Galaxies with properties that disagree between those obtained from the random forest and from \lephare{} likely include discoveries that would be missed by a random forest.

\end{itemize}

\section{\label{sc:formalism}Galaxy SMF formalism}

Below, the mathematical formalisms for inferring the intrinsic galaxy SMF used in this work are described. 

\subsection{\label{subsec:vmax}Consideration of volume}

Compared to their fainter counterparts, bright galaxies are more rare at every redshift. However, in a flux-limited survey, intrinsically faint galaxies are comparatively more difficult to observe at every redshift and so can appear underrepresented. A correction to this \enquote{Malmquist} bias \citep{Malmquist1922,Malmquist1925} was presented by \cite{Schmidt1968} and is now commonly employed in galaxy demographic studies, including those targeting the galaxy luminosity function and the galaxy SMF due to its simplicity. This correction is referred to as the 1/$V_{\rm max}$ correction and prescribes that every galaxy is weighted by the maximum volume (i.e., $V_{\rm max}$) in which it could be observed. In this work, galaxies are first binned according to their redshift and binned again according to their \Ms{}. Following \cite{Schmidt1968}, every galaxy ($i$) is weighted by
\begin{equation}
    V_{\mathrm{max}, i}=\frac{4 \pi}{3} \frac{\Omega_{\mathrm{survey}}}{\Omega_{\mathrm{sky}}}\left[d^3_{\rm c}\left(z_{{\rm high},i}\right)-d^3_{\rm c}\left(z_{{\rm low},i}\right)\right],
\end{equation}
where $\Omega_{\mathrm{survey}}$ is the solid angle spanned by the survey (in square degrees, 10.12 deg$^{2}$), $\Omega_{\mathrm{sky}}$ is the solid angle of the entire sky, i.e., \num{41 253} deg$^{2}$, and $d_{\rm c}$ is the co-moving distance. The maximum volume a within which a galaxy could be observed is bounded on the low end ($z_{\rm low}$) by the lower edge of the redshift bin and on the high end ($z_{\rm high}$) by the maximum redshift at which the galaxy could be observed without falling below the detection limit. For bright galaxies, the high end is effectively the upper edge of the redshift bin. The redshift bins used throughout match those used by \citetalias{Davidzon2017} and \citetalias{Weaver2023SMF} and are provided in Table~\ref{tab:volumes}.

Hereafter, number densities in each redshift bin and mass show are reported after having applied the $V_{\rm max}$ corrections. It is noted that additional methods for characterising the stellar mass function are explored in the literature (see, e.g., \citealt{Weigel2016} and citations therein). However, discrepancies in comparing the results do not appear strong enough to warrant departing from the 1/$V_{\rm max}$ method (\citetalias{Davidzon2017}). 

\subsection{\label{subsec:schechter}The Schechter function}

\begin{figure*}
    \centering
    \includegraphics[width=0.8\textwidth]{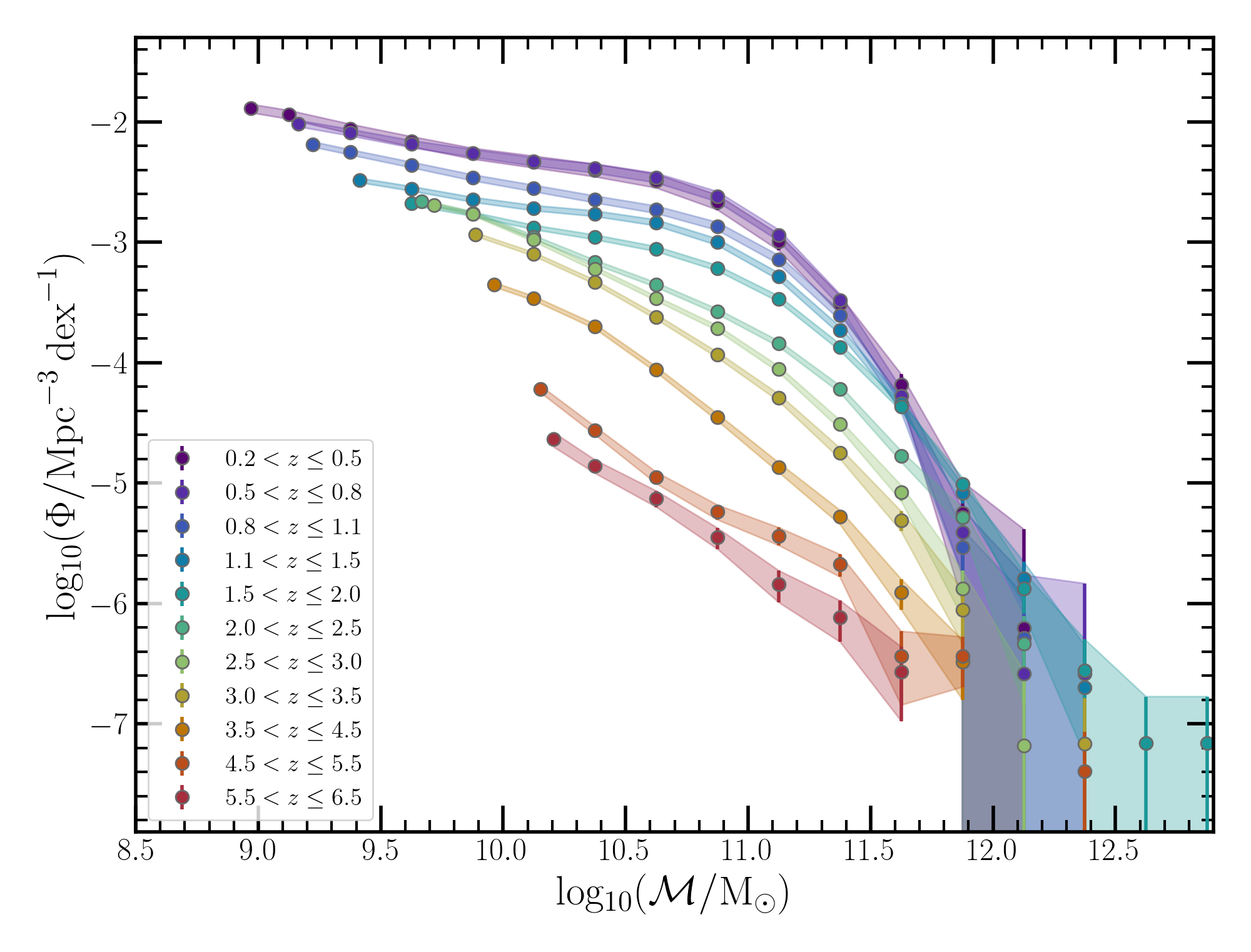}
    \caption{Evolution of the observed galaxy stellar mass function of the total sample across $0.2 < z \leq 6.5$. The \Ms{} bins are uniform except for the first bin, which is extended to the respective stellar mass limit. The \Ms{} bins below the stellar mass limit (Eq.~\ref{eq:tot_mcomp}) are not shown. Uncertainties are indicated by the solid error bars and shaded regions.}
    \label{fig:smf_totobs}
\end{figure*}

The number density of galaxies as a function of stellar mass (and luminosity) are frequently reported as following an analytical characterisation first introduced by \cite{Schechter1976} and since known as the \enquote{Schechter} function. The Schechter function can be written to describe the specific number density of galaxies as a function of stellar mass, $\Phi(\mathcal{M})$, by
\begin{equation}
    \Phi(\mathcal{M})\;\diff \mathcal{M}=\Phi^{\star}\left(\frac{\mathcal{M}}{\mathcal{M}^{\star}}\right)^\alpha \exp \left(-\frac{\mathcal{M}}{\mathcal{M}^{\star}}\right) \frac{\diff \mathcal{M}}{\mathcal{M}^{\star}}.
\end{equation}
Conceptually, the Schechter function describes the number density as a power law with slope $\alpha$ for galaxies below a characteristic mass, $\mathcal{M}^{\star}$ followed by an exponential decline. Both components are scaled by an overall normalisation $\Phi^{\star}$. The characteristic mass $\mathcal{M}^{\star}$ defines the point at which the Schechter function \enquote{turns over} and is sometimes referred to as the \enquote{knee} of the galaxy SMF. The evolution of the Schechter parameters across cosmic time is a topic of debate in the literature.

Although the Schechter function is predominantly used as an empirical description, \cite{Peng2010} presented a theoretical framework that gives rise to a Schechter function in the observed number counts. Further, \cite{Peng2010} predicted and demonstrated that depending on their types and environments, the distribution of galaxies as a function of their stellar mass may be described by a \enquote{double} Schechter function
\begin{equation}
    \Phi(\mathcal{M})\; \diff \mathcal{M}=\left[\Phi_1^{\star}\left(\frac{\mathcal{M}}{\mathcal{M}^{\star}}\right)^{\alpha_1}+\Phi_2^{\star}\left(\frac{\mathcal{M}}{\mathcal{M}^{\star}}\right)^{\alpha_2}\right] \exp \left(-\frac{\mathcal{M}}{\mathcal{M}^{\star}}\right) \frac{\diff \mathcal{M}}{\mathcal{M}^{\star}}.
\end{equation}
The double-Schechter function adds a second power-law component with its own normalisation $\Phi_2^{\star}$ and slope ${\alpha_2}$, though each power-law term is joined by the same characteristic mass ${\mathcal{M}^{\star}}$ above which there is an exponential decline. 

The physical meaning and the validity of the Schechter function (and the double-Schechter function), insofar that it describes the observed galaxy SMF in this work, is discussed further in Sect.~\ref{sc:results} and Sect.~\ref{sc:discussion}. Indeed, additional parametric descriptions have been suggested in the literature, for example, \citetalias{Davidzon2017} postulated that the SMF of galaxies at $z \sim 5$ is a power law. 

\begin{table} \centering
\footnotesize
\caption{Total co-moving volume, stellar mass limit (for the total sample), and number of galaxies with stellar masses above the stellar mass limit for each redshift bin.}
\begin{tabular}{lrrr}
\hline\hline
Redshift & Volume  & $\mathcal{M}_{\rm lim}$ & $N_{\rm gal}$ \\
interval & $10^{6}$ Mpc$^{3}$ & $\logten(\mathcal{M}/{\rm M}_{\odot})$ & ($\mathcal{M} >\mathcal{M}_{\rm lim}$) \\
\hline
$0.2 < z \leq 0.5$ & 6.37   & 8.94 & \num{81111} \\
$0.5 < z \leq 0.8$ & 15.35 & 9.08 & \num{164318} \\
$0.8 < z \leq 1.1$ & 23.35 & 9.20 & \num{140565} \\
$1.1 < z \leq 1.5$ & 39.84 & 9.33 & \num{140558} \\
$1.5 < z \leq 2.0$ & 57.48 & 9.47 & \num{122640} \\
$2.0 < z \leq 2.5$ & 60.61 & 9.58 & \num{89622} \\
$2.5 < z \leq 3.0$ & 60.51 & 9.68 & \num{68940} \\
$3.0 < z \leq 3.5$ & 58.80 & 9.77 & \num{40216} \\
$3.5 < z \leq 4.5$ & 109.89  & 9.93 & \num{22477} \\
$4.5 < z \leq 5.5$ & 98.76 & 10.05 & \num{2182} \\
$5.5 < z \leq 6.5$ & 88.36 & 10.16 & \num{219} \\
\hline
\end{tabular}
\label{tab:volumes}
\end{table}

\subsubsection{\label{subsubsec:eddington}Eddington bias}

Any observed distribution function that is both non-linear and has a non-zero second derivative of the observed variable will differ from the true underlying distribution by the so-called \enquote{Eddington} bias (\citealt{Eddington1913,Eddington1940}; see also \citealt{Teerikorpi2004}). The galaxy SMF is a distribution function of this type (as are apparent magnitude distributions and luminosity functions). The effect is most prominent for the most massive systems due to their rarity. For example, under the Schechter formalism, the exponential overturn after \Ms{}$^{\star}$ poses a strong second derivative. The consequence is that the observed number density of massive galaxies is larger than the true underlying number density due to less massive galaxies, that are far more abundant, being randomly scattered into higher mass bins because of their uncertainty in stellar mass. By contrast, massive galaxies being scattered into low-mass bins has a much less significant effect because of their relative abundance, providing fewer instances to scatter and causing a smaller relative change to the number of galaxies in the bin. 

Given a parametric representation of the true underlying galaxy SMF, such as the Schechter function or the double-Schechter function, the observed distribution may be described by a convolution of the parametric function with a kernel, $\mathcal{D}(\mathcal{M})$, that characterises the uncertainty in stellar mass. The intrinsic galaxy SMF may then be inferred via deconvolution. In this work, the same kernel utilised by \cite{Ilbert2013}, \citetalias{Davidzon2017}, and \citetalias{Weaver2023SMF} is used to characterise the stellar mass uncertainties
\begin{equation} \label{eq:kernel}
    \mathcal{D}(\mathcal{M_{\rm 0}}, z)=\frac{1}{2 \pi} \exp \left[\frac{-\mathcal{M}_{\rm 0}^2}{2 \sigma_{\text {Edd }}^2}\right] \frac{\tau_{\text {Edd }}}{2 \pi} \frac{1}{\left(\tau_{\text {Edd }} / 2\right)^2+\mathcal{M}_{\rm 0}^2}.
\end{equation}
The kernel includes a Gaussian component of standard deviation $\sigma_{\text {Edd}}$ that conveys the scatter in the stellar mass primarily due to photometric uncertainties. The Gaussian component is multiplied by a Lorentzian component that broadens the distribution to further account for the more drastic changes in stellar mass that result from uncertainty in the redshift. The Lorentzian component is characterised by the parameter $\tau_{\text {Edd }}$, which is redshift dependent, i.e., $\tau_{\text {Edd }} = \tau_{c}(1+z)$, reflecting the fact that generally photometric redshifts are less certain with increasing $z$. 

The exact values of $\tau_{\text {Edd }}$ and $\tau_{c}$ are determined by fitting the marginalised probability distributions of mass obtained from Sect.~\ref{subsubsec:sed}, following \cite{Ilbert2013} and \citetalias{Davidzon2017}. The distributions are well described by Eq.~(\ref{eq:kernel}) but larger values for $\sigma_{\text {Edd}}$ and $\tau_{c}$ are obtained compared to \citetalias{Davidzon2017}, i.e.,  $\sigma_{\text {Edd}} = 0.6$ and $\tau_{c} = 0.05$ compared to $\sigma_{\text {Edd}} = 0.35$ and $\tau_{c} = 0.04$, respectively. The increased uncertainty relative to \citetalias{Davidzon2017} is likely due to the fewer number of photometric filters considered in the present work, although the values are very similar to those obtained by \cite{Ilbert2013}.

\begin{figure*}
    \centering
    \includegraphics[width=\textwidth]{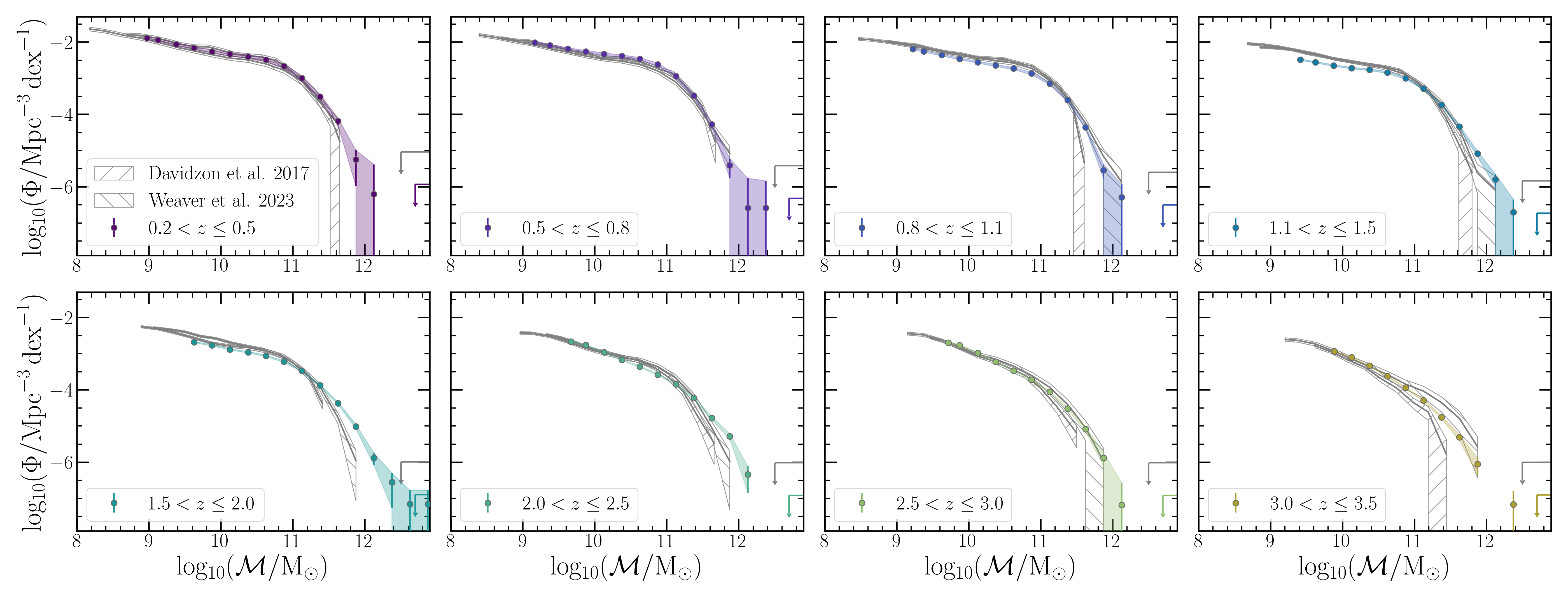}
    \caption{The observed total galaxy stellar mass function of DAWN PL (coloured points) and uncertainties (1$\sigma$, shaded) compared with two measurements from COSMOS, \citetalias{Davidzon2017} and \citetalias{Weaver2023SMF}, through to $3 < z \leq 3.5$. \Ms{} bins below the stellar mass limit (Eq.~\ref{eq:tot_mcomp}) are not shown. The volume limit according to the $N=0$ Poisson uncertainty is shown by the downward pointing coloured arrow for DAWN PL and in grey for \citetalias{Weaver2023SMF}.}
    \label{fig:smf_litcomp}
\end{figure*}

It is noted that the treatment of the Eddington bias may play a significant role in driving discrepancy between different measurements of the galaxy SMF in the literature \citep{Grazian2015,Davidzon2017}. \cite{Adams2021} explored the impact of several representations for the convolution kernel, including a Gaussian kernel, the kernel of Eq.~(\ref{eq:kernel}), and an empirically constructed kernel based on the posterior distribution resulting from varying galaxy redshifts and stellar masses. The authors found that utilising the kernel of Eq.~(\ref{eq:kernel}) obtained results essentially consistent with the empirically constructed kernel but differing from the results using the Gaussian kernel. Further consideration of the impact on the choice of convolution kernel is discussed in Sect.~\ref{sc:discussion}.

\section{\label{sc:results}Results}

The total observed galaxy SMF is presented, followed by the star-forming and quiescent components. The intrinsic galaxy SMF is obtained by modelling the observed SMF according to the formalisms defined above. Finally, the cosmic stellar mass density is calculated by integrating the intrinsic SMF. Throughout, the galaxy SMF is measured according to the redshift bins defined in Table~\ref{tab:volumes} and using \Ms{} bins of width 0.25\,dex. 

\subsection{\label{subsec:total_smf} Total stellar mass function}

The total galaxy SMF from $0.2 < z \leq 6.5$ is shown in Fig.~\ref{fig:smf_totobs}. The evolution of the total galaxy SMF is often discussed in the context of galaxies below and above the characteristic mass \Ms{}$^{\star}$, delineating \enquote{low-mass} and \enquote{massive} galaxies. Following the dark matter halo mass function (e.g., \citealt{Tinker2008}), low-mass galaxies overwhelmingly dominate the galaxy SMF at all redshifts but especially at high $z$ where large-scale structure has yet to form. Beyond a general description, the low-mass end of the galaxy SMF is mostly inaccessible through DAWN PL due to the stellar mass limit being greater than $\mathcal{M} >10^{9}$ M$_{\odot}$ at all redshifts and increasing to $\mathcal{M} >10^{10}$ M$_{\odot}$ by $z\sim3$.  However, the substantial volume of DAWN PL enables a statistically significant characterisation of the abundance of massive galaxies above $\mathcal{M} >10^{10.5}$ M$_{\odot}$ across the entire range of $0.2 < z \leq 6.5$. 

Before inferring the intrinsic galaxy SMF, a few conclusions can already be drawn. First, the early Universe was characterised by a period of rapid star formation. Several previous observations have shown that although star-formation rates peak at $z\sim2$, specific star-formation rates were significantly higher in the early Universe \citep{Behroozi2013,Madau2014}. The age of the Universe at redshift $z\sim 6$ corresponds to 0.92 Gyr after the Big Bang under the assumed cosmology. From $z\sim 6$ to $z\sim 3.25$, a timespan of 1 Gyr, the abundance of galaxies with mass $\mathcal{M} \sim10^{10.5}$ M$_{\odot}$ increased by a factor of at least one hundred. By contrast, the abundance of galaxies of similar mass grew only by a factor of ten between $z \sim 3.25$, and $z\sim 0.35$ despite spanning a significantly longer period of time (7.7 Gyr). This is essentially an extension of the \enquote{down-sizing} effect \citep{Cowie1996}. 
Additional features of the total galaxy SMF are better understood in view of a comparison with previous results from the literature (Sect.~\ref{subsubsec:litcomp}) and from an analysis of the intrinsic, rather than observed, galaxy SMF (Sect.~\ref{subsec:modeling}).  

It is emphasised that for now, only the observed galaxy SMF is presented, while the intrinsic galaxy SMF is presented in Sect.~\ref{subsec:modeling}. As such, seemingly unphysical features, such as higher-redshift bins containing a greater abundance of massive galaxies compared to lower-redshift bins, are almost certainly due to Eddington bias that has yet to be removed. The discussion in Sect.~\ref{subsec:massive} and Appendix~\ref{app:validation} further discuss the likely contamination of galaxies from the redshift bin $1.1 < z \leq 1.5$ to the next two higher redshift bins. As such, the galaxy SMF within this redshift interval should be considered with caution, as is discussed further below.

\subsubsection{\label{subsubsec:litcomp} Comparison with literature}

The evolution of the total galaxy SMF is compared with measurements from \citetalias{Davidzon2017} and \citetalias{Weaver2023SMF} in Fig.~\ref{fig:smf_litcomp} across $0.2 < z \leq 3.5$. Each work uses the same redshift binning, beginning at $0.2 < z \leq 0.5$, and the same \Ms{} binning. In Fig.~\ref{fig:smf_litcomp}, the coloured arrow indicates the $1\sigma$ Poisson upper limit for zero detections given the volume of DAWN PL, while the grey arrow indicates the same for the volume of \citetalias{Weaver2023SMF} (about a factor of 2 smaller than \citetalias{Davidzon2017}). Although galaxies detected at values of $\Phi$ lower than these limits may be real, they do not meaningfully constrain the SMF.

\begin{figure*}
    \centering
    \includegraphics[width=\textwidth]{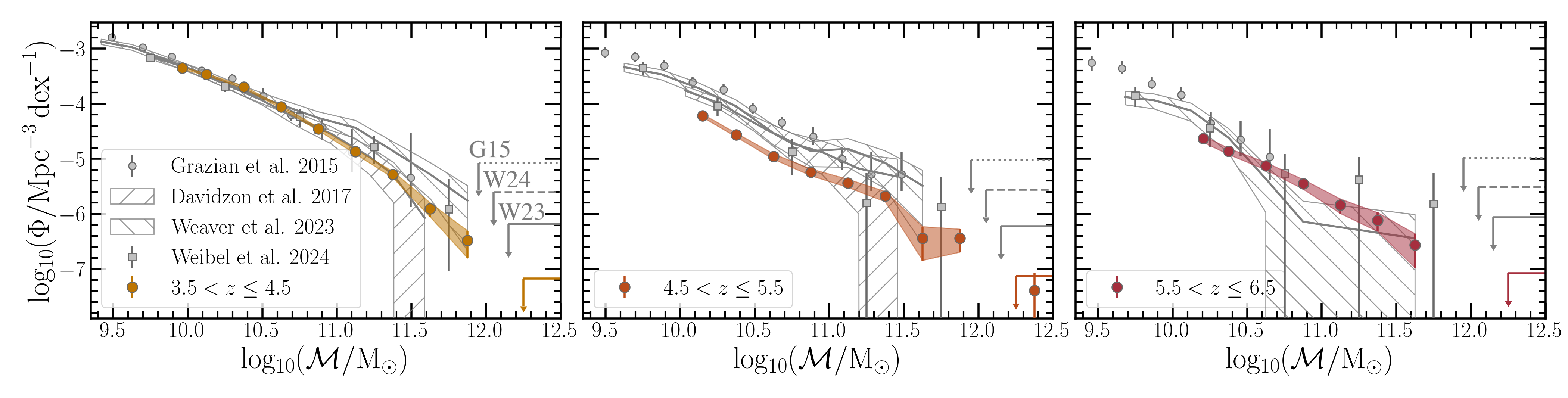}
    \caption{Same as Fig.~\ref{fig:smf_litcomp}, but at $z > 3.5$ and also showing measurements from \citetalias{Grazian2015} with HST (grey circles) and \citetalias{Weibel2024} with JWST (grey squares). }
    \label{fig:smf_litcomp2}
\end{figure*}

Figure~\ref{fig:smf_litcomp} confirms that at least through $z \leq 3.5$ the deep optical imaging of DAWN PL provides a sample of galaxies that is almost entirely consistent with a shallower NIR selection, such as the selection of \citetalias{Davidzon2017}, as suggested in Sect.~\ref{subsec:selection}.  In addition, the overall shape and normalisation of the total SMF is in agreement with both \citetalias{Davidzon2017} and \citetalias{Weaver2023SMF}. However, as noted in Sect.~\ref{subsec:SF_Q}, the selection function of DAWN PL precludes the detection of a sizeable fraction of intrinsically red objects above $z = 1.5$, which may explain the deviation of the PL SMF at $2 < z \leq 2.5$ in the vicinity of the knee ($\mathcal{M} \sim10^{10.9}$ M$_{\odot}$) compared to both \citetalias{Davidzon2017} and \citetalias{Weaver2023SMF}.  

A more significant point of disagreement is within the redshift interval $1.1 < z \leq 2.5$, which has been briefly discussed in Sect.~\ref{subsubsec:validation} and is treated more extensively in Appendix~\ref{app:validation}. Beginning at $1.1 < z \leq 1.5$ up until $z \sim 2$, there are too few galaxies per unit volume found by DAWN PL within the \Ms{} range that should be complete. The dearth of galaxies at $0.8<z\leq2.0$ may be related to the relative excess of $\mathcal{M}\gg\mathcal{M^*}$ galaxies at $z>2.0$. Without a strong observed constraint on the Balmer break, the high-redshift solutions during SED fitting outnumber low-redshift solutions, causing a systematic trend towards higher redshifts and correspondingly higher masses (due to the redshift-mass covariance). As demonstrated in Appendix~\ref{app:validation}, this systematic bias can be remedied through machine learning techniques, suggesting that the bias is caused by a degeneracy in the \lephare{} templates rather than the DAWN PL photometry. At this time, it is not clear how to best apply a correction that does not cause an inconsistency in other redshift bins. Ultimately, this is a strong example of Eddington bias at play (due to the uncertainty in stellar mass driven by the covariance with redshift), and as such, the biased redshift bins of $1.1 < z \leq 2.5$ provides a test-case to assess whether the treatment of Eddington bias (Sect.~\ref{subsubsec:eddington}) is sufficient. The subject is picked up again in Sect.~\ref{subsec:modeling}.

As previously discussed, the \photoz{} bias is not expected at $z > 2.5$ where the Lyman break is directly constrained (see Appendix~\ref{app:validation}). At $2.5 < z \leq 3.5$, the abundance of the most massive systems observed in COSMOS by both \citetalias{Davidzon2017} and \citetalias{Weaver2023SMF} is confirmed with significantly improved precision due to the improvement in Poisson uncertainty and cosmic variance by DAWN PL. Within this redshift range, DAWN PL also finds a handful of galaxies of greater mass than have been observed in COSMOS. Most of these galaxies are within the 1$\sigma$ volume limit, consistent with no detection. At $3 < z \leq 3.5$, a significant number of galaxies are found at masses greater than were observed by \citetalias{Davidzon2017}. \citetalias{Weaver2023SMF} report an even more significant excess of galaxies compared to \citetalias{Davidzon2017} at $3 < z \leq 3.5$, most noticeable above $\mathcal{M} >10^{11}$ M$_{\odot}$. However, the total galaxy SMF of DAWN PL is consistent with both works within their reported $1\sigma$ limits. By inspecting the COSMOS2020 catalogue, the galaxies at $3 < z \leq 3.5$ missed by DAWN PL appear intrinsically red and optically faint (e.g., with HSC $i \gg 26$), perhaps indicative of dusty star forming systems that are mostly undetected without near-infrared imaging. 

Beginning at $3.5 < z \leq 4.5$, the total galaxy SMF is further compared with results from \cite{Grazian2015}, hereafter referred to as \citetalias{Grazian2015}, and \cite{Weibel2024}, hereafter referred to as \citetalias{Weibel2024} in Fig.~\ref{fig:smf_litcomp2}. However, it is cautioned that both cover areas that are a factor of 70--100 smaller than DAWN PL (369 arcmin$^{2}$ and 500 arcmin$^{2}$, respectively), and a significant increase in cosmic variance should be expected. At $3.5 < z \leq 4.5$, DAWN PL continues to be consistent with \citetalias{Davidzon2017} while also finding galaxies at greater masses, further confirming the utility of deep optical imaging to detect significant samples of galaxies at least to $z\sim4$. The comparison further shows an agreement between DAWN PL with \citetalias{Grazian2015} that may at first be surprising given the difference in selection functions. \citetalias{Grazian2015} detects galaxies using the WFC3 $H_{160}$ band with a depth that significantly exceeds that of \citetalias{Davidzon2017}, yet their number densities are consistent with each other as well as with DAWN PL. Meanwhile, both DAWN PL and \citetalias{Davidzon2017} present number densities of massive galaxies (i.e., with masses $\mathcal{M} >10^{11}$ M$_{\odot}$) that are slightly lower than the measurements of both \citetalias{Weaver2023SMF} and \citetalias{Weibel2024}. \citetalias{Weibel2024} explains that their excess at $z\sim 4$ is largely driven by very red galaxies that are missed by HST and therefore should also be missed by ground-based optical surveys. However, the deep $K_{s}$ imaging utilised by \citetalias{Weaver2023SMF}, deeper than was available to \citetalias{Davidzon2017} by $\sim$0.5 mag, coupled with their significantly increased volume in comparison to \citetalias{Weibel2024}, appears to confirm a similar abundance.  

\begin{figure*}
    \centering
    \includegraphics[width=\textwidth]{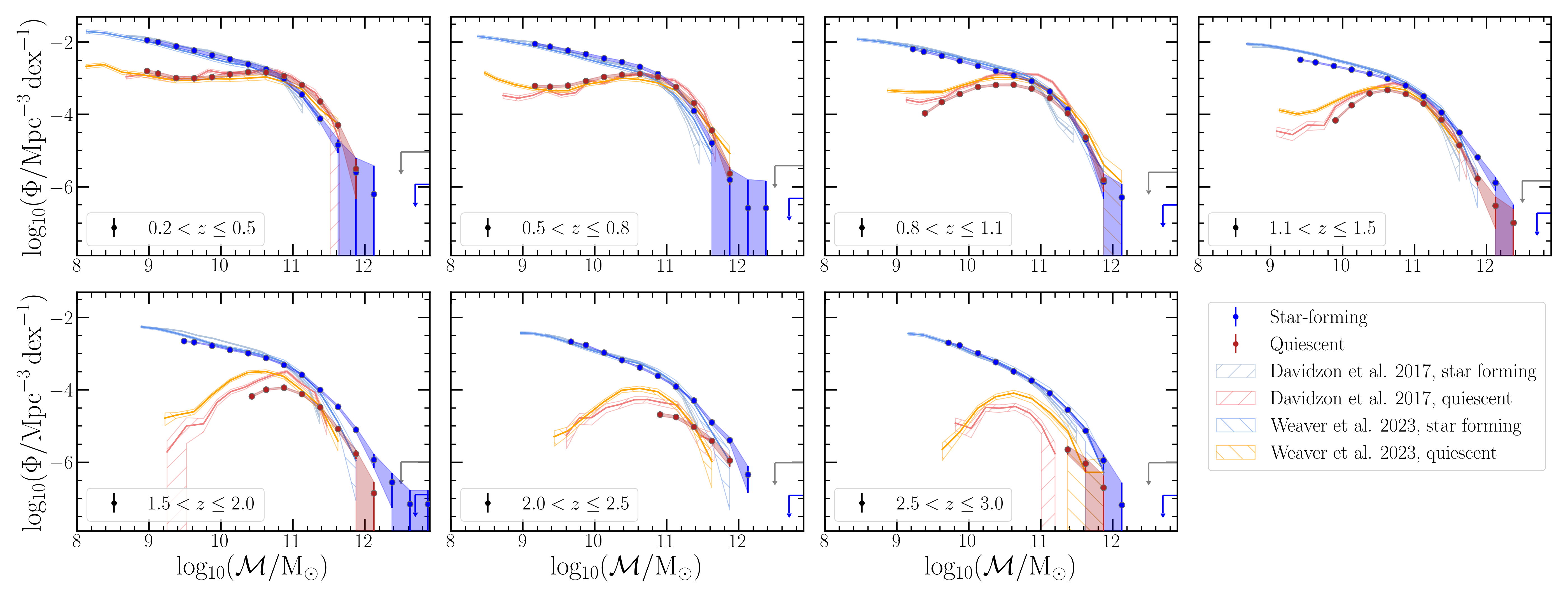}
    \caption{Same as Fig.~\ref{fig:smf_litcomp}, but comparing the SMFs of star-forming and quiescent galaxies until $z \leq 3$. }
    \label{fig:smf_litcomp_bytype}
\end{figure*}

The deep optical imaging of DAWN PL is sufficient to obtain a significant sample of galaxies at $4.5 < z \leq 5.5$. Above $\mathcal{M} >10^{11}$ M$_{\odot}$, the number density is in agreement with \citetalias{Davidzon2017} within the reported $1\sigma$ uncertainties and consistent with the abundance of \citetalias{Weaver2023SMF} within their $2\sigma$ uncertainties (not shown; see \citetalias{Weaver2023SMF}, Fig. 6). Further  comparison with \citetalias{Grazian2015} and \citetalias{Weibel2024}, both of which make use of deep space-based observations, shows similar agreement within the range of masses that overlap. However, the literature ultimately points towards an incompleteness in DAWN PL of galaxies with masses $10^{10} <\mathcal{M}/{\rm M}_{\odot} < 10^{11}$. The stellar mass limit of DAWN PL is compared to \citetalias{Weaver2023SMF} and explored in greater detail in Appendix~\ref{app:validation} in an analysis that suggests the completeness at $10^{10} < \mathcal{M}/{\rm M}_{\odot} < 10^{11}$ is between 40--60\% in DAWN PL. Notably, \citetalias{Weaver2023SMF} finds a sample of galaxies with masses greater than those observed by \citetalias{Grazian2015} and \citetalias{Davidzon2017}. DAWN PL provides a sample of galaxies at $4.5 < z \leq 5.5$ that are equally massive and have a number density that is consistent within $1\sigma$, although those of DAWN PL are more numerous and therefore more statistically significant assuming the uncertainties are well measured (Sect.~\ref{subsec:errors}). In addition, there are 13 galaxies in DAWN PL with masses exceeding the most massive one found in \citetalias{Weaver2023SMF}. Interestingly, with the assistance of deep JWST imaging, \citetalias{Weibel2024} finds two such galaxies with $\mathcal{M} >10^{11.6}$ M$_{\odot}$, but given their limited area, their number density is consistent within the Poisson 1$\sigma$ upper limit of zero detections.

Despite the inclusion of NIR imaging, \citetalias{Davidzon2017} was not able to find a significant sample of galaxies at $5.5 < z \leq 6.5$. By contrast, the deeper optical imaging of DAWN PL provides a selection of 219 galaxies above the estimated stellar mass limit. However, the comparison to \citetalias{Grazian2015}, \citetalias{Weaver2023SMF}, and \citetalias{Weibel2024} suggest mass-incompleteness below $\mathcal{M} \sim10^{10.5}$ M$_{\odot}$, similar to what is observed at $4.5 < z \leq 5.5$. Galaxies above $\mathcal{M} >10^{10.5}$ M$_{\odot}$ show a number density that is consistent with both \citetalias{Weaver2023SMF} and \citetalias{Weibel2024}, although both of these works only place upper limits at these masses and at this redshift. Meanwhile, \citetalias{Grazian2015} did not find such massive galaxies.

Beyond $z\sim6$, galaxies are not detected in DAWN PL with enough abundance to measure the galaxy SMF. Future work taking advantage of the imminent \Euclid imaging across the DAWN survey will provide samples of galaxies of similar statistical strength presently available at $z\sim3$ but to $z\sim 8$ \citep{McPartland2024}. 


\subsection{\label{subsec:SF_Q_smf}Star-forming and quiescent stellar mass functions}

The star-forming and quiescent components of the total galaxy SMF are separated using the NUV$rJ$ criteria described in Sect.~\ref{subsec:SF_Q}. Total uncertainties for each component are computed following Sect.~\ref{subsec:errors} for each galaxy type and redshift interval. Given the DAWN PL selection function, only the most massive ($\mathcal{M} > 10^{11}$ M$_{\odot}$) and brightest quiescent galaxies are detectable at $z > 2$. Consequently, the shape of the quiescent galaxy SMF is not well constrained at these redshifts. For example, at $2 < z \leq 2.5$, only 732 quiescent galaxy candidates above the mass completeness limit spanning five \Ms{} bins are found over the entire DAWN PL area. At $2.5 < z \leq 3$, the number drops by over an order of magnitude to 57 in three \Ms{} bins. Above $z > 3$, no quiescent galaxies are detected that satisfy the necessary criteria.

The observed star-forming and quiescent galaxy SMFs are compared with \citetalias{Davidzon2017} and \citetalias{Weaver2023SMF} in Fig.~\ref{fig:smf_litcomp_bytype}. At all redshifts, the star-forming galaxy SMF generally agrees well with the literature and shares many of the same features discussed in Sect.~\ref{subsec:total_smf}. Similarly, the shape and normalisation of the quiescent galaxy SMF is mostly in agreement with \citetalias{Weaver2023SMF}, despite being more sparsely sampled in mass due to the DAWN PL selection function. At $1.5 < z \leq 2$ and $2.0 < z \leq 2.5$, the DAWN PL catalog recovers fewer quiescent galaxies compared to \citetalias{Davidzon2017} and \citetalias{Weaver2023SMF}, potentially indicating an incompleteness related to the optical (rest-UV) selection function. Similar to the total SMF described in Sect.~\ref{subsec:total_smf}, the redshift range of $1.1 < z \leq 2.5$ is possibly compromised due to uncertain \photoz{}s. However, this appears to predominantly affect star-forming galaxies. As discussed previously, the cause of the uncertain \photoz{} estimates is the lack of an observed constraint on the Balmer break. Galaxies classified as quiescent are also red according to the emission observed in the \textit{Spitzer}/IRAC bands, which likely (and properly) restricts the \photoz{} solutions for quiescent galaxy templates but not for star-forming templates. See Appendix~\ref{app:validation} for further details. 

Of particular interest are the quiescent galaxy candidates found at $z > 2$. There are 19 galaxies are found with masses ($\mathcal{M} >10^{11.75}$ M$_{\odot}$) at $2 < z \leq 2.5$. If real, this sample includes some of the most massive quiescent galaxies to have been observed within this redshift range, exceeding the most massive ones found in \citetalias{Weaver2023SMF}. The 57 quiescent galaxy candidates at $2.5 < z \leq 3$ are similarly massive and appear to agree with the volume density suggested by \citetalias{Weaver2023SMF}. However, the DAWN PL sample is observed at a significance greater than $5\sigma$ compared to the Poisson volume limit in contrast to \citetalias{Weaver2023SMF} which only measures upper limits. 

Despite the difficulty of fully describing the evolution of the low-mass end of the quiescent galaxy SMF, a comparison of the rate of growth between the most massive systems ($\mathcal{M} >10^{11.75}$ M$_{\odot}$) and those of lower mass can still be made, at least at $z \leq 2.5$. Considering all quiescent galaxies up until $z \leq 2.5$, a significant fraction of the most massive quiescent galaxies appear to have already formed by $z \sim 2.5$, 2.9 Gyr after the Big Bang. Their observed evolution in number density suggests very little change from  $z \sim 2.25$ to $z \sim 0.35$, a period of time spanning 7.3 Gyr. This is similarly reported by \cite{Kawinwanichakij2020} but from $z \leq 1.5$. In contrast, galaxies of $\sim$$10^{11}$ M$_{\odot}$ increase in abundance by approximately a factor of $\sim$50. The contrast between the rate of growth of these systems is generally consistent with the observed quiescent SMF of \citetalias{Weaver2023SMF} within their reported uncertainties, although \citetalias{Weaver2023SMF} suggests an increase in abundance of quiescent galaxies with  $\mathcal{M} \sim10^{11}$ M$_{\odot}$ of $\sim$20. The impact of Eddington bias (Sect.~\ref{subsubsec:eddington}) is more drastic for these massive systems in comparison to their lower-mass counterparts. Nonetheless, estimates from the intrinsic quiescent galaxy SMF described below (Sect.~\ref{subsec:modeling}) suggest a similar discrepancy in the evolution of the most massive quiescent galaxies compared to those of more moderate mass, consistent with \citetalias{Weaver2023SMF}.  

It should be noted that the galaxies above $z > 2$ classified as \enquote{quiescent} by DAWN PL are essentially on the border of the colour-colour selection set that defines quiescent galaxies in Eq.~(\ref{eq:NUVrJ}) in Sect.~\ref{subsec:SF_Q} (see also Fig.~\ref{fig:NUVrJ}). Indeed, the rest-frame $r$ necessary for the NUV$rJ$ selection is not directly constrained by observation in DAWN PL and is instead predicted by the best-fit model and assisted by the HSC $z$ and IRAC \chOne{} bands between which the rest-frame $r$ wavelengths resides. Whether or not the systems identified above $z > 2$ are truly quiescent is difficult to confidently determine without the aid of near-infrared imaging, or better, spectroscopy. As a minimal test, quiescent galaxies found in DAWN PL are compared to those found by \citetalias{Weaver2023SMF} (i.e., in the COSMOS2020 catalogue). The comparison suggests that observed and rest-frame colours are consistent between the two independent samples and that their abundance agrees within the mass range that overlaps, despite \citetalias{Weaver2023SMF} having a lower mass completeness limit.

\subsection{\label{subsec:modeling}Modeling the stellar mass functions}

The total, star-forming, and quiescent galaxy SMFs are modelled independently according to the formalisms described in Sect.~\ref{subsec:schechter} and Sect.~\ref{subsubsec:eddington}. The data considered in the modeling includes, for each redshift bin, all \Ms{} bins above the completeness limit with the $1/V_{\rm{max}}$ correction applied and with uncertainties characterised by $\sigma_{\rm{tot}}$ (Sect.~\ref{subsec:errors}). A double-Schechter function is considered for both the total and star-forming SMFs at $z \leq 2$ and above a single Schechter function is used. At $z > 2$, the double-Schechter function does not provide an improved fit to the observations and the additional parameters are not warranted given the lack of structure observed in the vicinity of the knee. For the quiescent galaxy SMF, a single Schechter function is used starting at $0.8 < z \leq 1.1$ because the mass completeness limit prohibits the detection of any low-mass systems that would constrain the second component of the double-Schechter function. 

\subsubsection{\label{subsec:eddingtin_demo}The effects of Eddington bias}

\begin{figure}
    \centering
    \includegraphics[width=\columnwidth]{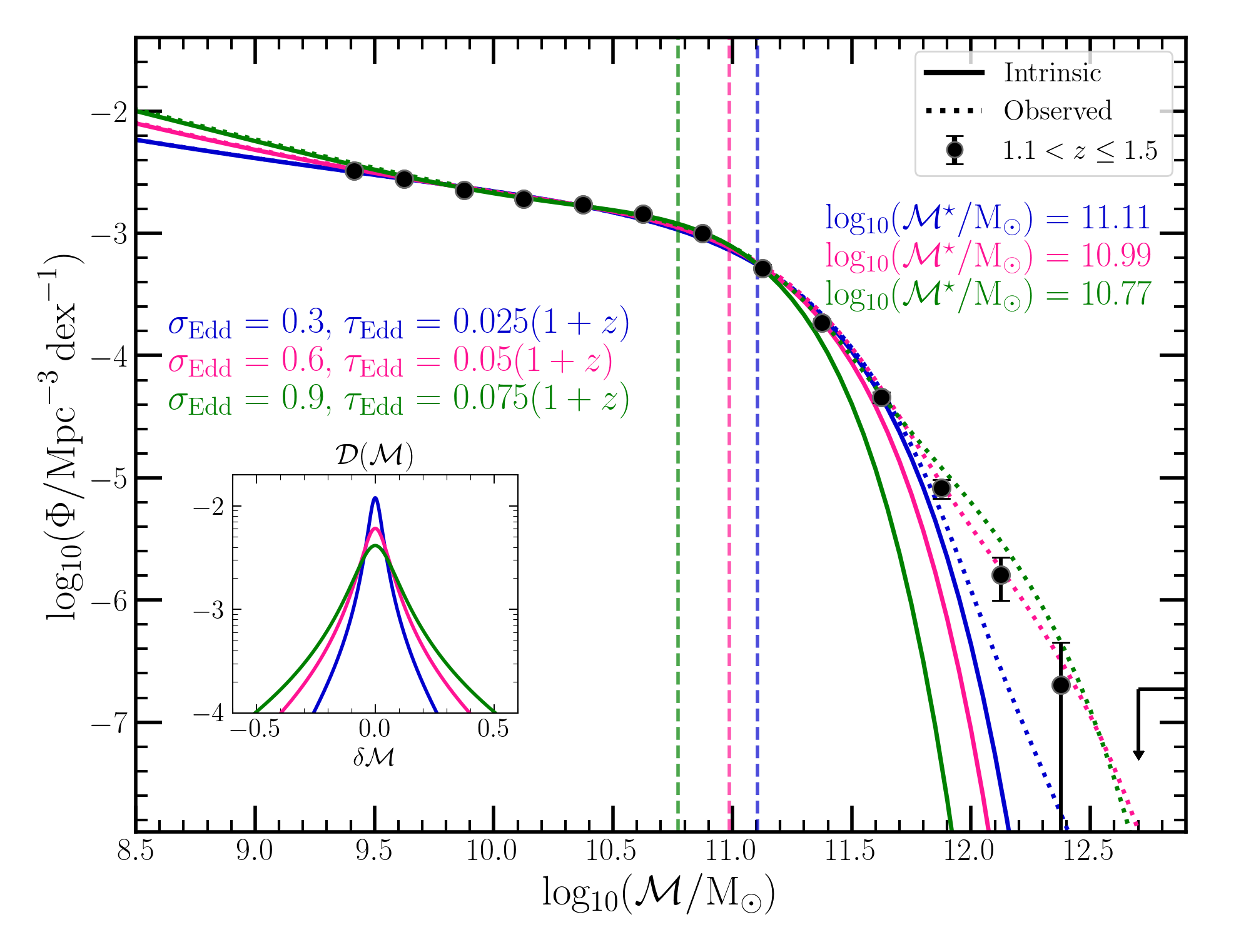}
    \caption{Variation in the best-fit double Schechter function resulting from three different parameterisations of the uncertainty in \Ms{} used for the Eddington bias kernel (Eq.~\ref{eq:kernel}) at $1.1 < z \leq 1.5$. The shape of the kernel according to each parameterisation is shown in the inset plot, with the corresponding parameter values annotated above in matching colours. The resulting intrinsic (solid) and observed (dotted) galaxy stellar mass function is shown for each case, with the value of the best-fit characteristic mass (\Ms{}$^{\star}$) annotated and given by the vertical dashed lines. The treatment of the Eddington bias significantly affects the inferred intrinsic galaxy stellar mass function.}
    \label{fig:tot_mcmc_examp}
\end{figure}

During modelling, the Schechter function (or double-Schechter function, where applicable) is convolved with the Eddington bias kernel from Eq.~(\ref{eq:kernel}) in Sect.~\ref{subsubsec:eddington}. Consequently, it is the convolved model representing the observed galaxy SMF that is fitted to the data, and the intrinsic stellar mass function is obtained by deconvolution. To demonstrate the role of the Eddington bias kernel, Fig.~\ref{fig:tot_mcmc_examp} provides a comparison at $1.1 < z \leq 1.5$ of the \enquote{observed} model of the galaxy SMF consisting of a double-Schechter function convolved with the Eddington bias kernel (dotted coloured line) of Eq.~(\ref{eq:kernel})  and the inferred intrinsic galaxy SMF (solid coloured line). The massive end is most dramatically affected by the Eddington bias kernel. It can be mathematically shown that the Eddington bias is most prominent where the magnitude of the second derivative of the underlying distribution function is greatest, which in the case of the Schechter function (or double-Schechter function) occurs around the characteristic mass \Ms{}$^{\star}$. Consequently, the selection of the Eddington bias parameters ($\sigma_{\rm{Edd}},\tau_{\rm{Edd}}$) significantly influences not only the best-fit value of \Ms{}$^{\star}$ but all of the best-fit Schechter parameters due to their covariance, as pointed out by \citetalias{Davidzon2017}. Three different characterisations of the Eddington bias parameters are compared in Fig.~\ref{fig:tot_mcmc_examp}, including the one used in this work, $\sigma_{\rm{Edd}} = 0.6,\tau_{\rm{Edd}} = 0.05(1+z)$, see Sect.~\ref{subsubsec:eddington}. The resulting best-fit values (from simple $\chi^{2}$ minimisation) for \Ms{}$^{\star}$ are indicated by the vertical dashed lines and corresponding annotations and vary by $\geq 0.3$\,dex solely due to changing the Eddington bias kernel. Although not annotated, the best-fit values of the other Schechter parameters vary similarly. For example, the normalisation of the first Schechter component, $\Phi^*_{1}$, varies by a factor of $\sim$5, the normalisation of the second Schechter component, $\Phi^*_{1}$, varies by a factor of $\sim$2, and the second slope parameter, $\alpha_{2}$, varies by a factor of $\sim$3.5. As such, any effort to compare the Schechter parameters obtained from different works must take into account the treatment of the Eddington bias, which is not uniform across the literature. 

Figure~\ref{fig:tot_mcmc_examp} also demonstrates that the treatment of the Eddington bias used herein is sufficient to characterise the observed excess of massive galaxies at $1.1 < z \leq 1.5$ (similar behaviour is seen in the next two redshift bins as well). Although the values of the Eddington bias parameters ($\sigma_{\rm{Edd}},\tau_{\rm{Edd}}$) were not selected by visual comparison, Fig.~\ref{fig:tot_mcmc_examp} nonetheless shows a very good agreement between the observed data and the intrinsic SMF that results from the fit that uses $\sigma_{\rm{Edd}} = 0.6$ and $\tau_{\rm{Edd}} = 0.05(1+z)$. By contrast, larger values over-predict the number density of observed massive galaxies, while smaller values under-predict. Indeed, comparing the $\chi^{2}$ of each fit, the smallest values achieve the worst reduced-$\chi^{2}$ value at 4.23. The other two achieve similar $\chi^{2}$ values at 1.72 for the values used herein and 1.68 for the largest kernel assumed. Despite the marginal improvement in the $\chi^{2}$, there appears to be no reason to prefer the larger kernel, as it is not supported by the measured posterior distributions of \Ms{} and it appears to over-predict the abundance of observed massive galaxies.

\subsubsection{\label{subsec:degeneracy}Degeneracy of model parameters}

Independent of the Eddington bias, it is well known that there exists a degeneracy between the low-mass slope $\alpha$, the turnover mass or knee, \Ms{}$^{\star}$, and the normalisation $\Phi^*$ (see, for example, Fig.~8 of \citealt{Stefanon2021} or Fig.~7 of \citetalias{Weibel2024}). The DAWN PL mass completeness limit is already nearly $\mathcal{M} \sim10^{10}$ M$_{\odot}$ by $z\sim2$, and therefore DAWN PL cannot constrain the galaxy SMF at low masses. Nonetheless, it should still be possible to constrain the knee or characteristic mass, $\mathcal{M} ^{\star}$, as well as the overall normalisation, $\Phi^*$. Consequently, the low mass slope is handled differently than the other Schechter parameters during the modelling step. 

It is emphasised that there is no solution that is completely free of assumption for modelling a parameter that is not well constrained by the data. One option is to freeze the low-mass slope throughout the modelling step and only fit for the characteristic mass and the normalisation, an option typically used in modelling the highest-redshifts included by the given selection function \citep{Davidzon2017,Weaver2023SMF,Weibel2024}. Alternatively, a prior may be used to restrict the allowable values of the low-mass slope while still accounting for some of the covariance between the Schechter parameters and the resulting uncertainty of their inferred values. The latter approach is used herein. In each redshift bin, a flat prior motivated by results from the literature is used, and the range of allowed values evolves with redshift. 

\begin{figure*}
    \centering
    \includegraphics[width=0.8\textwidth]{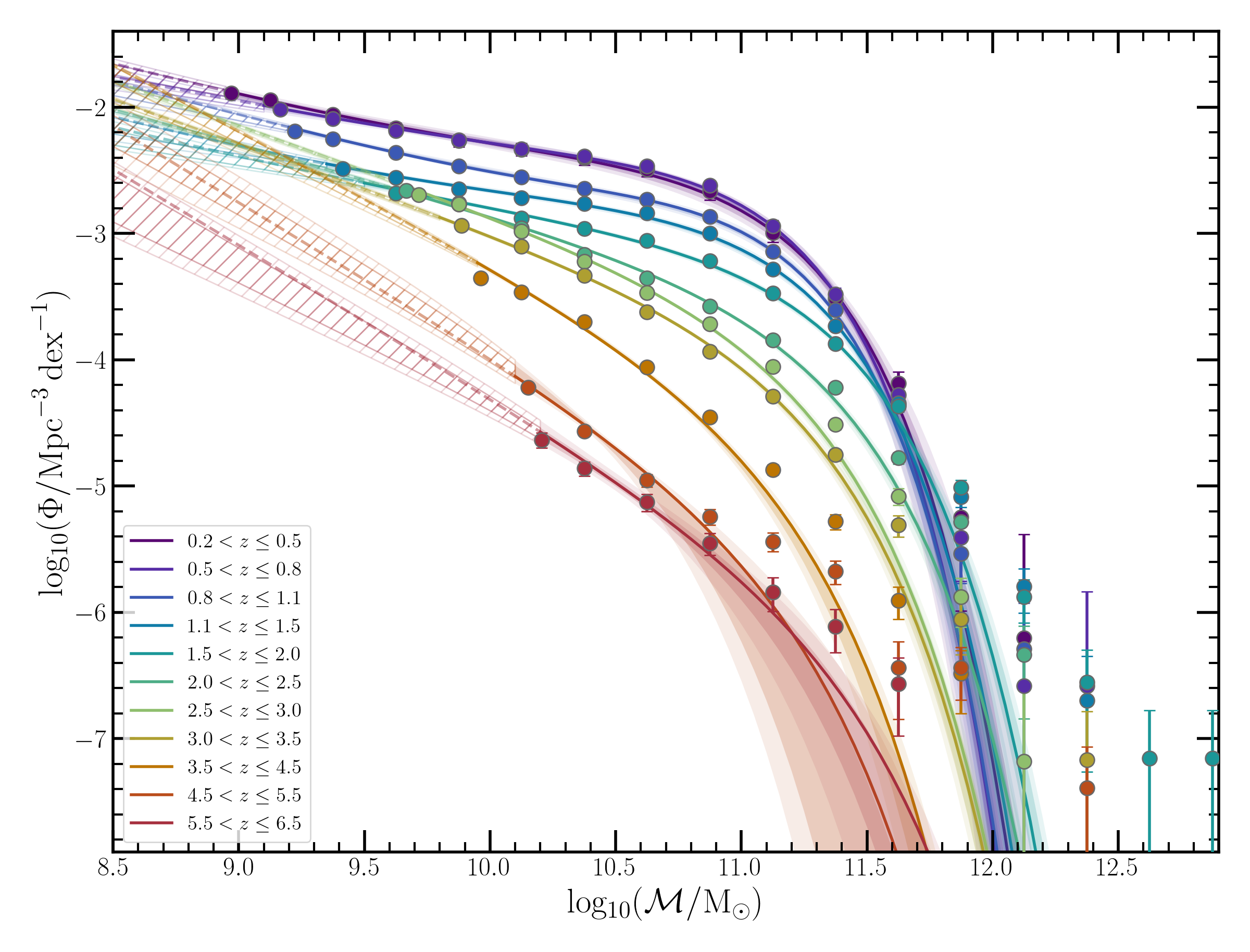}
    \caption{Evolution of the intrinsic galaxy stellar mass function of the total sample, shown as solid lines, across $0.2 < z \leq 6.5$. Maximum a posteriori values are assumed, as listed in Table~\ref{table:fit_total}. The 1$\sigma$ (2$\sigma$) uncertainty according to the variation in the MCMC chains is shown by the dark (light) shaded regions. Extrapolations below the stellar mass limit (Eq.~\ref{eq:tot_mcomp}) are shown by the dashed lines, and the uncertainties are indicated by the hatched regions. Observed values for corresponding redshift bins are shown in the matching colour as data points with error bars. }
    \label{fig:tot_mcmc}
\end{figure*}

At $z \leq 2.5$, the low-mass slope is bounded by the interval $-1.55 < \alpha < -1.3$, consistent with both \citetalias{Davidzon2017} and \citetalias{Weaver2023SMF}. At $z > 2.5$, \citetalias{Weaver2023SMF} chose to freeze $\alpha$ at the value found at $2.0 < z \leq 2.5$, although \citetalias{Davidzon2017} left the parameter free to vary throughout all redshift ranges and found the value to decrease (i.e., become steeper) at $z > 2.5$. In this work, the range of allowed values at $2.5 \leq z < 3.5$ is $-1.7 < \alpha < -1.45$, which is consistent with both works. At $3.5 \leq z < 4.5$, the range of allowed values is $-1.85 < \alpha < -1.65$, now departing from \citetalias{Weaver2023SMF} but remaining consistent with \citetalias{Davidzon2017} as well as the results from \citetalias{Grazian2015} and \citetalias{Weibel2024} that are well suited for characterising the low-mass slope at high $z$. Finally, at $z \geq 4.5$, the range of allowed values is $-2.2 < \alpha < -1.65$, allowing for the steeper slopes found by each of \citetalias{Grazian2015}, \citetalias{Davidzon2017}, and \citetalias{Weibel2024}. 

For double-Schechter models, the above prior is only imposed on the low-mass slope $\alpha_{1}$, while the slope of the second Schechter component, $\alpha_{2}$, is free to vary. As the low-mass end of the galaxy SMF is dominated by star-forming galaxies, the star-forming and total galaxy SMFs are modelled using the same priors on the low-mass slope, keeping in mind that the star-forming galaxy SMF is considered separately only below $z = 3$. For the same reason, the low-mass slope of the quiescent SMF is handled differently, using different ranges of allowable values for $\alpha$. At $z < 1.1$, the range of allowed values is $-2.5 < \alpha < 0$, while at $z\geq 1.1$, the range of allowed values is $0 < \alpha < 2.5$. In both cases, the range of allowed values is quite broad but nonetheless provides useful boundaries that are consistent with both \citetalias{Davidzon2017} and \citetalias{Weaver2023SMF}. Effectively, the prior requires that the slope is negative at $z < 1.1$ and positive at $z > 1.1$.\footnote{Note that at $z \geq 0.8$, a single Schechter function is assumed, and the meaning of $\alpha_{1}$ changes.}

The above handling of $\alpha$ encourages a physically plausible evolution of the galaxy SMF to result from the fitting procedure. An additional requirement is imposed following \citetalias{Weaver2023SMF}, in which an upper limit is placed on the normalisation $\Phi^*$ of the galaxy SMF. For every bin above the first redshift bin ($0.2 < z \leq 0.5$), the normalisation of the galaxy SMF is required to be less than or equal to the 84\% percentile (corresponding to $+1\sigma$ for a Gaussian distribution) of the posterior distribution of $\Phi^*$ from the previous redshift bin, $\Phi^*_{84}$. For double-Schechter models, both $\Phi^*_{1}$ and $\Phi^*_{2}$ are required to be less than $\max(\Phi^*_{1,84}$,$\Phi^*_{2,84})$. This additional prior, like the prior on $\alpha$, is physically motivated and also discourages an evolution of the intrinsic galaxy SMF in which the total stellar mass density (Sect.~\ref{subsec:SMD}) decreases as the amount of time since the Big Bang grows. In all cases, $\Phi^*$ is found to decrease with redshift as expected. 

\subsection{\label{subsec:MCMC}Markov chain Monte Carlo analysis}

In order to consider a range of possible values for the Schechter parameters and determine useful uncertainties on the representative values, the observed galaxy SMF is fitted in a Markov chain Monte Carlo (MCMC) approach based upon \texttt{emcee} \citep{Foreman-Mackey2013} following the process outlined in \citetalias{Weaver2023SMF}. Initial values are determined by first fitting each of the total, star-forming, and quiescent galaxy SMFs via a $\chi^{2}$ minimisation with the same priors discussed above. As \texttt{emcee} enables ensemble sampling, a collection of 500 MCMC \enquote{walkers} are initialised based on the best-fitting values from the $\chi^{2}$ minimisation but scattered within $2\%$. The walkers explore the parameter space simultaneously via multiprocessing and for as many steps as required to each beyond 10$\times$ their autocorrelation length. An additional requirement is that the estimated autocorrelation length is not changing more than 1\%. These criteria measure ensures that each Markov chain has satisfactorily converged and that the total area explored is not dictated by the initial conditions. 

Two sets of Schechter parameters are obtained from the MCMC analysis. The first correspond to the median of the marginalised posterior distributions of each parameter. The second corresponds to the maximum a posteriori parameters. As described by \citetalias{Weaver2023SMF}, these two sets of parameters provide different descriptions of the intrinsic galaxy SMF. The former is often used in the literature and describes parameter values that are frequently encountered during the sampling, implying that they may be representative due to their prevalence. However, the set of parameters defined by the median of each marginalised posterior distribution does not necessarily correspond to parameter values that are encountered simultaneously with each other, and therefore they are not guaranteed to provide a good fit (high likelihood) to the data. One of the appeals of using the median of the marginalised posterior distributions is that the uncertainty on a given parameter may be meaningfully determined from the percentiles of the posterior distribution. By contrast, the maximum a posteriori parameters provide a model that most closely resembles the data given the likelihood function and any assumed priors. The downside is that there is no single method to obtain uncertainties on their values. Accordingly, caution is advised when considering either set of Schechter parameters. Each set of parameters is presented in Table~\ref{table:fit_total} for the total SMF, Table~\ref{table:fit_sf} for the star-forming SMF, and Table~\ref{table:fit_q} for the quiescent SMF.

\begin{figure*}
    \centering
    \includegraphics[width=0.45\textwidth]{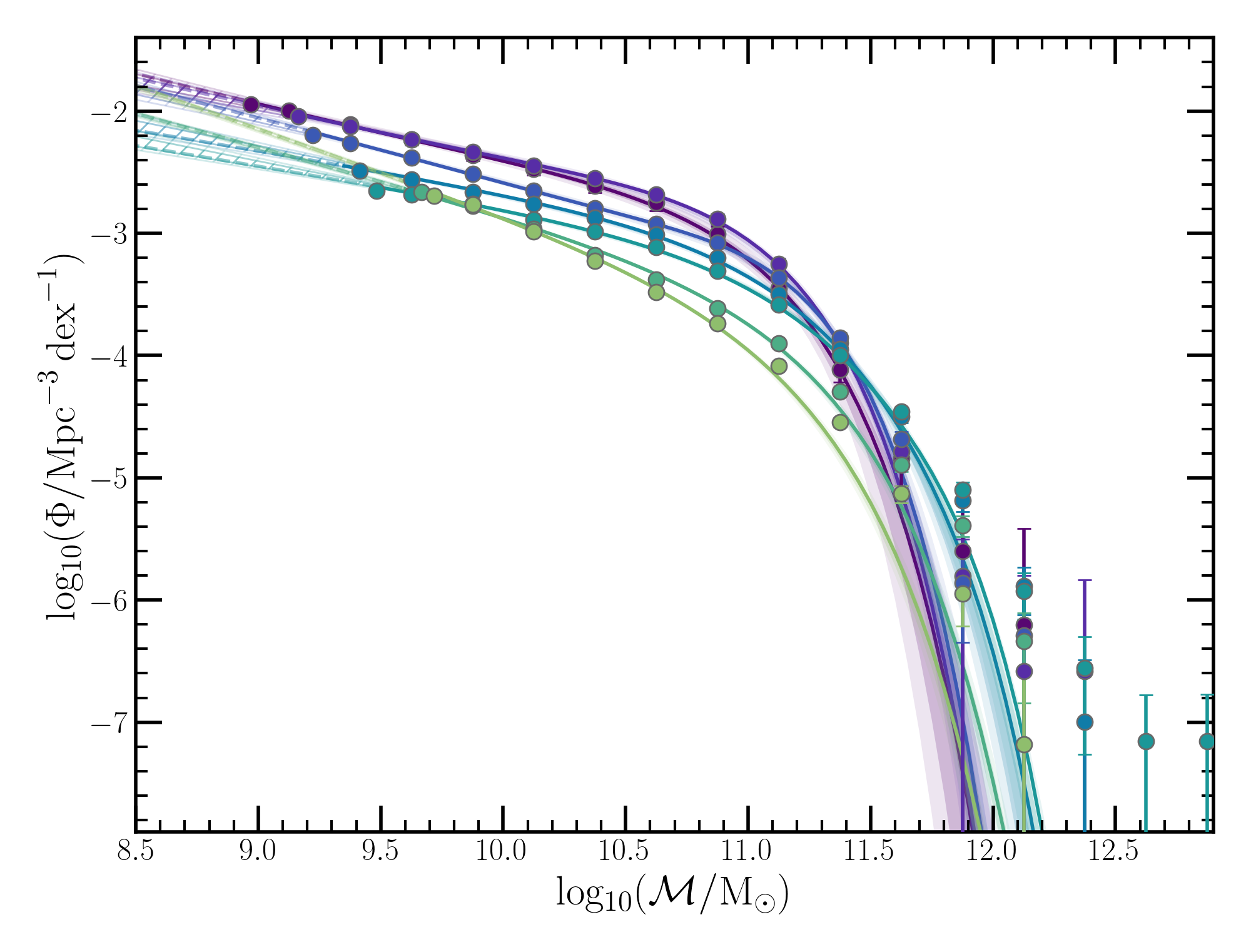}
    \includegraphics[width=0.45\textwidth]{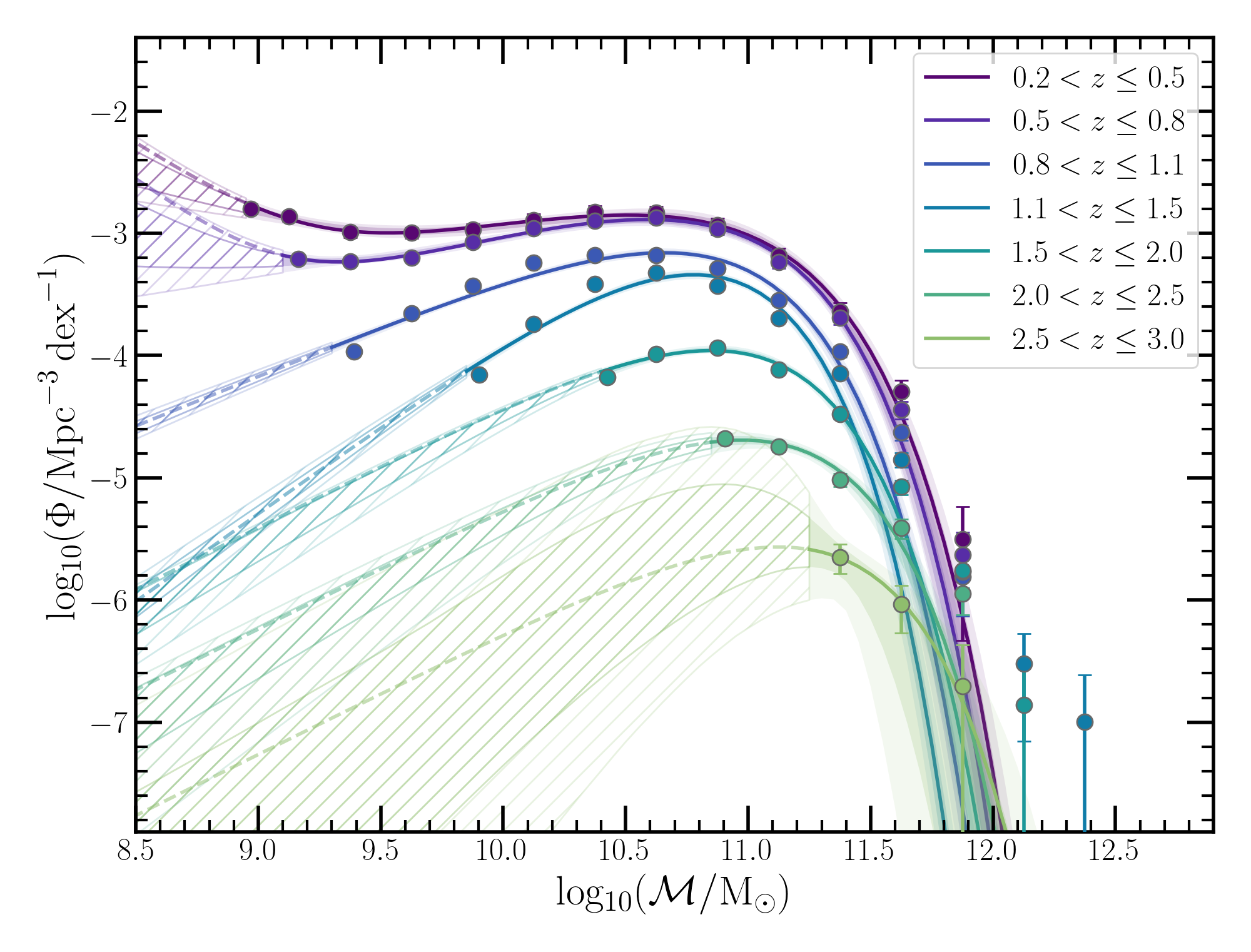}
    \caption{Same as Fig.~\ref{fig:tot_mcmc}, but separated into star-forming (left) and quiescent galaxies (right) at $z \leq 3$.}
    \label{fig:sfq_mcmc}
\end{figure*}

The intrinsic total galaxy SMF for all redshift bins is shown in Fig.~\ref{fig:tot_mcmc} according to the maximum a posteriori parameter values, while Fig.~\ref{fig:sfq_mcmc} shows the same for the intrinsic star-forming and quiescent SMF but restricted to $z \leq 3$. Dark (light) shaded regions correspond to the 1$\sigma$ (2$\sigma$) variation of the intrinsic SMF according to the MCMC chains of each parameter. In both figures, hatched regions of the plot indicate regions of parameter space that are extrapolated from the fit and not constrained by observation. As should be expected, the variation is greatest at $z > 4.5$ where the total uncertainties ($\sigma_{\rm{tot}}$) are weakest and the mass completeness limit is highest ($\mathcal{M} >10^{10}$ M$_{\odot}$). In comparison to the observed SMF, the intrinsic SMF differs most strongly on the massive end. This is due to the effect of the Eddington bias kernel, which results in an intrinsic SMF that does not necessarily trace the most massive bins, as demonstrated by Fig.~\ref{fig:tot_mcmc_examp}. Whether the abundance of such massive galaxies is driven by the uncertainty in \Ms{}, as assumed by the Eddington bias kernel, or the abundance of the most massive systems is accurate, is further discussed in Sect.~\ref{subsec:massive}. 

The goal of this work is not to investigate a possible evolutionary trend in the Schechter parameters that describe the intrinsic SMF. Instead, the Schechter function formalism is primarily useful for empirically describing the shape of the intrinsic SMF by enabling a straightforward accounting of the Eddington bias through forward modelling.  Nonetheless, a brief comparison to the literature of the representative values found at each redshift is warranted because it may validate the modelling procedure and the decisions regarding the treatment of the Eddington bias. This comparison is provided in Appendix~\ref{app:schechter}. Overall, the comparison shows broad agreement with the literature, with disagreement mostly explained by the covariance of the Schechter parameters and stellar mass ranges explored by each data set. Future works that seek to investigate an evolution in the Schechter parameters or perform detailed comparisons with the literature should take effort to account for both the differences in assumptions of the Eddington bias and for the covariance between the Schechter parameters. 

The star-forming component of the intrinsic SMF is depicted in the left panel of Fig.~\ref{fig:sfq_mcmc}, while the quiescent component is shown in the right panel. Both the star-forming and quiescent components are shown at $z \leq 3$, keeping with the separation defined in Sect.~\ref{subsec:SF_Q}. Generally, the star-forming component resembles the total SMF. The quiescent SMF, on the other hand, is quite different, perhaps most uniquely characterised by a low mass slope that increases with redshift. In contrast to the star-forming and total SMF, the intrinsic quiescent SMF suggests that the quiescent galaxies below the characteristic mass are rare in the early Universe, but become increasingly more common at late times. Despite the lack of constraints on the low-mass end of the quiescent SMF, the inferred Schechter parameters are mostly supported by \citetalias{Davidzon2017} and \citetalias{Weaver2023SMF}. The primary difference is in the normalisation, where a smaller value is typically found in DAWN PL in accordance with the differences in completeness discussed in Sect.~\ref{subsec:SF_Q}. In addition, the value of \Ms{}$^{\star}$ is higher in DAWN PL at $z > 1.5$. This is driven by both the stellar mass limit, which is $\logten(\mathcal{M}/{\rm M}_{\odot})>10.5$ at $z > 1.5$, and the abundance of massive quiescent galaxies that exceeds both \citetalias{Davidzon2017} and \citetalias{Weaver2023SMF} at these redshifts. As discussed in Sect.~\ref{subsec:SF_Q}, the intrinsic quiescent SMF also suggests that a significant fraction of the most massive quiescent galaxies, e.g., $\logten(\mathcal{M} /{\rm M}_{\odot})>11.5$, had already formed by $z\sim$\,2--3, just 2--3 Gyr after the Big Bang. Massive quiescent galaxies at $z > 2$ are discussed in greater detail in Sect.~\ref{subsec:massive}.

\section{\label{sc:discussion}Discussion}

The galaxy SMF is commonly used as a tool to constrain the cosmic stellar mass density. This section includes a validation of the evolution of the galaxy SMF measured and implied cosmic stellar mass density through a comparison with similar efforts in the literature. However, perhaps the greatest difference between DAWN PL and previous works is the volume explored to high redshift. As seen in Sect.~\ref{sc:results}, the increased volume leads to smaller uncertainties regarding the abundance of the most massive galaxies, especially at $z > 3.5$. The implications of these galaxies and the connection to their host dark matter halos and their environments is explored in this section. Additional systematics and interloper populations are further discussed, along with a presentation of a few select massive galaxies.

\subsection{\label{subsec:SMD}Cosmic stellar mass density}

The cosmic stellar mass density (SMD; $\rho_{\star}$) describes the total stellar mass, provided by all galaxies, per unit volume. A determination of the implied value of $\rho_{\star}$ at each redshift can be obtained by integrating the intrinsic SMF at that redshift, multiplied by \Ms{}, over an appropriate range of stellar mass.  A related quantity is the cosmic star-formation rate density (SFRD), which describes the star-formation rate over all galaxies per unit volume. The SFRD can be determined by multiple independent methods. For example, the SFRD may be measured from UV measurements \citep{Cucciati2012}, far-infrared measurements \citep{Gruppioni2013}, millimeter and radio measurements \citep{LeBorgne2009,Dunne2009}, and even gamma ray bursts \citep{Kistler2013}. The SFRD can be converted to the corresponding SMD by assuming a particular initial mass function for star formation and by characterising the fraction of stellar mass that is lost due to stellar evolutionary processes. A notable example of this procedure is demonstrated by \cite{Madau2014}. Consequently, measuring the SMD provides an opportunity to compare the various methods for measuring stellar mass growth in galaxies across cosmic time.

\begin{figure}
    \centering
    \includegraphics[width=\linewidth]{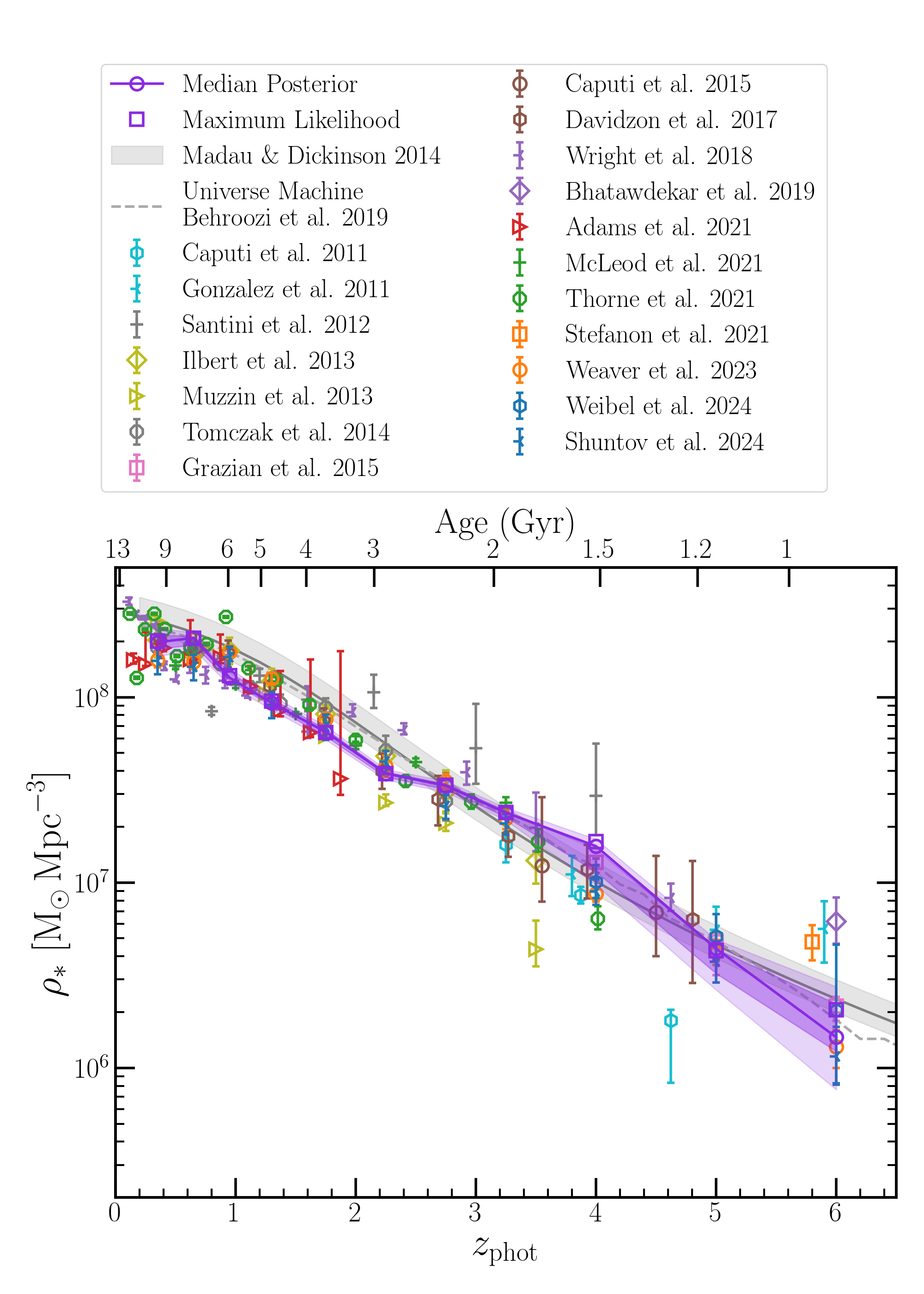}
    \caption{Stellar mass density, $\rho_{\star}$, obtained from integrating the intrinsic galaxy stellar mass function (from $\mathcal{M} =10^{8}$ M$_{\odot}$ to $\mathcal{M} =10^{13}$ M$_{\odot}$), shown for both the median posterior values (circle symbols and solid line) and the maximum a posteriori values (square symbols). Values are compared with measurements from the literature also derived from galaxy stellar mass functions \citep{Caputi2011, Gonzalez2011, Santini2012, Ilbert2013, Muzzin2013, Tomczak14, Grazian2015, Caputi2015,Wright2018,Bhatawdekar2019, Adams2021, McLeod2021, Thorne2021,Stefanon2021,Weaver2023SMF,Weibel2024,Shuntov2024} as well as integrations of the cosmic star-formation rate density functions \citep{Madau2014,Behroozi2019}. For the latter, a return fraction of 41\% (based on the \citealt{Chabrier2003} IMF, see Section~6.1 of \citealt{Ilbert2013}) is assumed. For \citet{Madau2014}, a shaded area is shown corresponding to return fractions between 25--30\% (where the latter value is similar to the \citealt{Salpeter1955} IMF).}
    \label{fig:smd}
\end{figure}

As described above, a measurement of $\rho_{\star}$ can be obtained directly from the inferred intrinsic SMF obtained in Sect.~\ref{subsec:modeling}. In the literature, the most commonly used range of integration is $10^{8}<\mathcal{M} /{\rm M}_{\odot}<10^{13}$. Therefore, the same interval is considered in this work to provide an estimate of the implied evolution of $\rho_{\star}$. A comparison with values from the literature is provided in Fig.~\ref{fig:smd}. Circular data points represent the value of $\rho_{\star}$ obtained from the median posterior values in Table~\ref{table:fit_total} while square-shaped data points are associated with the maximum a posteriori values. Dark (light) shaded regions correspond to $1\sigma$ (2$\sigma$) uncertainties obtained from randomly sampling the MCMC posteriors. All reported values from the literature have been converted to a Chabrier IMF and integrated across $10^{8}< \mathcal{M}/{\rm M}_{\odot} < 10^{13}$. 

From $0.2 < z \lesssim 6$, the evolution of $\rho_{\star}$ agrees well with the literature (\citetalias{Grazian2015}, \citetalias{Davidzon2017}, \citetalias{Weaver2023SMF}, \citetalias{Weibel2024}). At $z\sim4$, the same is true, although \citetalias{Weaver2023SMF} measures a slightly lower value, likely caused by their lower value of the low mass slope, $\alpha$. Given the power-law shape of the low-mass end of the SMF and the exponential decline at high-masses, $\rho_{\star}$ is significantly more sensitive to the characterisation of low-mass galaxies compared to massive galaxies. Thus, despite the abundance of massive galaxies that require extreme star-formation efficiencies at $z > 3.5$, their contribution to the SMD is dwarfed by galaxies with lower stellar masses. At $z \lesssim 4$, DAWN PL remains consistent with both \citetalias{Davidzon2017} and \citetalias{Weaver2023SMF}, except at the redshift bin corresponding to $z \leq 1.5$. At these lower redshifts, the difference in the selection of quiescent galaxies like causes a small discrepancy of less than a factor of 2. Note that at $z \leq 0.8$, DAWN PL finds a slightly greater value of $\rho_{\star}$, which may be due to the larger volume of DAWN PL that includes a great sampling of large-scale structure compared to COSMOS.

The broad agreement between the values of $\rho_{\star}$ measured from DAWN PL and the literature at-large validates the modelling procedure of Sect.~\ref{subsec:modeling}, and in particular, the treatment of the low-mass slope $\alpha$. Interestingly, at $z > 2.5$, comparison across the different works appears to show an agreement in the value of $\rho_{\star}$ that increases with publication date. In other words, the measurements which demonstrate the greatest deviations at $z > 2.5$ are also some of the earliest (e.g., before 2014). This likely reflects the improvement in data quality and may also suggest that observational methods for selecting and characterising galaxies at high redshift since improved or at least have become more uniform. At $z \leq 2.5$ a substantial fraction of the datasets, including this work, provide estimates of $\rho_{\star}$ that are lower than the values reported by \cite{Madau2014} and \cite{Behroozi2019}, at least until $z\sim0.5$.  Both \cite{Madau2014} and \cite{Behroozi2019} measure the SMD by converting measurements of the SFRD across redshifts using an assumed return fraction. Consequently, the origin of this discrepancy is not obvious, although there are many possible explanations explored by \cite{Madau2014}.

\subsection{\label{subsec:DM}Galaxy-dark matter connection}

A pillar of modern studies of galaxy evolution holds that galaxies form and evolve within dark matter halos. Consequently, the evolution of galaxies is directly linked to the evolution of their host dark matter halos \citep{Wechsler2018}. This relationship is generally referred to as the galaxy-halo connection. Numerous techniques are used in the literature to connect galaxies to dark matter haloes, for example, hydrodynamical simulations \citep{Vogelsberger2014,Schaye2015,Schaye2023}, semi-analytic models \citep{Guo2011,Henriques2015}, empirical forward modelling \citep{Behroozi2013,Behroozi2019}, abundance matching \citep{Marinoni2002,Stefanon2021}, and halo occupation distribution modelling \citep{Harikane2018,Shuntov_2022}. The mass of the host dark matter halo appears to be the dominant intrinsic property with which many aspects of evolution are correlated \citep[e.g.,][]{Contreras2015, Man2019}. Consequently, it is often a goal to measure the stellar-to-halo mass relationship (SHMR), which describes the stellar mass of galaxies as a function of the dark matter halo mass. While a detailed analysis involving the methods described above is beyond the scope of the present work, a qualitative comparison can easily be made by comparing the intrinsic SMF with the dark matter halo mass function (HMF). 

\begin{figure*}
    \centering
    \includegraphics[width=\textwidth]{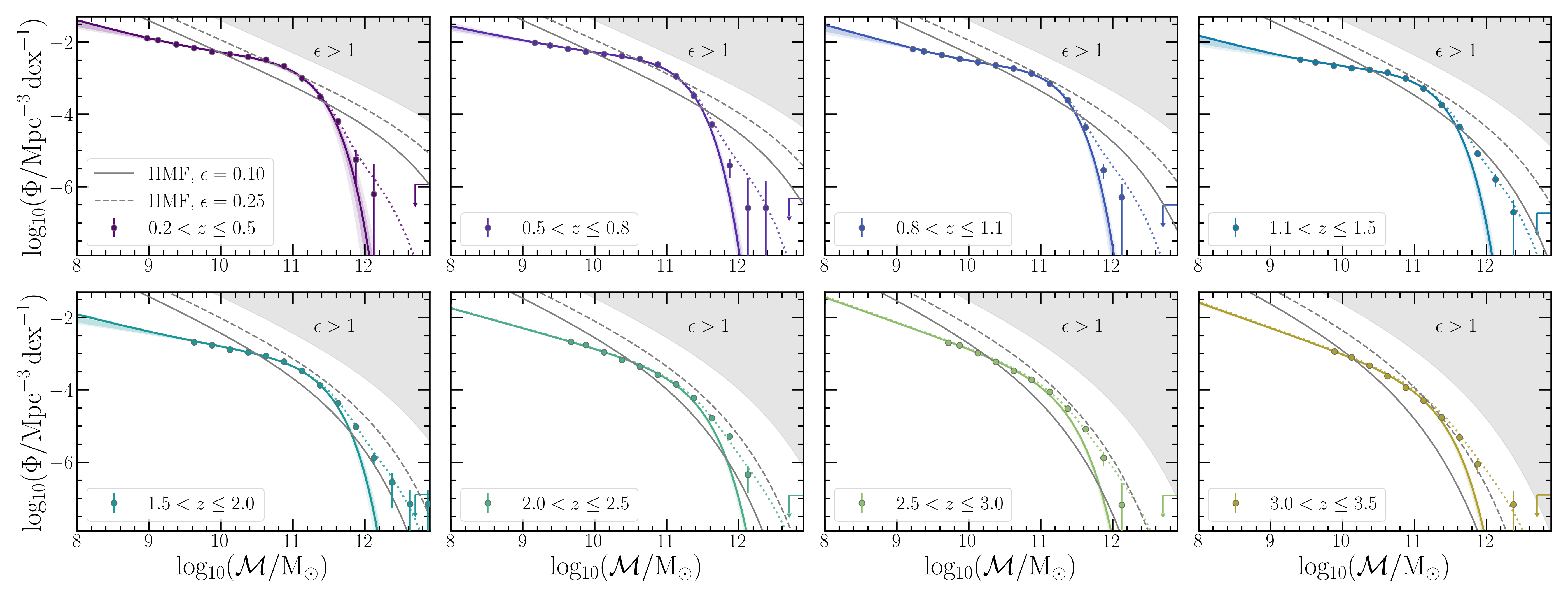}
    \caption{The intrinsic galaxy stellar mass function of the total sample through to $3.0 < z \leq 3.5$ shown as solid lines and observed values shown as data points with error bars in matching colour compared to the \cite{Tinker2008} dark matter halo mass function assuming a universal baryon fraction and various integrated star-formation (baryon-to-star conversion) efficiencies, $\epsilon$. \Ms{} bins below the stellar mass limit are not shown. Maximum a posteriori values for the intrinsic galaxy stellar mass functions are assumed. The uncertainty ($1\sigma$ dark, $2\sigma$ light) according to the variation in the MCMC chains is shown by the shaded regions. The dotted coloured line represents the \enquote{observed} model of the galaxy stellar mass function, i.e., including the Eddington bias (Sect.~\ref{subsubsec:eddington}). The shaded grey region corresponds to the number density of galaxies that would require 100\% conversion (or greater) of baryons into star.}
    \label{fig:hmf_lowz}
\end{figure*}

Like the galaxy SMF, the dark matter HMF describes the co-moving number density of dark matter halos as a function of dark matter halo mass. The HMF can be roughly translated into a predicted SMF by assuming a baryon fraction and an integrated star-formation efficiency, $\epsilon$. In SHMR studies, the value of a galaxy's ratio of stellar mass to dark matter halo mass defines the integrated star-formation efficiency, i.e., $\mathcal{M}/(\mathcal{M}_{\rm h}f_{\rm b}) \equiv \epsilon$, where $f_{\rm b}\equiv\Omega_{\rm b}/\Omega_{\rm m} = 0.166$ is the fraction of baryonic matter in the Universe. Consequently, the integrated star-formation efficiency should be thought of as the time-averaged efficiency by which baryons have been converted into stars over the lifetime of the host halo, in contrast to an instantaneous star-formation efficiency (SFR/$f_{\rm b} \dot{\mathcal{M}_{\rm h}}$) or a gas-depletion efficiency (SFR/$\mathcal{M}_{\rm gas}$). 

Figure~\ref{fig:hmf_lowz} shows the intrinsic SMF and observations at $0.2 < z \leq 3.5$ compared to the HMF of \cite{Tinker2008}. In each panel, the coloured solid curve represents the maximum a posteriori SMF, while the coloured dotted curve represents the equivalent for the observed SMF, i.e., the Schechter function (or double-Schechter function) convolved with the Eddington bias kernel of Eq.~(\ref{eq:kernel}). The HMF is scaled assuming a constant baryon fraction and three different values of $\epsilon$: $\epsilon = 0.10$, $\epsilon = 0.25$, and $\epsilon > 1$. The parameter space spanned by $\epsilon > 1$ indicates the stellar mass of galaxies that are disallowed by the assumptions of the HMF and a universal value of $f_{\rm b}$, as the formation of any galaxy with such a stellar mass would have required more baryonic matter than is available. The comparison with the scaled HMF at $z \leq 3$, shown in Fig.~\ref{fig:hmf_lowz}, suggests that galaxies at low-stellar masses and very high-stellar masses both suffer very low integrated star-formation efficiencies, as has been found in virtually all recent characterisations of the SHMR (\citealt{Wechsler2018} and citations therein). At low stellar masses, star formation can be suppressed by internal processes regulated by stellar feedback including photoionisation, supernovae, and stellar winds (e.g., \citealt{Hopkins2012}). At high stellar masses, feedback mechanisms driven by AGN and virial shock heating can suppress star formation (e.g., \citealt{Vogelsberger2013}). Galaxies near the characteristic \Ms{}$^{\star}$ of the Schechter function achieve the maximum $\epsilon$ between 10\% and 25\%. In SHMR analyses, the \Ms{} at which galaxies achieve the maximum $\epsilon$ is referred to as the \enquote{pivot-mass}. Galaxies at the pivot-mass are those that have been the most efficient at converting baryons into stars over the lifetime of the halo. Their host dark matter halos were massive enough to withstand stellar feedback but not yet massive enough to have developed dominant AGN components or to shock-heat in-falling gas. 

At $z \leq 3$, massive galaxies with $\mathcal{M} > \mathcal{M}^{\star}$ are characterised by lower values of $\epsilon$ than those with $\mathcal{M} \sim \mathcal{M}^{\star}$. However, at $3 < z \leq 3.5$, the intrinsic SMF suggests a change in the SHMR compared to low redshift. At $3 < z \leq 3.5$, essentially all galaxies observed above \Ms{}$^{\star}$ exhibit integrated star-formation efficiencies greater than or equal to 25\%. The maximum a posteriori model of the observed SMF, shown by the coloured dotted line, suggests that the abundance of the most massive systems is over-estimated due to uncertainties in their stellar masses. Even so, the area enclosed at $\epsilon > 0.10$ by the intrinsic SMF (in which the Eddington bias has been removed) is significantly greater than at lower-redshifts and includes galaxies of greater stellar mass. The increased average $\epsilon$ achieved by massive galaxies would therefore manifest in a broader peak in the SHMR, where a range of galaxy stellar masses are characterised by the maximum integrated star-formation efficiency. The work best-suited to validate this suggestion is \cite{Shuntov_2022}, wherein the authors performed a detailed halo-occupation distribution (HOD) modelling of galaxies selected from the COSMOS2020 catalogue. In this way, it is naturally complimentary to \citetalias{Weaver2023SMF}, which agrees very well with the observations from DAWN PL at $3 < z \leq 3.5$ (Fig.~\ref{fig:smf_litcomp}). Notably, \cite{Shuntov_2022} is the only work to have measured the SHMR to high $z$ ($z \leq 5.5$) using a large survey area ($> 1$ deg$^{2}$), reliable \photoz{}s, and galaxy stellar masses constrained by IR observations. Indeed, the authors find that the peak of the SHMR becomes substantially broader at $z > 2.5$, albeit with non-negligible uncertainty. A similar result was obtained by \cite{Harikane2018} from an even larger area ($\sim$100 deg$^{2}$), although the redshifts and stellar masses of the galaxies are significantly more uncertain due to a lack of rest-frame optical constraint from IR data. 

\begin{figure*}
    \centering
    \includegraphics[width=\textwidth]{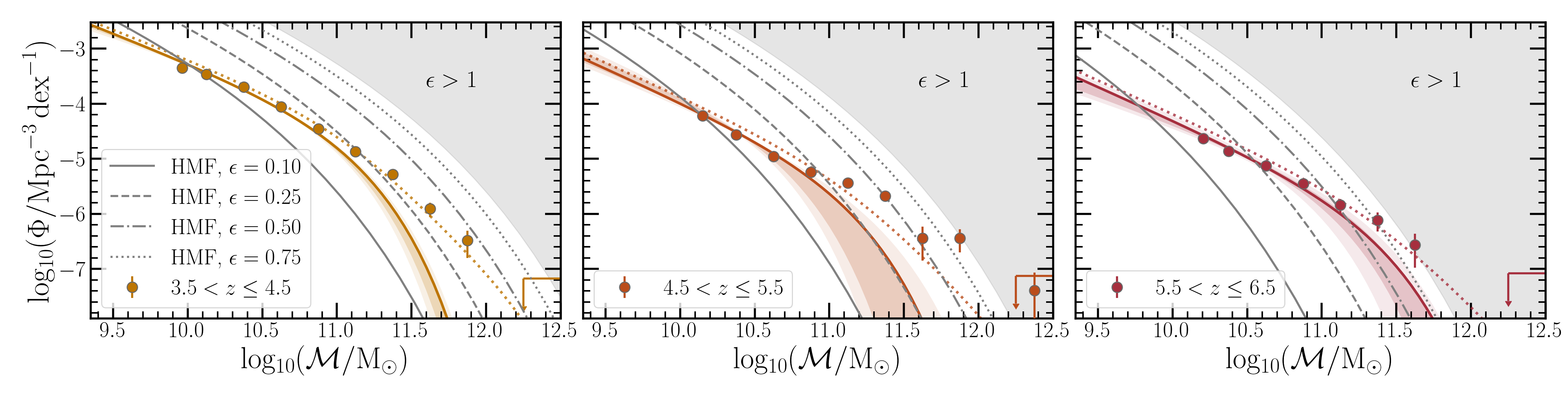}
    \caption{Same as Fig.~\ref{fig:hmf_lowz}, but at $z > 3.5$ and also showing additional values of $\epsilon$.}
    \label{fig:hmf_highz}
\end{figure*}

Another result of \cite{Shuntov_2022} suggests that at $z>3$, the most massive galaxies have larger ratios of stellar-to-dark matter halo mass compared to those at lower-redshift (i.e., they achieve greater overall integrated star-formation efficiencies). Although a careful determination of the host dark matter halos is required to say conclusively, this scenario is consistent in principle with the comparison of the SMF from DAWN PL at $3 < z \leq 3.5$ and the HMF. However, \cite{Shuntov_2022} further suggests that both effects, an increased fraction of stellar mass compared to dark matter halo mass, and a broader range of stellar masses achieving high integrated star-formation efficiencies, become stronger with increasing redshift. Figure~\ref{fig:hmf_highz} compares the intrinsic SMF and observations at $3.5 < z \leq 6.5$ compared to the \cite{Tinker2008} HMF. In Fig.~\ref{fig:hmf_highz}, two additional scalings of the HMF are shown, $\epsilon = 0.50$ and $\epsilon = 0.75$. The comparison of the intrinsic SMF and the scaled HMF at $3.5 < z \leq 4.5$ is mostly similar to $3.0 < z \leq 3.5$ but now includes galaxies that achieve an even greater integrated star-formation efficiency, in excess of $\epsilon = 0.25$ and approaching $\epsilon = 0.50$. As demonstrated by the agreement between the present observations and both \citetalias{Grazian2015} and \citetalias{Davidzon2017} in Fig.~\ref{fig:smf_litcomp2}, such galaxies have been observed before. Indeed, although their abundance may seem excessive, it is emphasised that \citetalias{Weaver2023SMF} found an even greater abundance of galaxies requiring large integrated star-formation efficiencies (see \citetalias{Weaver2023SMF} Fig. 17). The difference between the present work and these references is the greater statistical significance of their abundance, which is discussed further below. For comparison, \cite{Shuntov_2022} suggests that massive galaxies at $3.5 < z \leq 4.5$ may achieve integrated star-formation efficiencies that are a factor of $\sim2$ greater than those at $z \leq 3$, which is consistent with the $\epsilon$ required by the most massive galaxies observed in DAWN PL. 

The comparison between the intrinsic SMF and the HMF becomes more difficult at $z > 4.5$, where the inability to detect galaxies with low stellar mass weakens the constraints of the Schechter parameters. Consequently, there is some uncertainty in the shape of intrinsic galaxy SMF above $\logten(\mathcal{M} )/{\rm M}_{\odot})>11$. In any case, at $4.5 < z \leq 5.5$, there are galaxies observed that require $\epsilon > 0.50$, and even $\epsilon \geq 0.75$. This remains consistent within the 1$\sigma$ upper bound of the SHMR of \cite{Shuntov_2022}, where the authors found that the SHMR does not change significantly between $z\sim4$ and $z\sim 5$. Galaxies of approximately $\logten(\mathcal{M}/{\rm M}_{\odot}$)$\sim11.5$ require $\epsilon \sim 0.50$, according to the observations shown in Fig.~\ref{fig:hmf_highz}. The abundance of galaxies around this stellar mass is supported by \citetalias{Weaver2023SMF} and \citetalias{Weibel2024}, and \citetalias{Weaver2023SMF} actually suggests an even greater abundance of these galaxies (see again, Fig. 17 of \citetalias{Weaver2023SMF} and their Section 6.3, and also \citetalias{Weibel2024} Fig. 11) therefore requiring an even greater integrated star-formation efficiency. However, these works do not find galaxies of greater stellar mass. One galaxy, with a statistical significance that does not exceed the 1$\sigma$ upper bound for $N=0$ detections, is found with $\epsilon > 1$. This galaxy is discussed Sect.~\ref{subsec:massive}. The trend continues $5.5 < z \leq 6.5$ where again there are galaxies that require $\epsilon \geq 0.75$. Generally, the abundance of such galaxies is consistent with both \citetalias{Weaver2023SMF} and \citetalias{Weibel2024}, though it is noted that this comparison is highly uncertain, as neither of these two works can do more than provide upper limits on the abundance of galaxies with these stellar masses $\logten(\mathcal{M}/{\rm M}_{\odot}$)~$>10.75$ (see Fig.~\ref{fig:smf_litcomp2}).

The existence of a substantial population of galaxies at $z \sim 5$--$6$ with integrated star-formation efficiencies of $\epsilon \gg 0.25$ implies that the feedback mechanisms that inhibit star formation in massive galaxies at lower-redshifts are not as effective in the early Universe. To start, it is expected that these massive systems reside in dark matter halos in which feedback from stellar evolution (e.g., supernovae) is not efficient. From a theoretical perspective, \cite{Bassini2023} developed simulations in the Feedback In Realistic Environments (FIRE) project \citep{Hopkins2014,Hopkins2018} to assess the possible impact of stellar feedback on star formation in massive galaxies ($\mathcal{M}_{\rm h}\sim10^{12}\mathcal{M}_{\odot}$) at $z\geq 5.5$. Indeed, the authors found that stellar feedback is extremely inefficient for regulating star formation due to the significant gas surface densities common at these redshifts ($10^{2}$--$10^{3}$\,M$_{\odot}$ pc$^{-2}$), and that gas is consumed by star formation on shorter timescales (gas depletion time $\sim 20$ Myr) than it can be expelled. However, in less massive galaxies ($\mathcal{M} <10^{10}$ M$_{\odot}$), stellar feedback regulates star formation. Although perhaps coincidental, this is very nearly the value at which the intrinsic and observed SMF goes below $\epsilon = 0.10$ in Fig.~\ref{fig:hmf_highz}.

As discussed above, AGN activity is commonly cited as the primary factor inhibiting star formation in massive galaxies (\citealt{Wechsler2018} and references therein). However, a quantitative and detailed understanding of AGN feedback at high redshift that is empirically confirmed is not yet developed. To measure the impact of AGN feedback at lower-redshift ($z \leq 2.5$), \cite{Fiore2017} used a combination of measurements from molecular, ionized, broad absorption line, and X-ray winds and found that only galaxies with $\mathcal{M} \geq 10^{11}$ M$_{\odot}$ at $z \leq 2$ are capable of driving mass outflows rates that exceed the star-formation rate with a trend that decreases with increasing redshift. Through observation, it is difficult to fully assess the effects of outflows in galaxies at $z\sim 5$--6. At $z\geq 5$ far-IR observations with high-resolution and depth are necessary to detect outflows and measure their spatial distributions \citep{Fujimoto2019}, making statistical arguments difficult due to small sample sizes. \cite{Valentini2021} produced simulations of massive galaxies ($\mathcal{M}_{\rm h} = 10^{12}$ M$_{\odot}$) at $z = 6$ using GADGET3, a private successor to GADGET2 \citep{Springel2005}, with detailed characterisation of AGN activity. Ultimately, the authors found that AGN feedback could not significantly halt the inflow of cold gas or influence the star-formation history of the galaxy, even when manually increasing the efficiency of AGN feedback by a factor of 10.

Observations at lower-redshifts have historically implied a period of enhanced star-formation efficiency for the most massive galaxies in the early Universe \citep{Cowie1996}. For example, should massive galaxies at $z > 3.5$ undergo a period of enhanced star-formation efficiency, then it should be expected that a substantial abundance of massive quiescent galaxies is observed some time thereafter. Indeed, the evolution of both the observed quiescent SMF (Fig.~\ref{fig:smf_litcomp_bytype}) and the intrinsic quiescent SMF (Fig.~\ref{fig:sfq_mcmc}) suggest that a significant fraction of the most massive quiescent galaxies are already in place by $z\sim2.5$, just 2.57 Gyr after the Big Bang. Quiescent galaxies at this mass do not appear to substantially grow in abundance from then onward, nor does it appear that more massive quiescent systems arise later. Accordingly, the effects of AGN feedback, which were not yet strong enough to inhibit star formation at high $z$, may be delayed (e.g., \citetalias{Davidzon2017},\citealt{Valentini2021}). By $z\sim2$ at least, these systems are more massive than the minimum stellar mass found by \cite{Fiore2017} that is needed for AGN feedback to completely \enquote{clean} galaxies of their molecular gas. Assuming feedback is strong and gas supplies are absent, the only evolutionary pathway for such systems to grow in abundance is through merging events. \cite{Mundy2017} showed that on average, galaxies with $\mathcal{M} >10^{11}$ M$_{\odot}$ undergo approximately 0.5 major mergers across $0 < z \leq 3.5$, a process which adds typically 1--4\,$\times10^{10}$ M$_{\odot}$. Minor mergers should also be considered and are important at low $z$ \citep{Jespersen2022_merger_tree, Chuang2024_merger}, but it is difficult to measure their impact at high $z$ \citep{Lotz2010}. \cite{Conselice2022} found a slightly greater major merger rate at $\sim0.84$ than \cite{Mundy2017} and also found that the growth in stellar mass for the same galaxies is approximately a factor of 2 (0.3\,dex). The evolution in the shape of the observed quiescent SMF (Fig.~\ref{fig:smf_litcomp_bytype}) and total SMF with respect to the scaled HMF is consistent with this picture of little to no growth in the abundance of the most massive galaxies. This was found by \cite{Kawinwanichakij2020} this results extends the finding to $z\sim2.5$ and tentatively higher, considering the total sample.

Recent observations of exceptionally massive galaxies from JWST at $z > 8$ and above \citep{Harikane2023,Casey24,Weibel2024} imply the existence of similarly massive galaxies at later times. For example, \citetalias{Weibel2024} finds several galaxy candidates at $z \sim 8$--9 that would require equal or greater integrated star-formation efficiencies to those found in DAWN PL. In addition, \cite{Casey24} reports the discovery of three candidate galaxies at $z \sim 12$ with $\mathcal{M} \sim  4$--$10\times10^{9}$ M$_{\odot}$. 
These systems require extreme evolutionary histories acting on short time scales. Several possible theories have been put forth to explain them, including \enquote{feedback free} bursts of star formation \citep{Dekel2023,Li2023} and positive feedback from AGN \citep{Silk2024}. Without a measure of current star-formation activity in the massive galaxies found in DAWN PL, there is a degeneracy among evolutionary histories that would yield similar values of $\epsilon$ required for them to form, given that $\epsilon$ is a quantity that is integrated over the entire evolutionary timescale of the host halo. Spectroscopic confirmation, and a spectro-photometric analysis, is ultimately required to understand the nature of these massive galaxies in detail.

\subsection{\label{subsubsec:env_dep}Galaxy environment}

The results discussed in Sect.~\ref{subsec:DM} motivate investigation of a possible correlation between galaxy stellar mass and the local density field, or environment. The SHMR \citep{Wechsler2018,Behroozi2019,Shuntov_2022} and other observations, e.g., galaxy bias \cite{Beck2019}, show that the galaxies with the greatest stellar mass must also reside in the most massive dark matter halos. In the early Universe, the most massive dark matter halos are also the sites of large-scale structure formation \citep{Springel2006,Porqueres2019}. Consequently, there is reason to suspect a relationship between stellar mass and environment, even at high $z$.

Determining the role of the environment in affecting galaxy evolution has been a topic of study for decades (e.g., \citealt{Oemler1974}; see also \citealt{Conselice2014} and citations therein). Although environments may be characterised in different ways, it is common to describe environment in relation to the local density by measuring the relative \enquote{overdensity}, which is the difference in the local density compared to the average density; this is the definition used herein. Advancements in galaxy surveys across the past two decades have enabled studies seeking to link the role of environment in galaxy evolution, specifically through its impact on the galaxy SMF \citep{Bundy2006,Bolzonella2010,Peng2010}. These early works showed a greater abundance of massive galaxies in high-density environments. \cite{Peng2010}, in particular, introduced the concepts of \enquote{environmental quenching} and \enquote{mass quenching} which are now commonly used to describe two different ways galaxies are thought to transition from star-forming to quiescent. This has also been investigated in low-$z$ simulations, where \cite{Wu_Jespersen_env_2023, Wu2024_env} find strong environmental dependencies in galaxy properties.

At redshifts $z > 1$ it has generally been a challenge to measure the galaxy SMF as a function of environment because the high-redshift surveys were either too small in area to sample cosmologically distinct environments or did not have the necessary infrared wavelength coverage (e.g., from \textit{Spitzer}/IRAC) to reliably measure stellar masses at high $z$. Nonetheless, efforts to measure the impact of environment on the resulting galaxy SMF at higher redshifts are continually improving. As a few examples, \cite{Bolzonella2010} used some ten thousand galaxies from zCOSMOS \citep{Lilly2007} sampling a 1 deg$^{2}$ area to $z \sim 1$; \cite{Etherington2016} used several million galaxies across a few hundred deg$^{2}$ of the Dark Energy Survey (DES; \citealt{DES}) to $z \sim 1$; \cite{Davidzon2016} used \num{50000} galaxies from the VIMOS Public Extragalactic Redshift Survey (VIPERS; \citealt{Guzzo2014}) spanning $0.5 < z < 1$; \cite{Tomczak2017} analysed galaxies selected from 8 individual fields, each spanning between $\sim$150--450 arcmin$^{2}$, of the Observations of Redshift Evolution in Large-Scale Environments survey (\enquote{ORELSE}; \citealt{Lubin2009}) at $0.6 < z < 1.3$; \cite{Kawinwanichakij2017} used galaxies selected from the $\sim$300 arcmin$^{2}$ FourStar galaxy evolution survey (ZFOURGE; \citealt{Straatman2016}) out to $z\sim2$; \cite{Papovich2018} used the same galaxies at \cite{Kawinwanichakij2017} as well as an additional sample selected from the Newfirm Medium Band Survey (NMBS; \citealt{Whitaker2011}) spanning $\sim$1500 arcmin$^{2}$ again to $z\sim2$; and \cite{Forrest2024} considered two individual protoclusters at $z\sim2$ and at $z\sim3$ selected from COSMOS2020 \citep{Weaver2021}. 

Future investigations will undoubtedly make use of DAWN PL's significant volume, the imminent grism spectroscopy from \Euclid \citep{Scaramella22}, and reliable stellar masses provided by the deep \textit{Spitzer}/IRAC data to explore many facets of galaxy evolution in relation to the environment. Although such detailed analyses are currently beyond the scope of this work, enough physical parameters have already been measured and formalisms defined to obtain a simple estimate of the abundance of galaxies as a function of their stellar mass and environment. It should be noted that despite the uniquely large area and rest-optical coverage at these redshifts, the limiting factor in this preliminary investigation is the photometric redshift precision, relative to similar, spectroscopically-based studies in the literature.

The local density of every galaxy is estimated using a simplified implementation of the procedure described by \cite{Kovac2010} for measuring the density field of the zCOSMOS spectroscopic sample \citep{Lilly2007} over a 1 deg$^{2}$ area. The same measurements were used by \cite{Peng2010}. Following \cite{Kovac2010}, the local density around each galaxy is estimated using an aperture of variable size. The radius of the aperture is determined by the distance to the $N$th nearest neighbour, where $N=5$ \citep{Kovac2010,Peng2010,Davidzon2016}. The local density measurement is thus $D = 5/(\pi d_{5}^{2})$, where $d_{5}$ is the distance to the $5$th nearest neighbour. Galaxies are each given equal weight such that the resulting density measurement simply describes the counts of objects within the aperture to avoid biasing the overdensity measurement around central objects of significant stellar mass \citep{Kovac2010}. For the tracer sample, all galaxies above the stellar mass limit (Sect.~\ref{subsec:m_completeness}) are used for simplicity and consistency with the above selections (Sect.~\ref{subsec:selection}) of galaxies used to construct the SMFs. Galaxies are also binned according to the redshift binning used throughout (i.e., Table~\ref{tab:volumes}). Note that this redshift spacing is approximately similar to the spacing used by \cite{Papovich2018}. Each galaxy is assigned an overdensity value, $\delta$, which is the difference between the density measured at the position of the galaxy and the average density considering all galaxies and their respective density values, divided by the average density. Finally, galaxies are grouped into quartiles of overdensity based on the distribution of $\logten$(1+$\delta$). 

\begin{figure}
    \centering
    \includegraphics[width=\linewidth]{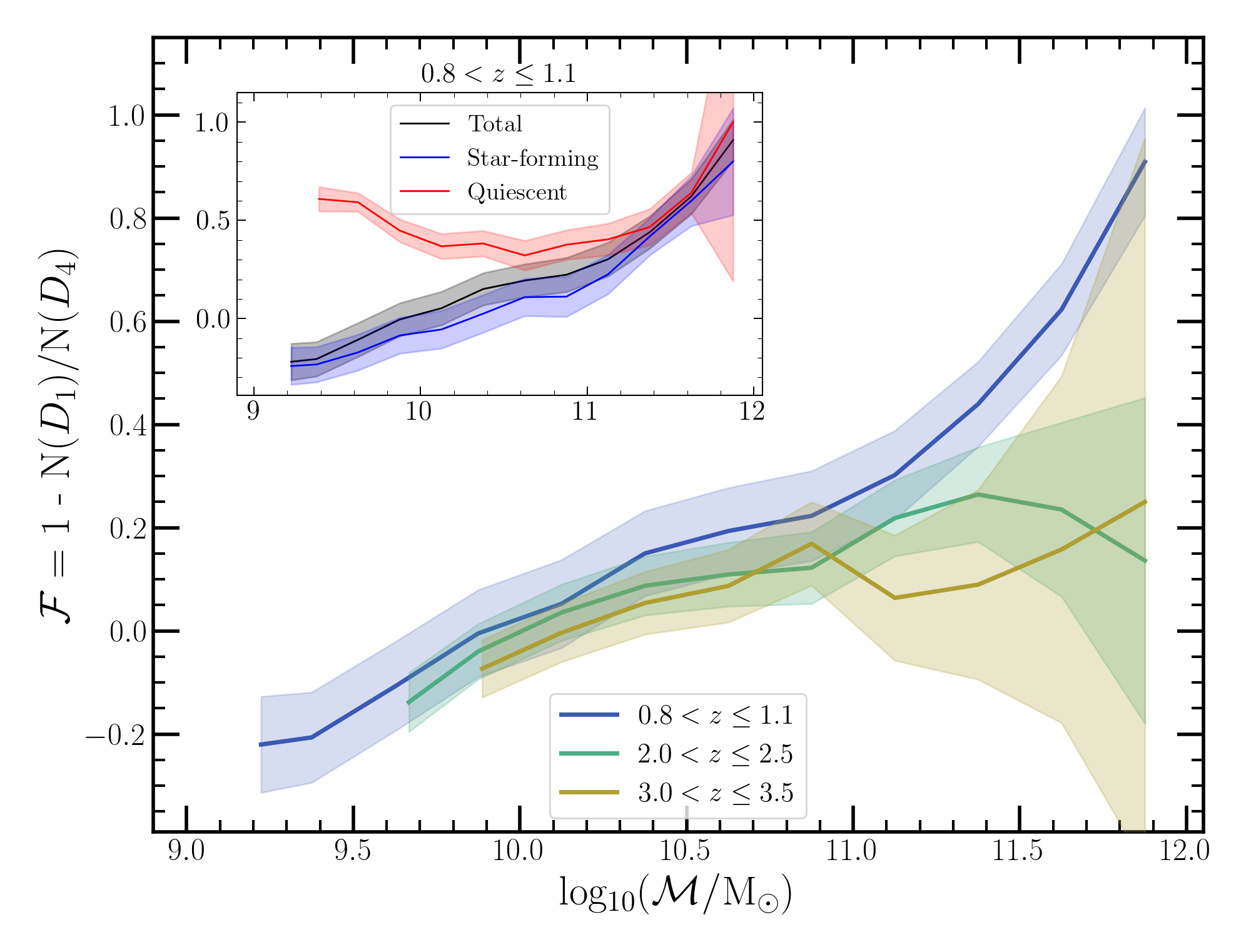}
    \caption{The fraction of all galaxies contributed to each \Ms{} bin by the most dense environments, $D_{4}$, compared to the least dense environments, $D_{1}$, relative to unity, $\mathcal{F} = 1 - N(D_{1})/N(D_{4})$. See text for a description of the density estimates. For $\mathcal{F} < 0$, more galaxies are contributed by the least-dense environments; $\mathcal{F} = 0$ implies the two quartiles contribute equally; and $\mathcal{F} > 0$ indicates that more galaxies are contributed by the most dense environments. The solid coloured lines indicates the ratio of the counts in each quartile (subtracted from 1), and the shaded regions correspond to the $1\sigma$ uncertainties propagated from $\sigma_{\Phi}$ (Sect.~\ref{subsec:errors}). At low masses ($\mathcal{M} <10^{10}$ M$_{\odot}$), more galaxies are found in low-density environments, while galaxies above are preferentially located in high-density environments. The inset plot shows the same relationship for $0.8 < z \leq 1$ separated into star-forming and quiescent galaxies. The over-abundance of low-mass quiescent galaxies in high-density environments is indicative of \enquote{environment quenching} \citep{Peng2010,Peng2012}.}
    \label{fig:smf_env}
\end{figure}

The galaxy SMF is measured across each redshift bin and for each density quartile. Uncertainties $\sigma_{\Phi}$ in each \Ms{} bin are computed as described in Sect.~\ref{subsec:errors}, and volumes treated as described in Sect.~\ref{sc:formalism}. As is commonly done in the literature, the change in the shape of the galaxy SMF is measured by comparing the resulting SMFs from the most dense quartile, $D$4, to the least dense quartile, $D$1. Similar to \cite{Tomczak2017}, the ratio of each SMF is measured here for each redshift bin considered. The result is summarised in Fig.~\ref{fig:smf_env}. The $y$-axis is given by 
\begin{equation}
\mathcal{F} = 1 - N(D_{1})/N(D_{4}),
\end{equation}
where $\mathcal{F}$ is the fraction of galaxies contributed to each \Ms{} bin by the most dense quartile relative to unity. For example, at $\mathcal{F} < 0$, more galaxies are contributed by the least-dense environments; $\mathcal{F} = 0$ implies the two quartiles contribute equally; and $\mathcal{F} > 0$ indicates that more galaxies are contributed by the most dense environments. The solid coloured lines represent the ratio of the counts in each quartile (subtracted from 1), and the shaded regions correspond to the $1\sigma$ uncertainties propagated from $\sigma_{\Phi}$ (Sect.~\ref{subsec:errors}). The two lowest redshift bins $0.2 < z \leq 0.5$ and $0.5 < z \leq 0.8$ are not shown as they show no considerable change compared to $0.8 < z \leq 1.1$. However, the increased abundance of massive galaxies in dense environments is obvious at these low redshifts and is generally reminiscent of \cite{Peng2010} and the other low-redshift works. Thus, Fig.~\ref{fig:smf_env} begins with $0.8 < z \leq 1.1$ to provide a point of comparison with recent works \citep{Etherington2016,Tomczak2017,Papovich2018} reaching to higher redshifts. In the primary area of the figure, the value of $\mathcal{F}$ is given for the total SMF for each of the redshift interval shown. 

Figure~\ref{fig:smf_env} shows that the abundance of galaxies as a function of stellar mass depends on the local environment, at least from $0.8 < z \leq 1.1$ out to $3.0 < z \leq 3.5$. The analysis does not go above $z > 3.5$ because it is not possible to measure a reliable signal given the current uncertainty in the data. Each redshift interval shows that the least massive galaxies are more abundant in environments with low-density, i.e., $\mathcal{F} < 0$. Note that the shape of the total SMF becomes steeper with increasing redshift. Consequently, even though the range of stellar masses with $\mathcal{F} < 0$ decreases with increasing redshift, their relative contribution to the total fraction of galaxies in the entire redshift bin grows. The result appears consistent with many previous works, in particular \cite{Davidzon2016} and \cite{Tomczak2017} which both found obvious increases in the abundance of massive galaxies of all types in high-density regions, at least at $z\sim1$. In the inset, the $0.8 < z \leq 1.1$ comparison is further broken down into star-forming and quiescent galaxies (using the selection criteria of Sect.~\ref{subsec:SF_Q}). Unsurprisingly, the quiescent galaxies show a strong dependence on environment at all stellar masses. Notably, low mass quiescent galaxies, e.g., $\logten(\mathcal{M} /{\rm M}_{\odot}) < 10$, show an increased abundance in dense regions compared to galaxies of intermediate mass, e.g., $10 < \logten(\mathcal{M} /{\rm M}_{\odot}) < 11$. This may be understood as a result of \enquote{environmental} quenching \citep{Peng2010,Peng2012}. The trend reverses at $\logten(\mathcal{M}/{\rm M}_{\odot}) > 11$, and at $\logten(\mathcal{M}/{\rm M}_{\odot}) > 11.5$ the environmental dependence becomes statistically indistinguishable from the relationship shown by the star-forming galaxies.

The redshift bin $0.8 < z \leq 1.1$ shows the strongest environmental dependence and the trend appears to continue at $2.0 < z \leq 2.5$, although with a weaker slope at high masses, e.g., $\logten(\mathcal{M}/{\rm M}_{\odot}) > 11$. By contrast, \cite{Papovich2018} found no significant dependence on environment at these redshifts. However, it should be noted that the area explored by \cite{Papovich2018} was approximately 20$\times$ smaller than DAWN PL and therefore likely sampled a smaller range of physical structures and the effect of cosmic variance is higher. The role of environment appears weak through to $3.0 < z \leq 3.5$, although the correlation appears significant, at least until $\logten(\mathcal{M}/{\rm M}_{\odot}) \sim 11$, above which the uncertainty is consistent with no-environmental dependence ($\mathcal{F} = 0$). No other work has yet been able to place constraints on the environmental dependence of the galaxy SMF at these early times, approximately 2 billion years after the Big Bang.  

The exact interpretation of Fig.~\ref{fig:smf_env} requires caution, given the very basic estimation of local density as well as the uncertainties in the \photoz{}s. In particular, a definitive evolutionary trend, or change in the dependence on environment over cosmological time, is likely inaccessible at present. Although each redshift bin includes plenty of galaxies to calculate the density field (e.g., Table~\ref{tab:volumes}), there has not been an attempt to account for any expected evolution in the tracer sample from one redshift bin to another, and consequently the meaning of the density field may not be uniform across redshifts. That said, it should still be reasonable to compare high and low-density regions consistently within each redshift bin. Moreover, the result appears reasonable in comparison to some of the results in the literature at $z\sim 1$ and is also supported by the discussion in Sect.~\ref{subsec:DM}.

\subsection{\label{subsec:massive}Systematic uncertainties in the abundance of massive galaxies}

The discussion in Sect.~\ref{subsec:DM} argued that the abundance of massive galaxies at $z > 3.5$ requires integrated star-formation efficiencies in excess of $\epsilon = 0.25$. Although such systems may be above what is conventionally expected (e.g., from \citealt{Behroozi2019}),  systems with similar stellar masses and abundance have been observed for nearly a decade, at least at $3.5 < z \leq 4.5$ \citep{Grazian2015,Davidzon2017} and again more recently \citep{Weaver2023SMF,Weibel2024}. The difference now is that such galaxies are observed in DAWN PL with a statistical significance that should be addressed. In particular, the abundance of the galaxies that require great integrated star-formation efficiencies cause the characteristic mass of the Schechter function to grow at high redshift. Such an excess may imply one of three scenarios: (1) the Eddington bias is underestimated, (2) the massive galaxies are dramatically affected by some systematic bias, or (3) that, after a full accounting of the most massive galaxies, the Schechter formalism does not apply at $z > 3.5$. 

Beyond the excess of massive galaxies at $z > 3.5$, there does not appear to be evidence of scenario (1). As described in Sect.~\ref{subsubsec:eddington}, the shape of the Eddington bias kernel (Eq.~\ref{eq:kernel}) was derived from the PDF(\Ms{}) obtained from SED fitting. This is a common procedure in the literature \citep{Ilbert2013,Grazian2015,Davidzon2017,Adams2021,Weaver2023SMF,Weibel2024}. Moreover, the resulting shape is significantly broader than what was used by both \citetalias{Davidzon2017} and \citetalias{Weaver2023SMF}. Although scenario (3) may be true, especially given the discussion presented in Sect.~\ref{subsec:DM} regarding the influence of feedback mechanisms at $z > 3.5$, it is not possible to definitively conclude one way or the other without spectroscopic characterisation of a statistically significant sample of the massive galaxy candidates. 

At present, it appears worthwhile to further consider scenario (2), the possible impact of systematic biases on the most massive galaxies. Perhaps the most obvious source of systematic bias may be from SED modeling \citep{Marchesini2009}. Various works have investigated the consistency of stellar mass estimates obtained from different codes \citep{Mobasher2015,Pacifici2023}, finding variation at the order of $\sim0.15$\,dex depending on the method. Variation of this scale appears to be a reasonable characterisation according to tests that have investigated the constraining power of the DAWN PL filter set using COSMOS2020 presented in \cite{Chartab2023} and \citetalias{Zalesky2024} (as well as Appendix~\ref{app:validation}). In addition, the consistency in the SED modelling procedure between \citetalias{Davidzon2017}, \citetalias{Weaver2023SMF}, and this work should at least enable straightforward comparison of galaxies and their stellar masses. It is possible that contribution from dust and metallicity are significant \citep{Mitchell2013} for some galaxies, however. The setup of \lephare{} used here, and also in \citetalias{Davidzon2017} and \citetalias{Weaver2023SMF}, allows for a range of reddening due to dust though only a few discrete values of metallicity to limit degeneracy \citep{Ilbert2013,Laigle2016}. That said, \cite{Mitchell2013} showed that uncertainty in metallicity most significantly affects galaxies of low to intermediate stellar mass ($\mathcal{M}<10^{10.5}$ M$_{\odot}$). The stellar masses of the most extreme galaxies may be best understood after obtaining spectroscopic confirmation of their redshift.

\subsubsection{\label{subsubsec:interlopers}Galaxy misclassification}
Massive galaxies at high redshift may be mistaken for other objects. For example, high-$z$ galaxies are at times difficult to distinguish from dwarf stars (e.g., \citealt{Stanway2008}), in particular L- and T-dwarfs, which are faint in the optical but bright in infrared wavelengths. Although more easily identified by their near-infrared fluxes, these systems can also be based on their $z-$\chOne{} colours and \chOne{}$-$\chTwo{} colours. In addition, the high-resolution of Subaru HSC also allows for the consideration of morphology. To this end, all of the galaxy candidates with excessive integrated star-forming efficiencies (i.e., $\epsilon \geq 0.25$) were visually inspected and obvious point sources flagged. The total number of objects that appear point-like form a small fraction and their removal from the sample has no significant impact on the results here. Bright foreground stars of other types can also be a problem by biasing the fluxes measured in their vicinities. The impact of bright ($i < 14$ mag) foreground sources on resulting stellar masses is investigated by measuring the nearest neighbour to every galaxy binned by stellar mass and redshift. There is no statistically significant correlation between bright foreground objects and massive galaxies. 

Another systematic bias may be introduced by the unaccounted for presence of AGN. Using \lephare{}, a set of AGN/QSO templates are fit to every object alongside galaxy and stellar templates following \cite{Weaver2021} (see also \citealt{Laigle2016}). There do not appear to be any massive galaxies that have SEDs that cannot be explained predominantly by stellar emission. However, it is difficult to observationally distinguish AGN without mid/far-infrared imaging, and therefore some of the best-fit AGN templates are degenerate with the galaxy solution. As pointed out by \citetalias{Weaver2023SMF}, it is not clear whether AGN at $z > 3$ are bright enough in the rest-frame optical/NIR to significantly skew stellar mass measurements and resulting constraints on number densities. For example, the authors point to \cite{Ito2022}, which showed that although AGN appear common at these redshifts, their optical emission is more than an order of magnitude less than what is emitted in stellar light from their host galaxies. Recently, \cite{Guo2024} measured the abundance of AGN at $z\sim5$ compared to ordinary galaxies and found that their abundance is between 0.5--1\,dex smaller than Lyman break galaxies. As such, it is unlikely for a randomly selected galaxy to suffer severe AGN contamination. 

The unaccounted for contribution of an AGN is likely to have a more significant effect at low redshift. Using a sample of $\sim$900 X-ray luminous AGN at $0.5 < z \leq 3$, \cite{Florez2020} showed that stellar mass estimates become increasingly overestimated with increasing $f_{\rm AGN}$, where $f_{\rm AGN}$ is the fraction of emission from 8--1000$\,\mu$m contributed by AGN relative to stars. At $f_{\rm AGN} > 0.6$, stellar masses can be overestimated by up 0.5\,dex (with some scatter). This result could significantly impact the massive end of the SMF if galaxies with very luminous AGN are ubiquitous in massive galaxies. However, the authors compare the abundance of luminous AGN to galaxies without X-ray counterparts and find that the latter are 2 orders of magnitude more abundant at $\logten(\mathcal{M}/{\rm M}_{\odot}) > 11$. While AGN contribution is an important consideration in individual objects, the abundance of luminous AGN (or lack thereof) suggests their cumulative effect on the SMF may not be not substantial. Indeed, regardless of redshift, removing all galaxies that have smaller $\chi^{2}$ values in the best-fit AGN templates compared to the best-fit galaxy template has a negligible impact on the galaxy SMF.  

Recently, \cite{Forrest2024B} showed that the number density of massive galaxies, e.g., $\logten(\mathcal{M}/{\rm M}_{\odot}) > 11.5$, may be overestimated at $3 < z < 4$ due to the misclassification of very dusty galaxies at low $z$. Generally, their result likely has some applicability to the massive galaxies found here at $3 < z < 4$, although the extent is unclear. In some cases, the $P(z)$ measured by \lephare{} shows two peaks, and a secondary \photoz{} solution can be inferred where the secondary solution is implies a low-redshift dusty galaxy. However, in detail, it is not immediately clear that the galaxies targeted by \cite{Forrest2024B} are detected in DAWN PL. First, the galaxies are selected from near-infrared imaging and photometry rather than from optical. More importantly, both their photometric- and spectroscopic-inferred rest-frame colours (e.g., Figs. 1 and 5 of \citealt{Forrest2024B}, respectively) indicate a parameter-space that is preclued by DAWN PL's optical selection function. The former partially overlap with the $UVJ$ selection boundary for quiescent systems, while the latter shows that these galaxies are, in fact, extremely dusty. Although there are differences between the $UVJ$-selections and NUV$rJ$-selections (Sect.~\ref{subsec:SF_Q}), there are virtually zero quiescent galaxy candidates found at $z > 3$ in DAWN PL, and even the \enquote{dusty} quadrant of the NUV$rJ$ diagram is essentially empty above $z > 3$. Consider further the comparison between this work, \citetalias{Davidzon2017}, and \citetalias{Weaver2023SMF} at $3.0 < z \leq 4.5$ in Fig.~\ref{fig:smf_litcomp} and Fig.~\ref{fig:smf_litcomp2}. The excess of \citetalias{Weaver2023SMF} over \citetalias{Davidzon2017} and this work appears to be driven by galaxies that are faint and intrinsically redder than what can be detected from the HSC $r+i+z$ selection. Consequently, it is possible that the interloper population identified by \cite{Forrest2024B} does not imply a drastic reduction in the number-densities of massive galaxies presented here.

\subsubsection{\label{subsubsec:images}Ultra-massive galaxies}

\begin{figure*}
    \centering
    \includegraphics[width=0.65\textwidth,trim={30mm 5mm 25mm 0},clip]{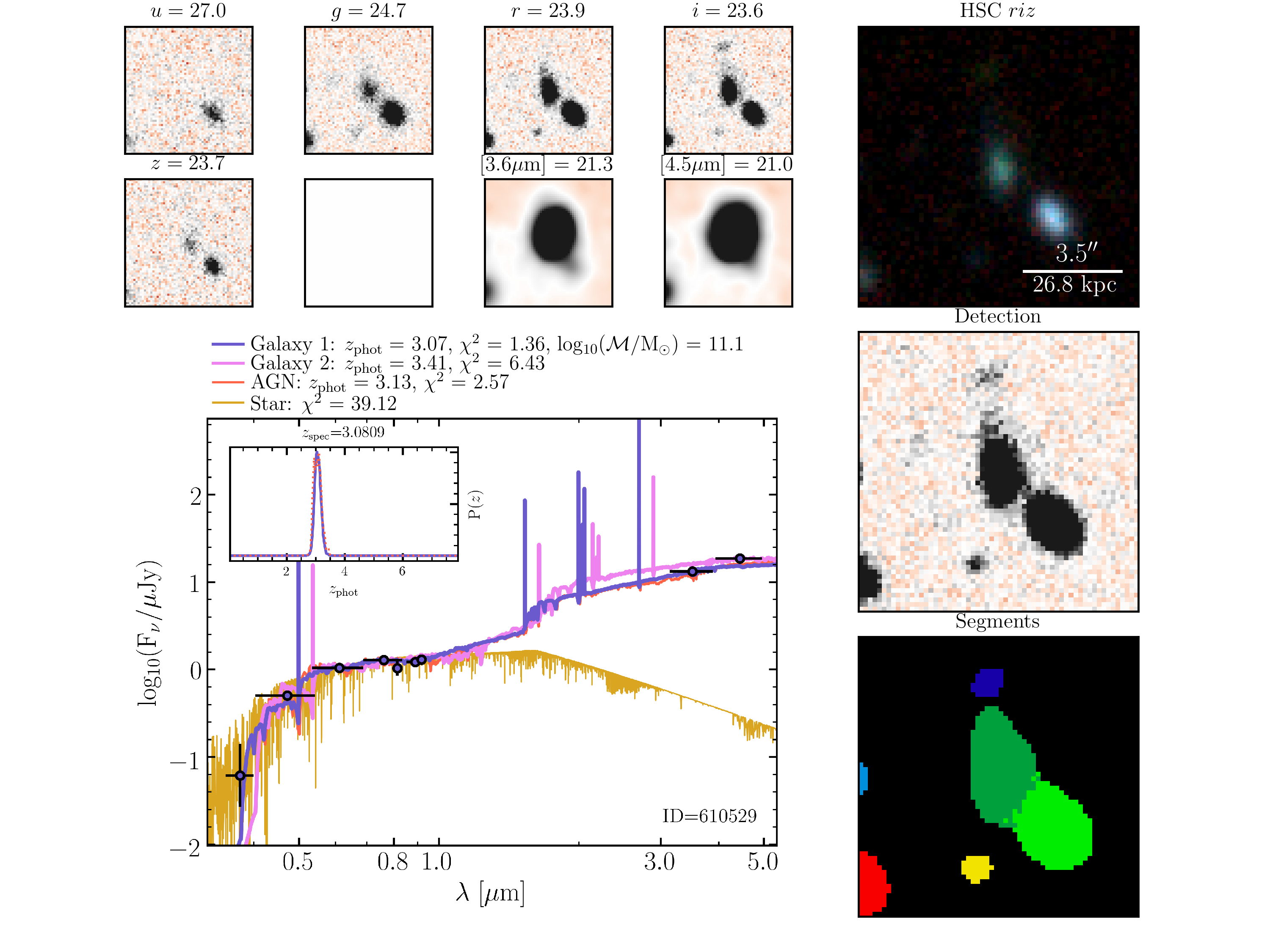}
    \includegraphics[width=0.65\textwidth,
    trim={30mm 5mm 25mm 0},clip]{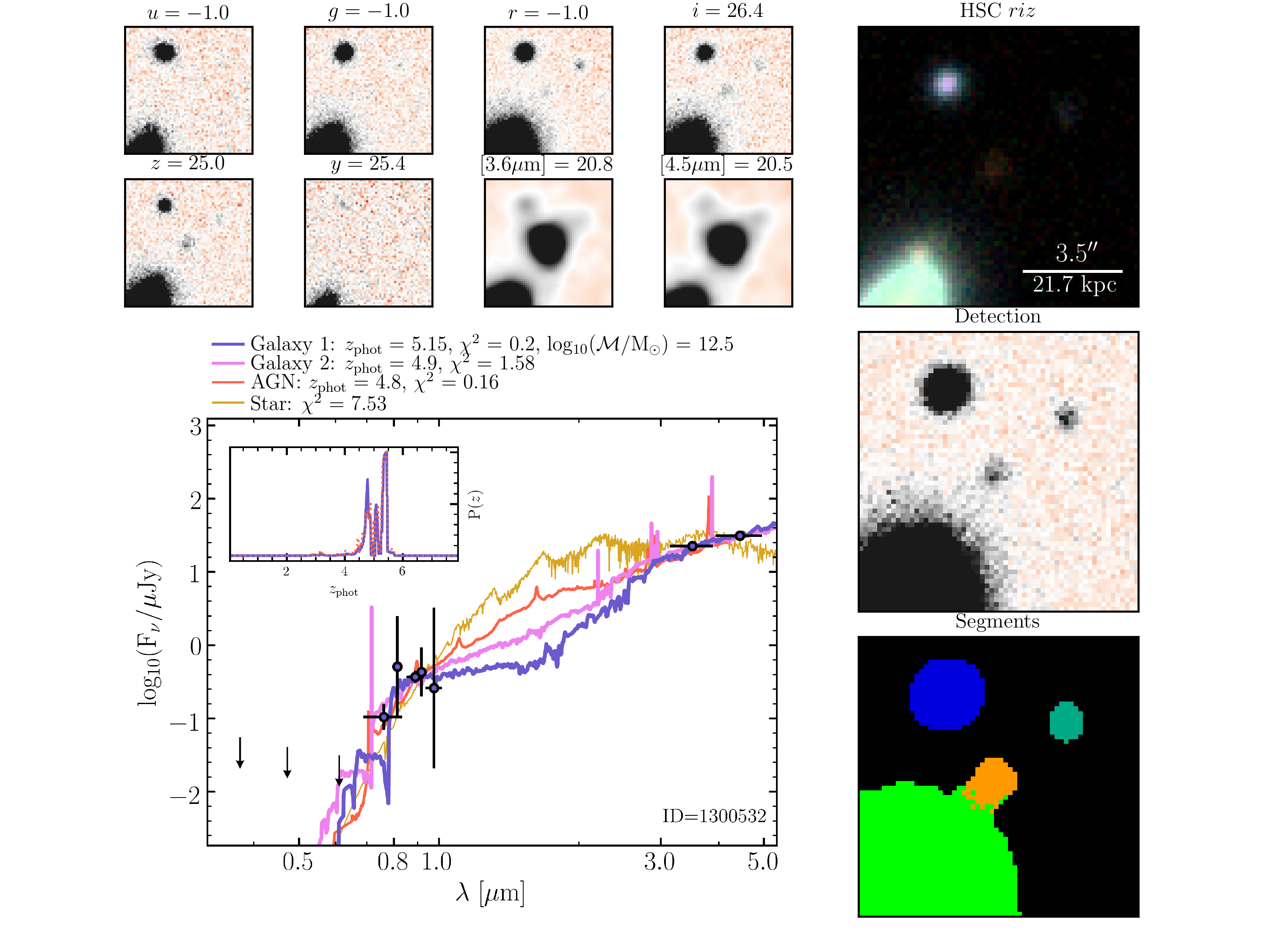}
    \caption{Illustrative data for two galaxies observed by DAWN PL. See text for details. Top: The most massive galaxy from DAWN PL with a reliable spectroscopic redshift at $z > 3$, with $\logten(\mathcal{M}/{\rm M}_{\odot})=11.1\pm0.1$. The galaxy is located in EDF-F (which does not include HSC $y$ imaging) and the reported ID corresponds to the DAWN PL EDF-F catalogue. The spectroscopic redshift is obtained from cross-matching to with GOODS-S catalogue \citep{Garilli2021,Kodra2023}. Bottom: The most massive galaxy observed at $4.5 < z \leq 5.5$. The stellar mass is highly uncertain at $\logten(\mathcal{M}/{\rm M}_{\odot})=12.5^{+0.1}_{-0.9}$, likely due to uncertainty in the dust content. The galaxy is located in EDF-N and the reported ID corresponds to the DAWN PL EDF-N catalogue. }
    \label{fig:specz}
\end{figure*}

This work introduces a hitherto unexplored region of parameter space occupied by ultra-massive galaxies. Seen only in smaller surveys by chance, their impact on the bright-end of the SMF can be enormous (see Sect.~\ref{subsec:eddingtin_demo}). This seciton presents some of the most convincing candidates yet found.


Figure~\ref{fig:specz} presents the most massive galaxy at $z > 3$ with a reliable spectroscopic confirmation, and the galaxy discussed in Sect.~\ref{subsec:DM} with a stellar mass requiring and integrated star-formation efficiency $\epsilon > 1$ at $z\sim 5$. The first galaxy, shown in the left panel of Fig.~\ref{fig:specz}, is located in EDF-F with $z_{\rm spec} = 3.08$ ($z_{\rm phot} = 3.07$) and $\logten(\mathcal{M}/{\rm M}_{\odot})=11.1\pm0.1$. The spectroscopic redshift is obtained from the GOODS-S catalogue \citep{Garilli2021,Kodra2023}, which overlaps EDF-F. In addition to showcasing a resolved morphology and large physical size, the galaxy is characterised by a relatively flat SED and weak Balmer break. This galaxy is likely forming stars efficiently and may have a young stellar population; indeed, referring back to Fig.~\ref{fig:hmf_lowz}, this galaxy should be approximately at the peak of the SHMR. Galaxies at this redshift are easily disentangled from stellar interlopers, in particular based on their $g-z$ and $z-[3.6\mu{\rm m}]$ colours. Note that there is not any HSC $y$ coverage over EDF-F, thus the blank cutout. Generally, this galaxy highlights the ability to properly identify galaxies known to be at high redshift using the DAWN PL photometry, though \citetalias{Zalesky2024} presents further detail regarding the performance of the DAWN PL \photoz{} in comparison to \num{3300} spectroscopic galaxies matched to the GOODS-S catalogue. Although there is no way to directly validate the stellar mass of this particular galaxy at present, \citetalias{Zalesky2024} also provides a demonstration that shows that when the redshift of the galaxy is accurate, the resulting \Ms{} obtained from DAWN PL is fully consistent with what would be inferred from photometry including near-IR (and additional) constraints.

The second galaxy, shown in the right panel of Fig.~\ref{fig:specz} is located in EDFN with $z_{\rm phot} = 5.15$ and $\logten(\mathcal{M}/{\rm M}_{\odot})=12.5^{+0.1}_{-0.9}$. The asymmetric uncertainty is indicative of the parameter space and the $P(\mathcal{M})$ for the galaxy: the probability distribution is broad but includes a local maximum near $\logten(\mathcal{M}/{\rm M}_{\odot})=12.5$, with few models capable of producing greater stellar mass. The cutouts and reported magnitudes show that the galaxy is $\sim$4.5 magnitudes brighter in \textit{Spitzer}/IRAC compared to HSC $z$, thus driving the substantial stellar mass. Stellar population ages in the galaxy templates are required to be younger than the age of the Universe at the assumed redshift, though such a significant excess beyond \num{4000}\,\AA\,(rest frame) implies an advanced age, assuming the redshift is correct. The $P(z)$ shows multiple peaks, and any lower value would lower the stellar mass while providing additional time to grow an old stellar population. The uncertainty in the stellar mass may be driven by an uncertainty in the dust content, which is not well constrained due to the lack of near-IR photometry. In any case, the 4.5 magnitude difference between $z$ and \chOne{} is extreme. Although no emission lines are included in the best-fit template, the secondary redshift solution includes emission lines expected for a star-forming galaxy. Again, it must be emphasised that the physical properties, including \photoz{} and \Ms{}, are not inferred from any one template but from the posterior distributions, consistent with \citetalias{Davidzon2017} and \citetalias{Weaver2023SMF}. It is not expected that the foreground object visible in the lower left corner of the cutouts impacts the photometry of this galaxy, despite its proximity. The segmentation of the image looks reasonable, and \cite{WeaverFarmer} provides several demonstrations of \farmer{} photometry package that show that the light from objects in such proximity to one another can be properly distinguished. Moreover, the stellar mass is primarily driven by the \chOne{} and \chTwo{} fluxes, which are significant and do not appear contaminated by the neighbouring object.

The degeneracy among the galaxy, AGN, and star templates is clearly evident in the second galaxy (right panel, Fig.~\ref{fig:specz}). As discussed in Sect.~\ref{subsubsec:interlopers}, it is not clear whether an AGN at this redshift would be bright enough at the observed wavelengths to impact the stellar mass estimate, though a sizeable contribution from AGN light would bring the \Ms{} estimate down. To reiterate the above, removing all galaxies that have smaller $\chi^{2}$ values in the best-fit AGN template compared to the best-fit galaxy template is inconsequential to the galaxy SMF, but doing so would have removed this source, which appears interesting and worthy of discussion. Comparing the $\chi^{2}$ values, the stellar template is rejected with high confidence, although visually it appears to be a reasonable fit. In the short-term, NIR observations from \Euclid may provide clarity on the physical properties of this galaxy, and at least provide further confidence in ruling out the stellar solution. However, determining the true nature of this galaxy ultimately requires spectroscopic follow up. 


\section{\label{sc:summary}Summary and conclusions}

The evolution of the galaxy SMF is characterised across $0.2 < z \leq 6.5$ using the DAWN PL catalogues. The DAWN PL catalogues span an effective area of 10.13 deg$^{2}$, exceeding COSMOS \citep{Scoville2007,Weaver2021} and the recent analyses of the galaxy SMF therein of \citetalias{Weaver2023SMF} by an order of magnitude and of \cite{Shuntov2024} by a factor of twenty. Such a unique, tenfold increase in volume samples cosmologically diverse structure and environments, while also exploring new parameter space of the most massive galaxies that cannot be found in smaller surveys. This work constitutes a benchmark for the \enquote{pre-launch} analysis of the galaxy SMF. 

Generally, the abundance of massive systems observed by \citetalias{Weaver2023SMF} is confirmed and upper limits of number density are replaced by firm estimates. A few dozen galaxy candidates are also observed with even greater stellar mass. This is enabled by a significant reduction in both Poisson uncertainty and cosmic variance, (e.g., a factor of $\sim5$ for galaxies with $\mathcal{M} \sim10^{11}$ M$_{\odot}$ at $z\sim5$). All galaxies are detected from a deep composite stack of Subaru HSC $r+i+z$ that rivals shallower near-infrared selection (e.g., \citetalias{Davidzon2017}) and reaches stellar masses above $\mathcal{M} \sim10^{10}$ M$_{\odot}$ at $z\sim6$. However, there is evidence that the most significantly dust-attenuated are not detected above $z \sim 1.5$. 

The number density of galaxies is described by a total uncertainty that includes Poisson uncertainty, cosmic variance, and uncertainty due to SED fitting with a zeroth-order correction for the covariance of stellar mass with redshift. To $z\sim3$, the SMF can be separated into a star-forming component and a quiescent component. A Schechter function, or double Schechter function where appropriate, is used to infer the intrinsic total SMF as well as the star-forming and quiescent components with Eddington bias accounted for following the procedure of \cite{Ilbert2013}, \citetalias{Davidzon2017}, and \citetalias{Weaver2023SMF}. The observed and intrinsic SMFs are compared to the most recent measurements in the COSMOS field as well as representative space-based measurements from HST and JWST \citep{Grazian2015,Weibel2024}, finding good agreement across all redshifts but with minor differences at $z\sim 5$--6 caused by the selection functions. Selection function aside, the most significant difference with measurements from space-based facilities is the difference in volume explored, where DAWN PL samples a volume between 70--100 times larger and thus the improvements in Poisson uncertainty, cosmic variance, and access to massive galaxies over \citetalias{Weaver2023SMF} are yet greater compared over \citetalias{Grazian2015} and \citetalias{Weibel2024}.

Additional key results of this analysis are as follows:
\begin{enumerate} 
    \item The intrinsic and observed SMF is compared with the \cite{Tinker2008} dark matter HMF at each redshift under various integrated star-formation efficiencies, $\epsilon$. Up until $z\sim3$, the abundance relative to the HMF of galaxies with stellar masses above and below the characteristic stellar mass, \Ms{}$^{\star}$, is indicative of inefficient star-formation ($\epsilon < 0.1$). However, massive galaxies requiring $\epsilon >0.3$--0.5 are observed across $3.5 < z \leq 6.5$. Their abundance provides substantial evidence that feedback mechanisms in massive galaxies are not strong enough to regulate their star formation at this epoch or before, as suggested by recent numerical simulations \citep{Valentini2021,Bassini2023} and recent measurements of the stellar-to-halo mass relation (SHMR; \citealt{Shuntov_2022}).
    \item Exceptionally massive quiescent galaxies ($\mathcal{M} >10^{11.7}$ M$_{\odot}$) are already fully formed as early as $z\sim2.5$ with a number density $\Phi \sim 10^{-6}$ Mpc$^{-3}$. Their number density is consistent with little to no growth from $z\sim 2.5$--3 to $z\sim 0.2$, suggesting that their gas supplies are completely consumed or otherwise disrupted and that feedback mechanisms may prevent further growth \citep{Fiore2017}. The inappreciable increase in their number density over time is consistent with the expected stellar mass growth from mergers in galaxies with $\mathcal{M} >10^{11}$ M$_{\odot}$ across these redshifts  \citep{Mundy2017,Conselice2022} and appears to extend the result of \cite{Kawinwanichakij2020} to even earlier times for quiescent systems.
    \item The galaxy SMF is investigated as a function of environment out to $z\sim 3.5$ following a rudimentary characterisation of the density-field based on \cite{Kovac2010}. At $0.8 < z \leq 1.1$, the quiescent SMF in the most dense regions shows clear signs of \enquote{environmental quenching} \citep{Peng2010}. In contrast to some previous works (e.g., \citealt{Papovich2018}), but supported by others \citep{Davidzon2016,Tomczak2017} massive galaxies of all types are preferentially found in high-density environments at all redshifts. The environmental dependence appears weaker at earlier times, though as early as $3 < z \leq 3.5$, massive galaxies are observed to be more abundant in high-density environments while less massive galaxies are preferentially located in low-density environments. 
    
\end{enumerate}

As emphasised above, DAWN PL has provided a powerful dataset uniquely suited to characterise the high-mass end of the galaxy stellar mass function due to the exceptionally deep \textit{Spitzer}/IRAC imaging over an unmatched area. Eventually, EDF-N and EDF-F (as well as EDF-S and the EAFs) will be observed by \Euclid to near-infrared depths of 26 mag \citep{McPartland2024,Mellier2024}, providing an opportunity for a \enquote{post-launch} analysis of the galaxy SMF. These data will be processed as part of the official \Euclid data releases and matched with \textit{Spitzer}/IRAC in future data releases from the Cosmic Dawn Survey. Upon their releases, a window of exploration will be opened through the epoch of reionisation to at least $z\sim8$ with greater statistical power than this work and with further potential for discovery of intrinsically red massive galaxies. In addition, an improvement in the \photoz{} performance is expected from the near-IR photometry, as well as more complete selection of quiescent galaxies at $z > 1.5$ and robust rejection of interloper populations. Complementing the imminent near-IR photometry, \Euclid will also obtain hundreds of thousands of spectroscopic redshifts in the EDFs and EAFs alone \citep{Mellier2024}, enabling a precise characterisation of the clustering of galaxies and therefore the connection to both their dark matter halos and their local environments The combination of the survey area, wavelength coverage, photometric depth, and complementary spectroscopy will establish a legacy of precision galaxy demographics and galaxy evolution studies that push ever further into the cosmic past.


%
%

\begin{acknowledgements}
\AckEC  
\end{acknowledgements}

%
%

\bibliography{lukas_bib_h20}

%

\begin{appendix}
  \onecolumn 
\section{\label{app:schechter}Schechter parameters}

Schechter parameters resulting from the MCMC analysis are presented in Table~\ref{table:fit_total} for the total sample, Table~\ref{table:fit_sf} for the star-forming sample, and Table~\ref{table:fit_q} for the quiescent sample. The median of the marginalised posterior distributions of each parameter are shown with uncertainties corresponding to the 16th and 84th percentiles (i.e., enclosing 68\% of the data). The maximum a posteriori parameter values are provided in brackets. Below, the comparison with the Schechter parameters of \citetalias{Grazian2015}, \citetalias{Davidzon2017}, \citetalias{Weaver2023SMF}, and \citetalias{Weibel2024} is discussed.

Starting with the low-mass slope, the maximum a posteriori value and the median of the posterior are each consistent with the one or more of \citetalias{Grazian2015}, \citetalias{Davidzon2017}, \citetalias{Weaver2023SMF}, or \citetalias{Weibel2024}, by construction, i.e., due to the priors that are assumed. As a whole, the representative values of the low-mass slope are most similar to \citetalias{Davidzon2017}. To compare the other Schechter parameters with the literature, it must be acknowledged that the treatment of the Eddington bias between this work and other works may be different, which will affect the inferred values of the Schechter parameters.

Of particular interest in the literature is the appropriate value of the characteristic mass, \Ms{}$^{\star}$, at each redshift. The value of \Ms{}$^{\star}$ is believed to coincide with the peak of the stellar-to-halo mass relation (SHMR) and thus represents that stellar mass at which a galaxy is most efficient at converting baryons into stars \citep{Behroozi2013,Wechsler2018}. Following similar lines of reasoning, it is thought that above \Ms{}$^{\star}$, quenching mechanisms become more effective.  The values of \Ms{}$^{\star}$ found here appear to be mostly consistent with the literature within the reported uncertainties, and especially with \citetalias{Weaver2023SMF}, the next largest survey by volume. Indeed, every value of \Ms{}$^{\star}$ through to and including $1.1 < z \leq 1.5$ is fully consistent with \citetalias{Weaver2023SMF}. Note that $1.1 < z \leq 1.5$ marks the onset of the increased uncertainty in \photoz{}s that leads to an excess of massive galaxies in DAWN PL. Nonetheless, median of the \Ms{}$^{\star}$ posterior distribution is $\logten(\mathcal{M}^{\star}/{\rm M}_{\odot})  = 10.98^{+0.05}_{-0.04}$, while \citetalias{Weaver2023SMF} reports a value of $\logten(\mathcal{M}^{\star}/{\rm M}_{\odot}) = 11.00^{+0.07}_{-0.11}$. This comparison,  may suggest that the adopted Eddington bias kernel is reasonable given the resulting representative values of \Ms{}$^{\star}$. Although the representative values at $1.5 < z \leq 2.0$ and $2.0 < z \leq 2.5$ are somewhat larger than what was found by \citetalias{Weaver2023SMF}, they are very nearly consistent within $1\sigma$. At $2.5 < z \leq 3.0$, where it is believed that the uncertainty in the \photoz{}s decreases due to the Lyman break being better constrained, the representative value of \Ms{}$^{\star}$ is again fully consistent with \citetalias{Weaver2023SMF} with a difference significantly less than the $1\sigma$ uncertainties. 

As previously discussed, \citetalias{Weaver2023SMF} fixes the value of the low-mass slope $\alpha$ at $-1.46$ at $z > 2.5$, while in this work the value of $\alpha$ is modelled with a prior that encourages a steeper (more negative) slope with increasing $z$. This is important to consider because $\alpha$ and \Ms{}$^{\star}$ are anti-correlated, with more negative (i.e., steeper) values of $\alpha$ leading to greater values of \Ms{}$^{\star}$ (see, again, Fig. 8 of \citealt{Stefanon2021} and Fig. 7 of \citetalias{Weibel2024}). As such, the representative values of \Ms{}$^{\star}$ should be generally greater than what was found by \citetalias{Weaver2023SMF} at $z \gg 2.5$ and they should agree more closely with works that found greater values of $\alpha$ at these redshifts \citep{Grazian2015,Davidzon2017,Weibel2024}. The former is confirmed throughout all remaining redshift bins. As for the latter, the representative value of \Ms{}$^{\star}$ is greater than what is found by \citetalias{Davidzon2017} at $3.0 < z \leq 3.5$ but consistent within the reported $1\sigma$ uncertainties. The lowest redshift bin of \citetalias{Grazian2015} and \citetalias{Weibel2024} is $3.5 < z \leq 4.5$. At these redshift, \citetalias{Davidzon2017} notably reports a value of \Ms{}$^{\star}$ that increases significantly. The representative value of \Ms{}$^{\star}$ found in this work is lower than each of \citetalias{Grazian2015}, \citetalias{Weibel2024}, and \citetalias{Davidzon2017}, but remains consistent within their reported uncertainties. Looking more closely, the difference with \citetalias{Davidzon2017} may be explained by their lower normalisation, where the two parameters are anti-correlated. Both \citetalias{Grazian2015} and \citetalias{Weibel2024} finds a higher value of \Ms{}$^{\star}$, while finding a smaller value for $\alpha$. The comparison with \citetalias{Grazian2015} is sensible, as they report a much lower normalisation than either this work or \citetalias{Weibel2024}. By contrast, \citetalias{Weibel2024} finds a similar normalisation to this work, and the source of the disagreement in the reported values of \Ms{}$^{\star}$ is not clear. At $z\sim4$, they assume an Gaussian-shaped Eddington bias kernel during their modeling with $\sigma=0.13$, which has a similar overall shape to the kernel used herein (Eq.~\ref{eq:kernel} with $\tau_{c}=0.05$ and $\sigma_{\rm Edd}=0.6$) but lacks the wings, which may be the cause.  

At $4.5 < z \leq 5.5$ and $5.5 < z \leq 6.5$, the value of \Ms{}$^{\star}$ shows significant variation, likely caused by the absence of low-mass constraints and the degeneracy with normalisation. As reported in Table~\ref{table:fit_total}, the maximum a posteriori values of \Ms{}$^{\star}$ are always larger than the median posterior values, suggesting that the likelihood space is peaked at the high-mass end but not strongly. Notably, \citetalias{Davidzon2017} also finds an even larger value of \Ms{}$^{\star}$ at $4.5 < z \leq 5.5$, though still consistent with the reported uncertainties. The difference may be explained by their lower normalisation. It is emphasised that \citetalias{Davidzon2017} is the only other survey of substantial cosmic volume that also allowed for variation in $\alpha$, which should make the comparison between the results more appropriate. Although the representative values of \Ms{}$^{\star}$ found at these redshifts in DAWN PL are larger than \citetalias{Grazian2015} and \citetalias{Weibel2024}, the SMFs at $z > 4.5$ of both \citetalias{Grazian2015} and \citetalias{Weibel2024} are dominated by low-mass galaxies far below the stellar mass completeness limits of DAWN PL. Both works find significantly higher normalisations, by contrast. Interestingly, the normalisation found by \citetalias{Weibel2024} at $4.5 < z \leq 5.5$ is higher than their value at $3.5 < z \leq 4.5$, and even at $z\sim7$ \citetalias{Weibel2024} finds a higher normalisation than at $3.5 < z \leq 4.5$. This is against the trend of decreasing normalisation of the galaxy SMF with increasing redshift found in this work and by \citetalias{Grazian2015}, \citetalias{Davidzon2017}, and \citetalias{Weaver2023SMF}. Consequently, the difference with \citetalias{Weibel2024} difference at $z > 4.5$ is likely due to the very small number of massive galaxies that can be found in their volume ($\sim70\times$ smaller than DAWN PL).

\renewcommand{\arraystretch}{1.4}

\begin{table*}[h]
\caption{Double ($z\leq2$) and single ($z>2$) Schechter parameters derived for the total mass complete sample.}
\footnotesize
\begin{threeparttable}
\resizebox{\textwidth}{!}{%
\begin{tabular}[]{cccccc}
\hline
\hline
$z$-bin  & $\logten\,\mathcal{M}^*$ & $\alpha_1$ & $\logten\,\Phi_1$ & $\alpha_2$ & $\logten\,\Phi_2$  \\
  & (M$_\odot$) &   & ($\mathrm{Mpc}^{-3}\,\mathrm{dex}^{-1}$) &   & ($\mathrm{Mpc}^{-3}\,\mathrm{dex}^{-1}$) \\
\hline
MCMC fit &&&& \\
\hline
$0.2 < z \leq 0.5$ & $10.92^{+0.09}_{-0.09}[10.94]$ & $-1.42^{+0.07}_{-0.08}[-1.55]$ & $-3.10^{+0.19}_{-0.16}[-3.38]$ & $-0.61^{+0.37}_{-0.32}[-0.84]$ & $-2.88^{+0.17}_{-0.17}[-2.79]$ \\
$0.5 < z \leq 0.8$ & $10.88^{+0.04}_{-0.05}[10.87]$ & $-1.37^{+0.05}_{-0.07}[-1.40]$ & $-3.03^{+0.12}_{-0.15}[-3.09]$ & $-0.54^{+0.27}_{-0.23}[-0.56]$ & $-2.78^{+0.03}_{-0.06}[-2.74]$ \\
$0.8 < z \leq 1.1$ & $10.86^{+0.04}_{-0.05}[10.89]$ & $-1.46^{+0.07}_{-0.06}[-1.54]$ & $-3.30^{+0.14}_{-0.10}[-3.46]$ & $-0.35^{+0.26}_{-0.20}[-0.55]$ & $-2.98^{+0.07}_{-0.07}[-2.93]$ \\
$1.1 < z \leq 1.5$ & $10.98^{+0.04}_{-0.04}[10.99]$ & $-1.44^{+0.09}_{-0.08}[-1.55]$ & $-3.62^{+0.22}_{-0.14}[-3.84]$ & $-0.77^{+0.19}_{-0.16}[-0.86]$ & $-3.18^{+0.08}_{-0.09}[-3.12]$ \\
$1.5 < z \leq 2.0$ & $11.15^{+0.04}_{-0.05}[11.14]$ & $-1.39^{+0.06}_{-0.10}[-1.55]$ & $-3.70^{+0.21}_{-0.21}[-4.05]$ & $-1.01^{+0.41}_{-0.25}[-1.10]$ & $-3.77^{+0.24}_{-0.29}[-3.56]$ \\
$2.0 < z \leq 2.5$ & $11.15^{+0.02}_{-0.02}[11.15]$ & $-1.55^{+0.00}_{-0.01}[-1.55]$ & $-3.83^{+0.02}_{-0.02}[-3.84]$ &  -- & -- \\
$2.5 < z \leq 3.0$ & $11.04^{+0.02}_{-0.02}[11.05]$ & $-1.70^{+0.00}_{-0.01}[-1.70]$ & $-3.93^{+0.03}_{-0.02}[-3.94]$ &  -- & -- \\
$3.0 < z \leq 3.5$ & $11.04^{+0.02}_{-0.02}[11.05]$ & $-1.69^{+0.01}_{-0.01}[-1.70]$ & $-4.07^{+0.03}_{-0.03}[-4.09]$ &  -- & -- \\
$3.5 < z \leq 4.5$ & $10.87^{+0.05}_{-0.06}[10.93]$ & $-2.01^{+0.06}_{-0.03}[-2.05]$ & $-4.48^{+0.12}_{-0.07}[-4.57]$ &  -- & -- \\
$4.5 < z \leq 5.5$ & $10.68^{+0.24}_{-0.18}[10.95]$ & $-2.14^{+0.10}_{-0.04}[-2.20]$ & $-4.99^{+0.49}_{-0.26}[-5.46]$ &  -- & -- \\
$5.5 < z \leq 6.5$ & $11.03^{+0.12}_{-0.16}[11.20]$ & $-2.05^{+0.10}_{-0.10}[-2.20]$ & $-5.75^{+0.32}_{-0.15}[-6.09]$ &  -- & -- \\
\hline
\hline

\end{tabular}
}
\begin{tablenotes} 
\item Values are given for the median posterior distributions with even-tailed $68\%$ range, and the values corresponding to the maximum likelihood solution in brackets.
\end{tablenotes}
\end{threeparttable}
\label{table:fit_total}
\end{table*}

\begin{table*}[h]
\caption{Double ($z\leq2$) and single ($z>2$) Schechter parameters derived for the star-forming mass complete subsample.}
\footnotesize
\begin{threeparttable}
\resizebox{\textwidth}{!}{%
\begin{tabular}[]{cccccc}
\hline
\hline
$z$-bin  & $\logten\,\mathcal{M}^*$ & $\alpha_1$ & $\logten\,\Phi_1$ & $\alpha_2$ & $\logten\,\Phi_2$  \\
  & (M$_\odot$) &   & ($\mathrm{Mpc}^{-3}\,\mathrm{dex}^{-1}$) &   & ($\mathrm{Mpc}^{-3}\,\mathrm{dex}^{-1}$) \\
\hline
MCMC fit &&&& \\
\hline
$0.2 < z \leq 0.5$ & $10.79^{+0.11}_{-0.13}[10.84]$ & $-1.42^{+0.06}_{-0.07}[-1.54]$ & $-3.08^{+0.14}_{-0.13}[-3.34]$ & $-0.49^{+0.62}_{-0.63}[-0.94]$ & $-3.26^{+0.29}_{-0.29}[-3.12]$ \\
$0.5 < z \leq 0.8$ & $10.81^{+0.06}_{-0.06}[10.80]$ & $-1.41^{+0.04}_{-0.06}[-1.46]$ & $-3.08^{+0.09}_{-0.12}[-3.16]$ & $-0.51^{+0.31}_{-0.32}[-0.56]$ & $-3.05^{+0.07}_{-0.12}[-2.97]$ \\
$0.8 < z \leq 1.1$ & $10.79^{+0.06}_{-0.05}[10.81]$ & $-1.49^{+0.05}_{-0.04}[-1.51]$ & $-3.29^{+0.09}_{-0.07}[-3.33]$ & $0.01^{+0.32}_{-0.27}[-0.10]$ & $-3.25^{+0.08}_{-0.08}[-3.21]$ \\
$1.1 < z \leq 1.5$ & $11.08^{+0.04}_{-0.06}[11.07]$ & $-1.37^{+0.04}_{-0.06}[-1.33]$ & $-3.47^{+0.09}_{-0.15}[-3.38]$ & $-0.82^{+1.00}_{-0.46}[0.97]$ & $-4.12^{+0.76}_{-0.33}[-5.02]$ \\
$1.5 < z \leq 2.0$ & $11.13^{+0.04}_{-0.06}[11.09]$ & $-1.36^{+0.03}_{-0.05}[-1.33]$ & $-3.58^{+0.08}_{-0.13}[-3.50]$ & $-0.44^{+1.11}_{-0.79}[1.00]$ & $-4.47^{+0.90}_{-0.28}[-4.87]$ \\
$2.0 < z \leq 2.5$ & $11.08^{+0.02}_{-0.02}[11.09]$ & $-1.55^{+0.00}_{-0.00}[-1.55]$ & $-3.79^{+0.02}_{-0.02}[-3.80]$ &  -- & -- \\
$2.5 < z \leq 3.0$ & $11.01^{+0.02}_{-0.02}[11.02]$ & $-1.70^{+0.01}_{-0.00}[-1.70]$ & $-3.90^{+0.03}_{-0.02}[-3.92]$ &  -- & -- \\
\hline
\hline

\end{tabular}
}
\begin{tablenotes} 
\item Values are given for the median posterior distributions with even-tailed $68\%$ range, and the values corresponding to the maximum likelihood solution in brackets.
\end{tablenotes}
\end{threeparttable}
\label{table:fit_sf}
\end{table*}

\begin{table*}[h]
\caption{Double ($z\leq 0.8$) and single ($z>0.8$) Schechter parameters derived for the quiescent mass complete subsample.}
\footnotesize
\begin{threeparttable}
\resizebox{\textwidth}{!}{%
\begin{tabular}[]{cccccc}
\hline
\hline
$z$-bin  & $\logten\,\mathcal{M}^*$ & $\alpha_1$ & $\logten\,\Phi_1$ & $\alpha_2$ & $\logten\,\Phi_2$  \\
  & (M$_\odot$) &   & ($\mathrm{Mpc}^{-3}\,\mathrm{dex}^{-1}$) &   & ($\mathrm{Mpc}^{-3}\,\mathrm{dex}^{-1}$) \\
\hline
MCMC fit &&&& \\
\hline
$0.2 < z \leq 0.5$ & $10.87^{+0.05}_{-0.05}[10.91]$ & $-1.85^{+0.29}_{-0.32}[-2.34]$ & $-4.88^{+1.18}_{-0.33}[-5.87]$ & $-0.48^{+0.14}_{-0.11}[-0.59]$ &  $-2.86^{+0.05}_{-0.05}[-2.88]$ \\
$0.5 < z \leq 0.8$ & $10.83^{+0.03}_{-0.03}[10.84]$ & $-1.66^{+0.45}_{-0.46}[-2.48]$ & $-5.00^{+2.28}_{-0.37}[-6.39]$ & $-0.39^{+0.09}_{-0.07}[-0.44]$ &  $-2.87^{+0.03}_{-0.04}[-2.86]$ \\
$0.8 < z \leq 1.1$ & $10.74^{+0.02}_{-0.02}[10.74]$ &  -- & -- & $-0.17^{+0.04}_{-0.04}[-0.18]$ & $-3.09^{+0.02}_{-0.02}[-3.09]$ \\
$1.1 < z \leq 1.5$ & $10.62^{+0.02}_{-0.03}[10.62]$ &  -- & -- & $0.46^{+0.07}_{-0.07}[0.45]$ & $-3.31^{+0.01}_{-0.01}[-3.30]$ \\
$1.5 < z \leq 2.0$ & $10.82^{+0.02}_{-0.03}[10.84]$ &  -- & -- & $0.10^{+0.11}_{-0.07}[0.02]$ & $-3.88^{+0.01}_{-0.01}[-3.89]$ \\
$2.0 < z \leq 2.5$ & $10.95^{+0.04}_{-0.05}[10.98]$ &  -- & -- & $0.05^{+0.18}_{-0.05}[0.00]$ & $-4.62^{+0.04}_{-0.04}[-4.62]$ \\
$2.5 < z \leq 3.0$ & $10.93^{+0.17}_{-0.23}[11.12]$ &  -- & -- & $0.45^{+0.85}_{-0.40}[0.00]$ & $-5.47^{+0.70}_{-0.23}[-5.49]$ \\
\hline
\hline

\end{tabular}
}
\begin{tablenotes} 
\item Values are shown for the median posterior distributions with even-tailed $68\%$ range, and the values corresponding to the maximum likelihood solution in brackets.
\end{tablenotes}
\end{threeparttable}
\label{table:fit_q}
\end{table*}

\pagebreak

\section{\label{app:validation}Validation of galaxy properties}



Section~\ref{subsubsec:validation} summarises additional validation tests that were performed to gain insight into systematic uncertainties and/or biases present in both the physical properties provided by the DAWN PL catalogues and the resulting galaxy SMFs. The first test (App.~\ref{subapp:refit_cosmos}) builds off a set of tests presented in \citetalias{Zalesky2024} involving measuring physical properties of galaxies from the COSMOS2020 catalogue \citep{Weaver2021} from SED fitting with \lephare{} using just the DAWN PL filter set. The second test (App.~\ref{subapp:random_forest}) builds off a framework presented by \cite{Chartab2023} involving measuring physical properties of galaxies from the COSMOS2020 catalogue using a random forest regressor trained on the DAWN PL filter set and the physical properties provided by COSMOS2020. Each test uses the same input catalogue, which includes photometry from the COSMOS2020 \farmer{} catalogue but restricted to the DAWN PL filter set and with uncertainties scaled to match the flux-flux error distributions from DAWN PL as described by  \citetalias{Zalesky2024}.

\subsection{\label{subapp:refit_cosmos}Re-fitting COSMOS2020}

As described in  \citetalias{Zalesky2024}, \photoz{}s and stellar masses are measured for COSMOS2020 galaxies using the modifications detailed above. For this test, a detection image with similar properties to DAWN PL is also created by adding noise to the COSMOS2020 HSC $r$, $i$, and $z$ images such that their noise properties (e.g., RMS) match DAWN PL; the images are then added together in the same way as the DAWN PL detection image. Galaxies are detected in the new-image and cross-matched to the COSMOS2020 coordinates to perform a selection according to the DAWN PL selection function. Finally, \photoz{}s and stellar masses are derived for each galaxy using \lephare{} in the same setup used for the creation of the DAWN PL catalogues. 

 \citetalias{Zalesky2024} demonstrated that the galaxy properties derived from re-fitting the COSMOS2020 catalogue in this way showed excellent agreement with the official values from \cite{Weaver2021} measured with over forty photometric bands. In particular,  \citetalias{Zalesky2024} showed that the stellar masses were highly accurate and unbiased in comparison to the official values provided by \cite{Weaver2021} when the \photoz{}s agreed. The \photoz{}s agreed for 95\% of objects brighter than $i \leq 25$ and for more than 80\% of objects that are fainter. Here, \enquote{agreement} is defined as not being classified as an outlier according to the standard definition when evaluating \photoz{}s, i.e., $|z_{1}-z_{2}|/(1+z_{2}) < 0.15$ \citep{Hildebrandt2012}, where $z_{1}$ is the \photoz{} obtained from re-fitting COSMOS2020 and $z_{2}$ is the \photoz{} of \cite{Weaver2021}. Using the new set of properties measured from the re-fitted COSMOS2020 catalogue with \lephare{}, the exact selections used to construct the galaxy SMFs from the DAWN PL catalogue (Sect.~\ref{subsec:selection}) can be applied to make an identical selection of galaxies consistent with the analysis above. In doing so, more than 95\% of all objects, regardless of their brightness, have \photoz{}s that are in agreement with those provided by \cite{Weaver2021}. These objects are used to measure the evolution of the stellar mass function using the same stellar mass bins and redshift bins as above. The result is shown in Fig.~\ref{fig:smf_litcomp_extra1} for $z \leq 3.5$ and Fig.~\ref{fig:smf_litcomp_extra2} for $z>3.5$. The results are shown alongside the observed galaxy SMF from DAWN PL, \citetalias{Davidzon2017} and \citetalias{Weaver2023SMF}. 

The agreement is excellent with DAWN PL within the associated uncertainties and volume limit of COSMOS2020, and strong with \citetalias{Davidzon2017} as well. This provides support for the assumption that the DAWN PL selection function is competitive with the near-IR selection of \citetalias{Davidzon2017} at every redshift. The agreement with \citetalias{Weaver2023SMF} is similar to what is achieved with DAWN PL. It is emphasised, however, the selection of galaxies used here is not identical to what was used by \citetalias{Weaver2023SMF}, so some minor difference is to be expected. The only redshift range indicative of significant incompleteness is $4.5 < z \leq 5.5$ and perhaps $5.5 < z \leq 6.5$, though the uncertainty in the literature makes the comparison difficult. Ultimately, it would appear that the lack of near-IR likely results in an incompleteness for the fainter galaxies at $z \geq 4.5$, which is perhaps not surprising. The values of the different selections described in Sect.~\ref{subsec:selection} were varied without resulting in a significant difference.  

\begin{figure}
    \centering
    \includegraphics[width=0.5\linewidth]{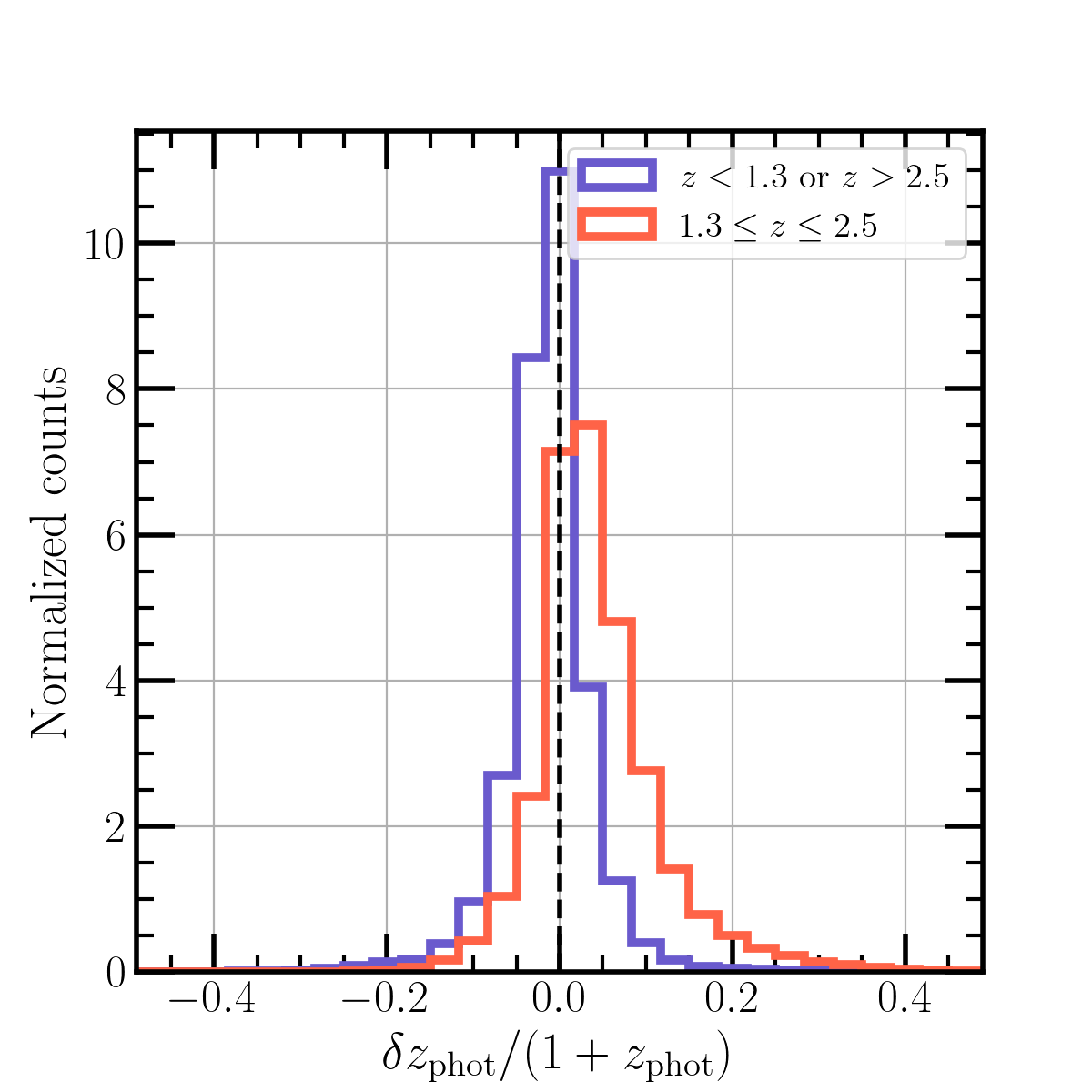}
    \caption{Difference in \photoz{} for galaxies from COSMOS2020 re-fitted with \lephare{} using just the DAWN PL bands compared to the original COSMOS2020 \citep{Weaver2021} values. The redshift range $1.3 \leq z \leq 2.5$ shows greater spread with a tendency towards higher redshifts, especially in the wings, due to lack of constraint on both the Lyman break and Balmer break. Galaxies outside this interval do not have the same problem and constrain at least one of the two breaks. A small number of galaxies scattering to higher redshifts can affect the observed galaxy SMF in this redshift interval.}
    \label{fig:cosmos_dz}
\end{figure}

It is also possible to use the re-fitted COSMOS2020 catalogue to investigate the source of the DAWN PL excess prominently observed at $1.5 < z \leq 2.5$. Interestingly, the feature is not observed in the re-fitted COSMOS2020 data. A closer inspection of the re-fitted \photoz{}s compared to the \cite{Weaver2021} is suggestive of the problem, however. The most likely cause is that low-mass galaxies at $z\sim1.1-1.5$ drift towards higher redshifts, artificially increasing their stellar masses due to the change in the luminosity distance. Recall that at $z \sim 1.5$, the Balmer break exits the HSC $z$ filter and sits between HSC $z$ and \chOne{}. Consequently, the degeneracy in redshift among the possible templates considered by \lephare{} during SED fitting increases significantly above $z = 1.1$ with a non-negligible bias towards higher-redshift solutions. The preferential drift towards higher-redshifts is simply due to there being more available templates that can satisfy the observed photometry. Galaxies are less likely to be placed at lower-redshift because then the Balmer break would have been observed, thus ruling out low-$z$ solutions. The effect is demonstrated in Fig.~\ref{fig:cosmos_dz}. 

Even without a bias towards higher-redshift solutions, the prospect of a low-redshift galaxies (e.g., at $z = 1.2$) being placed at higher redshift (e.g., to $z = 2$) presents an effect that is observed more often than the inverse. At fixed mass, galaxies are less abundant at high $z$ at all masses. Thus, the fractional change in the number density is always greater when a galaxy is moved from a low-redshift bin to a higher-redshift. In addition, the change in the luminosity distance implies a greater inferred mass for objects being moved to higher redshifts than they would have if placed to a lower redshift, where galaxies are more numerous. This is essentially the Eddington bias, but considered from the perspective of covariance with the redshift. Finally, a randomly selected low-redshift galaxy is more likely to be brighter in the observed-frame, further exacerbating the impact on the observed number density at higher redshift. It is important to emphasise that, for consistency with \citetalias{Davidzon2017} and \citetalias{Weaver2023SMF}, \lephare{} was not run with an apparent magnitude prior (or any prior). It is noted that this trend is visible even in Fig. C1 of \citetalias{Zalesky2024} in the comparison between the newly obtained \photoz{} and those presented by \cite{Weaver2021}. 

\subsection{\label{subapp:random_forest}Random forest predicted galaxy properties}

With a goal to further validate the physical properties of galaxies measured with \lephare{} in this work, an additional set of galaxy properties are obtained following the machine learning methods explored by \cite{Chartab2023}. Therein, the authors analysed the information content of multiwavelength galaxy surveys using a combination of an information theory and machine learning techniques. Specifically, the authors investigated the information content of the COSMOS2020 catalogue that is encoded by the photometry measured in the Hawaii Two-0 / DAWN survey PL (\citetalias{Zalesky2024}) filter set, namely, CFHT MegaCam $u$, Subaru HSC $griz$, and \textit{Spitzer}/IRAC \chOne{} and \chTwo{}. Among other results, the authors showed that for an optically selected catalogue, a random forest regressor \citep{Breiman2001} can be trained to predict the flux in near-infrared bands (e.g., UltraVista $YJH$), photometric redshifts, and stellar masses, each with impressive accuracy. In their work, they trained a random forest regressor on a training sample selected from the COSMOS2020 catalogue and then predicted galaxy properties from a test sample. Their analysis is extended here, where a random forest regressor is trained on the COSMOS2020 catalogue and galaxy properties are predicted from the DAWN PL catalogue. 

The basic approach for training the random forest described by \cite{Chartab2023} is followed here. One modification, however, is the selection of galaxies. Rather than splitting the COSMOS2020 catalogue into a training and a test sample, all galaxies that satisfy the basic requirements of signal-to-noise in the appropriate filters (Sect.~\ref{subsec:selection}) are used to train the random forest to predict \photoz{}s using photometry from the modified COSMOS2020 catalogue (i.e., just using the DAWN PL filters). Under this configuration, it was observed that high-redshift solutions were disfavoured too often. This bias can be seen at $z > 2$ in Fig.~11 of \cite{Chartab2023}. The bias is likely caused by the conservative nature of a random forest: by number, high-redshift galaxies are comparatively rare, and a random forest is prevented from predicting properties outside its training data by construction. To remedy this bias, an alternative selection was used to select an approximately even number of galaxies as a function of redshift. The downside to this approach is that it removes a significant portion of the training sample.

\begin{figure}[h]
    \centering
    \includegraphics[width=1\linewidth]{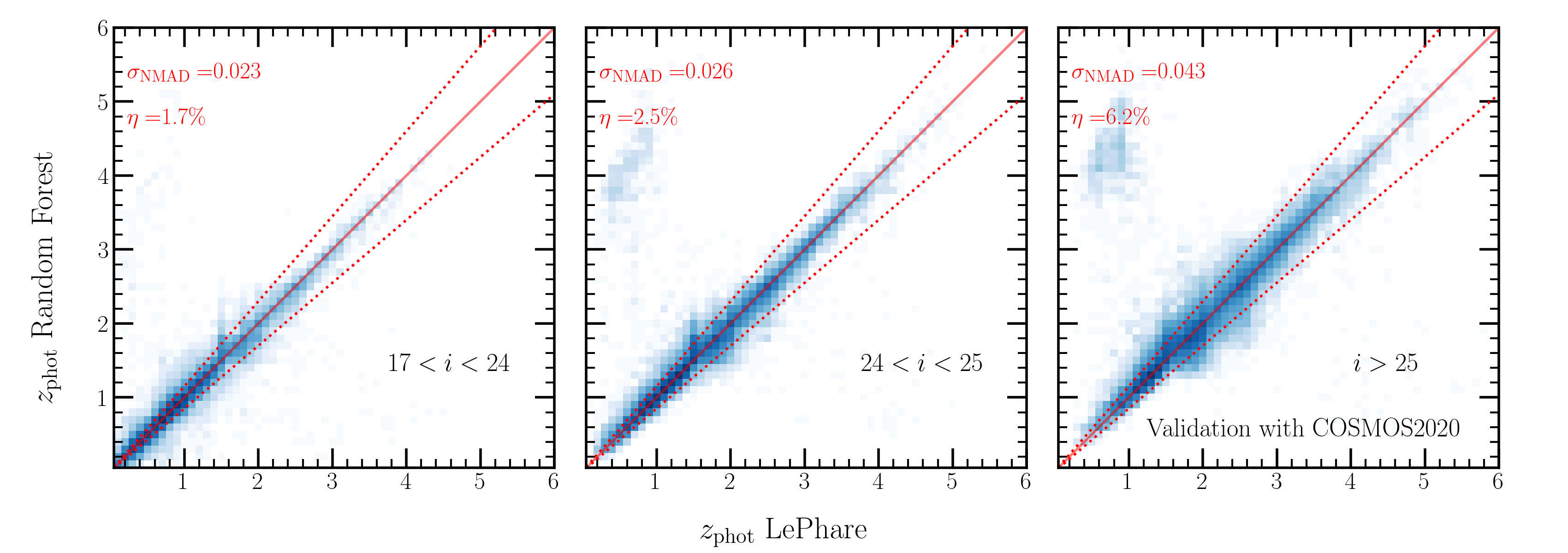}
    \includegraphics[width=1\linewidth]{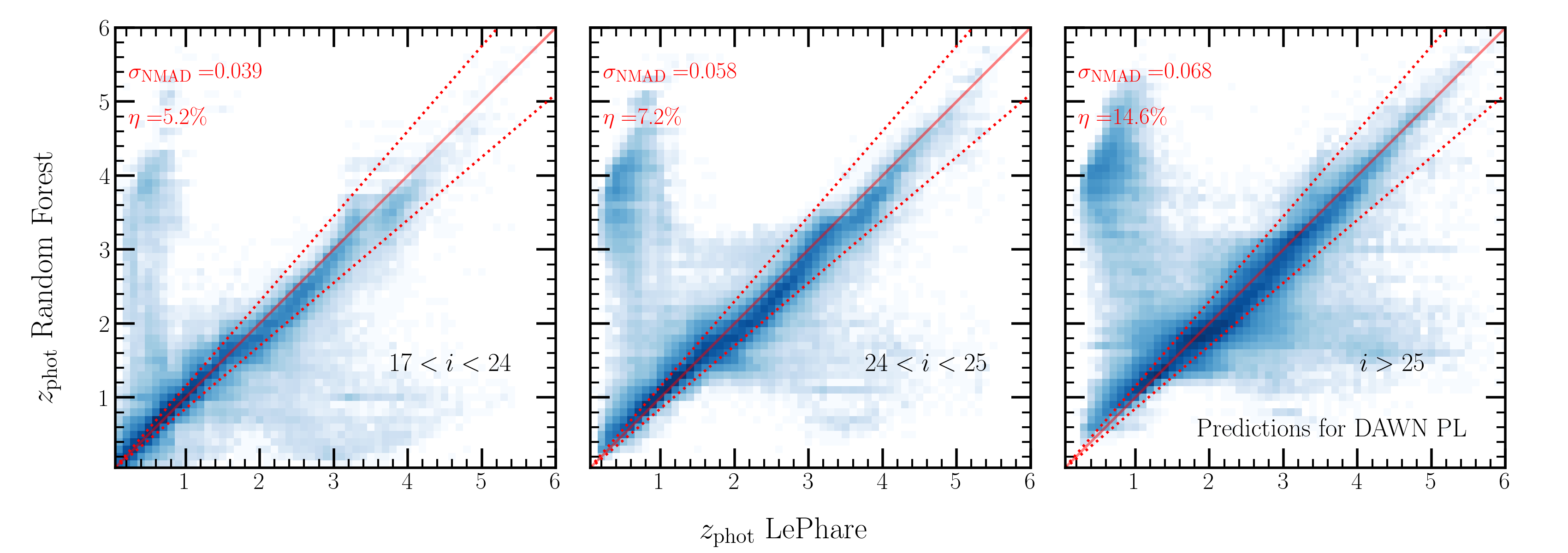}
    \caption{Top: Validation of photometric redshifts (\photoz{}) predicted from a random forest regressor trained on photometry from the COSMOS2020 in the DAWN PL filter set. The $y$-axis shows the random forest predicted redshifts after scattering the photometry within the photometric errors, while the $x$-axis shows the official \cite{Weaver2021} \photoz{} measured with \lephare{} and $>40$ photometric bands. Galaxies are selected for this comparison using the criteria described in Sect.~\ref{subsec:selection}. 
    Bottom: Comparison of \photoz{}s predicted from DAWN PL photometry using the random forest trained on COSMOS2020 as described above. The $y$-axis shows the random forest predicted redshifts from DAWN PL, while the $x$-axis shows the redshift measured from \lephare{} using the same photometry provided by \citetalias{Zalesky2024}. Galaxies selected for this validation also satisfy the requirements of Sect.~\ref{subsec:selection}. Shading corresponds to logarithmic density.}
    \label{fig:rf_zvalid}
\end{figure}

The random forest is first trained to predict redshifts from the modified COSMOS2020 catalogue. As a demonstration of the quality of the predictions, the photometry is scattered within the photometric uncertainties, and new predictions are obtained. The result is shown in the top row of Fig.~\ref{fig:rf_zvalid} for galaxies that have sufficient signal-to-noise in the necessary bands (Sect.~\ref{subsec:selection}) and that are above the nominal stellar mass limit (Sect.~\ref{subsec:m_completeness}). The $x$-axis is the official COSMOS2020 redshift (column name: \texttt{`lp\textunderscore zPDF'}) and the $y$-axis is the random-forest prediction. Colour scaling shows logarithmic density. Generally, the agreement is excellent, although it is emphasised that \photoz{}s obtained from re-fitting the modified COSMOS2020 catalogue with \lephare{} are better for the faintest galaxies, with 5.4\% outliers in comparison to 6.1\%. One area of improvement obtained by the random forest is in the redshift range $1.5 < z \leq 2.5$, in which there is no significant bias. Although there is scatter, it appears approximately equal in each direction, whereas the values from \lephare{} show a small bias towards higher redshifts (Fig.~\ref{fig:cosmos_dz}). 

The random forest trained on the modified COSMOS2020 catalogue is next used to predict galaxy properties from the DAWN PL photometry. Before doing so, photometric offsets are determined and applied to every photometric filter based on the median colours of stars in each field. Stars are identified from both COSMOS2020 and DAWN PL (with EDF-N and EDF-F treated separately) based on the $\chi^{2}$ of the best-fit \lephare{} stellar and galaxy templates. Only stars with signal-to-noise greater than 10 are used for this calibration. Then, median differences in each available colour are measured (i.e., with seven available filters, there are six independent colours) between DAWN PL and COSMOS2020. Finally, a system of equations is solved to determine a linear offset for each band that minimises the median difference in each colour measured between COSMOS2020 and DAWN PL. The resulting offsets are less than $\sim5\%$ with the exception of \textit{Spitzer}/IRAC \chTwo{}, which is $\sim15\%$. Overall, these are of similar scale to what is obtained from \lephare{} when fixing the redshifts of a spectroscopic sample to their spectroscopic values and minimising the differences in the photometry predicted by the best-fit \lephare{} galaxy templates (e.g., ~\citealt{Weaver2021}, Table 3; see also \citealt{Ilbert2013} and \citealt{Laigle2016} for further description of this operation). The random forest predicted redshifts for DAWN PL are shown in the bottom row of Fig.~\ref{fig:rf_zvalid} ($y$-axis) and compared to the \photoz{} measured with \lephare{} for the same galaxies ($x$-axis). The same selection criteria are applied as for the row above (i.e., those described in Sect.~\ref{subsec:selection}). It should not be expected that the redshifts from the random forest predictions and \lephare{} agree in DAWN PL as well as they do for COSMOS because random forest was trained to predict the exact redshifts used in the comparison. Nonetheless, the random forest predicted redshifts for DAWN PL agree well with the values obtained from \lephare{}, and for the entire sample generally. Indeed, the agreement is similar to what \cite{Weaver2021} showed when comparing the output from fitting the same exact photometry with the two separate SED-fitting codes, \lephare{} and \eazy{} (see, e.g., their Fig.~14). Interestingly, a number of bright galaxies ($17 < i < 24$) that are moved from higher-redshift solutions to low-redshift solutions by the random forest. As will be shown below, this largely solves the excess of massive galaxies seen in the DAWN PL galaxy SMF at $1.5 < z \leq 2.5$. 

\begin{figure}
    \centering
    \includegraphics[width=0.45\linewidth,trim={0 0 20mm 0},clip]{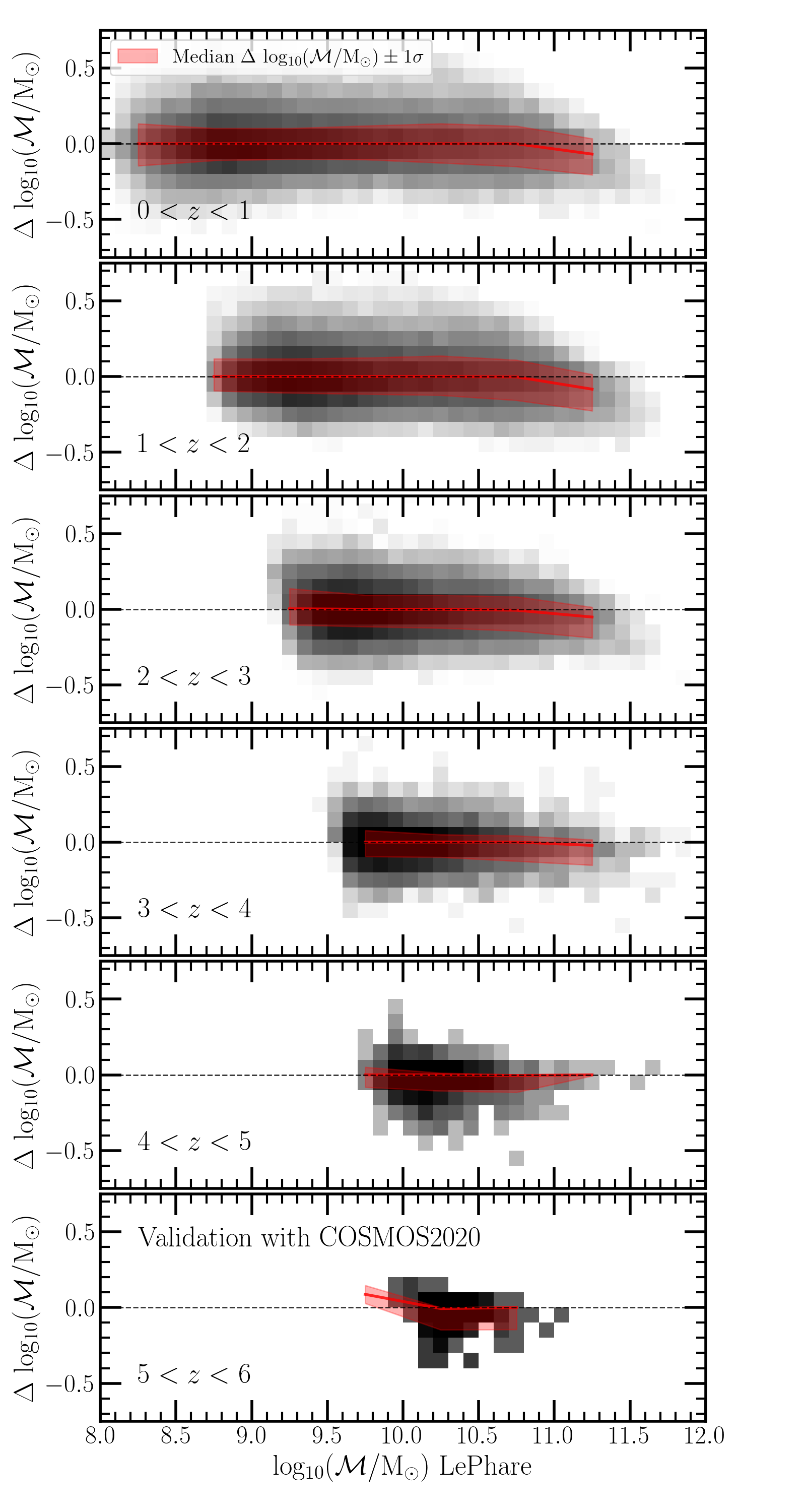}
    \includegraphics[width=0.45\linewidth,trim={20mm 0 0 0},clip]{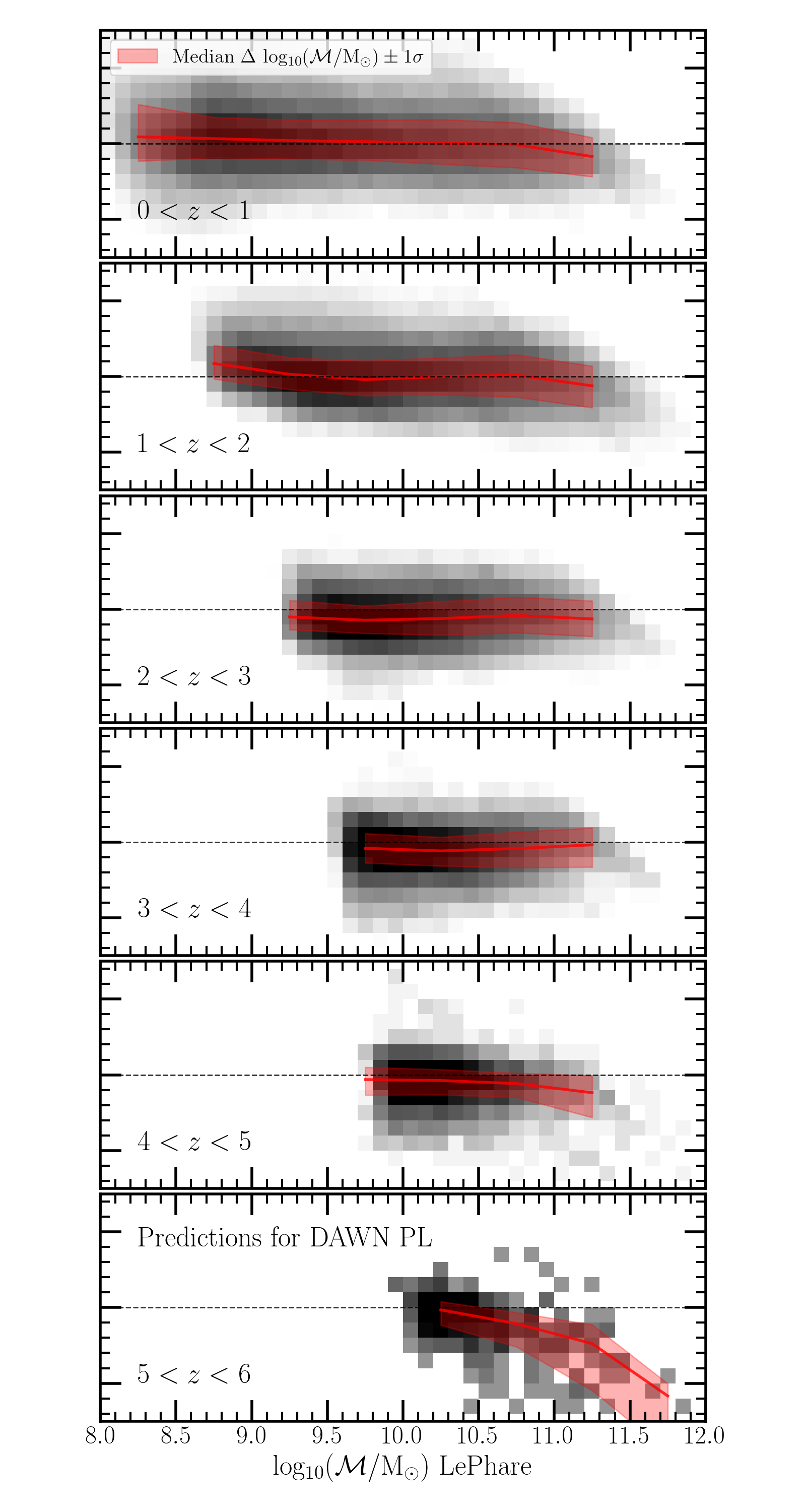}
    \caption{Left: Validation of stellar masses predicted from a random forest regressor trained on photometry from the COSMOS2020 in the DAWN PL filter set. The $y$-axis shows the difference in stellar mass between the random forest predicted values, after scattering the photometry within the photometric errors, and the official \cite{Weaver2021} stellar masses measured with \lephare{} and greater than $40$ photometric bands. The $x$-axis shows the official \cite{Weaver2021} stellar masses. Galaxies are selected for this comparison using the criteria described in Sect.~\ref{subsec:selection}. Right: Comparison of stellar masses predicted from DAWN PL photometry using the random forest trained on COSMOS2020 as described above. The $y$-axis shows the difference in stellar mass between the random forest predicted values from DAWN PL and the stellar masses measured from \lephare{} using the same photometry provided by \citetalias{Zalesky2024}. The $x$-axis shows the \lephare{} stellar mass. Galaxies selected for this validation also satisfy the requirements of Sect.~\ref{subsec:selection}. Shading corresponds to logarithmic density.}
    \label{fig:rf_Mvalid}
\end{figure}

Departing from \cite{Chartab2023}, stellar masses are determined separately from the \photoz{}. A second random forest is trained using the modified COSMOS2020 photometry as well as the official \cite{Weaver2021} \photoz{} values as inputs, and the training target is the stellar mass. The primary reason for separating these steps is to exploit the information in the redshift related to the stellar mass, i.e., the luminosity distance. This also ensures that the predicted stellar mass is consistent with the observed distribution of redshift vs. stellar mass provided by the COSMOS2020 catalogue, which is not as likely if the stellar mass is predicted without knowledge of the redshift. For the purposes of validation, stellar masses are predicted from the modified COSMOS2020 catalogue, again with photometry scattered within the photometric uncertainties, and the \photoz{}s predicted by the random forest in the previous step as inputs. The result is shown in the left-hand side of Fig.~\ref{fig:rf_Mvalid} for galaxies that have sufficient signal-to-noise in the necessary bands (Sect.~\ref{subsec:selection}) and that are above the nominal stellar mass limit (Sect.~\ref{subsec:m_completeness}). In addition, galaxies are selected to have redshifts that agree within $0.15(1+\delta z)$ to avoid differences that would arise simply from discrepant redshifts. The $x$-axis is the official COSMOS2020 stellar mass (column name: \texttt{`lp\textunderscore mass\textunderscore best'}) and the $y$-axis is the difference in stellar mass from the prediction and the official value. colour scaling shows logarithmic density, and the red shading shows the $1\sigma$ spread. The agreement is again strong. In this case, it appears that the random-forest performs equally well compared to SED fitting (e.g., \citetalias{Zalesky2024}, Fig.~11). However, without utilising the redshift as an input, the quality of the random forest prediction degrades. 

The random forest trained on the modified COSMOS2020 catalogue is next used to predict stellar masses from the DAWN PL photometry and from the random-forest predicted redshifts. The same linear offsets applied to the photometry before predicted the redshifts are used again. The random forest predicted stellar masses for DAWN PL are compared to the values measured from \lephare{} in the right-hand side of Fig.~\ref{fig:rf_Mvalid}. The same selection criteria are applied as for the row above, i.e., those described in Sect.~\ref{subsec:selection}, and the requirement that redshifts that agree within $0.15(1+\delta z)$. At $z < 4$, the agreement with the values determined with \lephare{} is nearly as good as in the COSMOS2020 validation shown in the left-hand side, without any bias. Starting at $4 < z < 5$, a bias begins to appear for the most massive galaxies, where the random forest predicts a smaller stellar mass than is determined with \lephare{}. As can be seen from the left-hand panel (as well as Fig.~\ref{fig:smf_litcomp2}), there are very few galaxies with stellar mass $\mathcal{M} \gtrsim 10^{11}$ M$_{\odot}$ in COSMOS2020 at $z>4$. As such, the random forest is biased against predicting such great stellar masses because such galaxies are largely absent from the training data. However, it is expected that over the entire DAWN PL volume, which is an order of magnitude larger than COSMOS2020, there should be many intrinsically bright galaxies at high redshift. Indeed, at every redshift interval, the most massive galaxies are found to be less massive according to the random forest, even in the training data. Care must therefore be taken when considering galaxy properties inferred from machine learning techniques.

\begin{figure*}[ht]
    \centering
    \includegraphics[width=\textwidth]{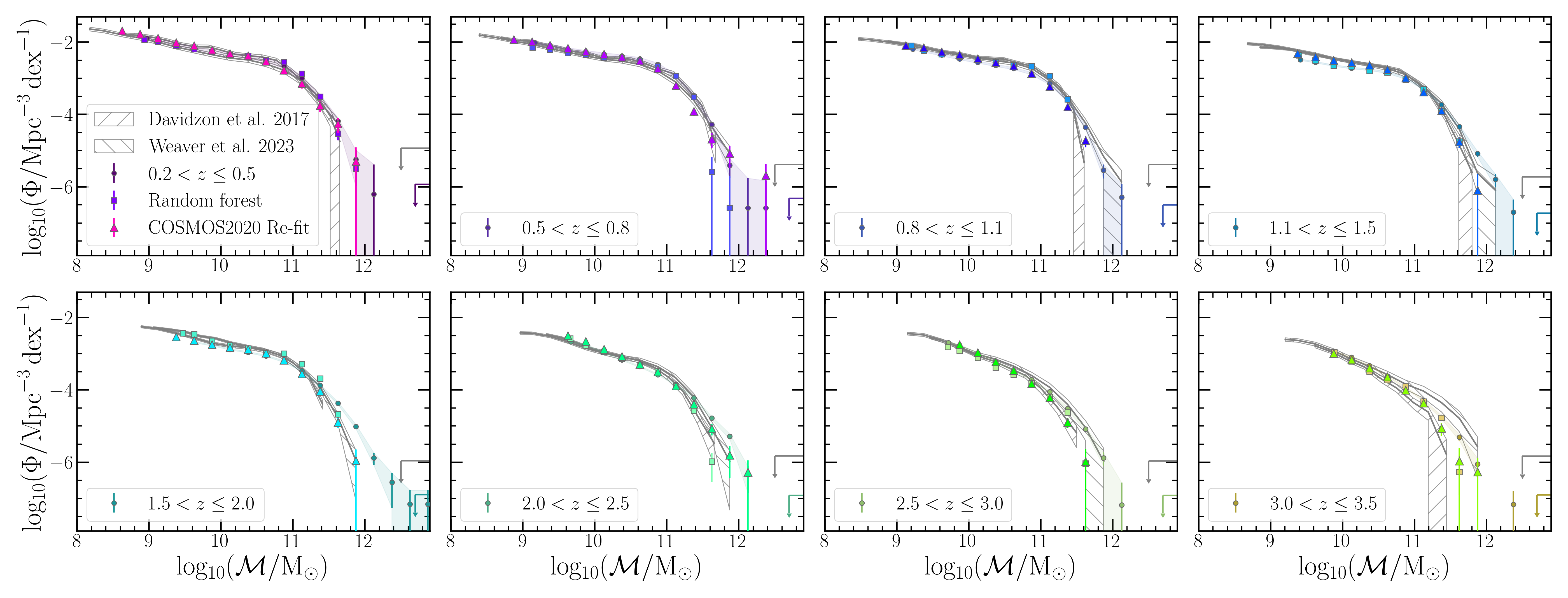}
    \caption{Same as Fig.~\ref{fig:smf_litcomp}, but with additional data points from the random forest (squares) and COSMOS2020 re-fit (triangles). Due to the conservative nature of random forests, large stellar masses are generally disfavoured at all redshifts. At $1.1 < z \leq 2.5$, this appears to correct for the bias observed from \lephare{} alone, but at $z > 2.5$, this appears to lead to an incompleteness. COSMOS2020 re-fitted using just the DAWN PL bands closely resembles the DAWN PL observed galaxy stellar mass function.}
    \label{fig:smf_litcomp_extra1}
\end{figure*}

\begin{figure*}[ht]
    \centering
    \includegraphics[width=\textwidth]{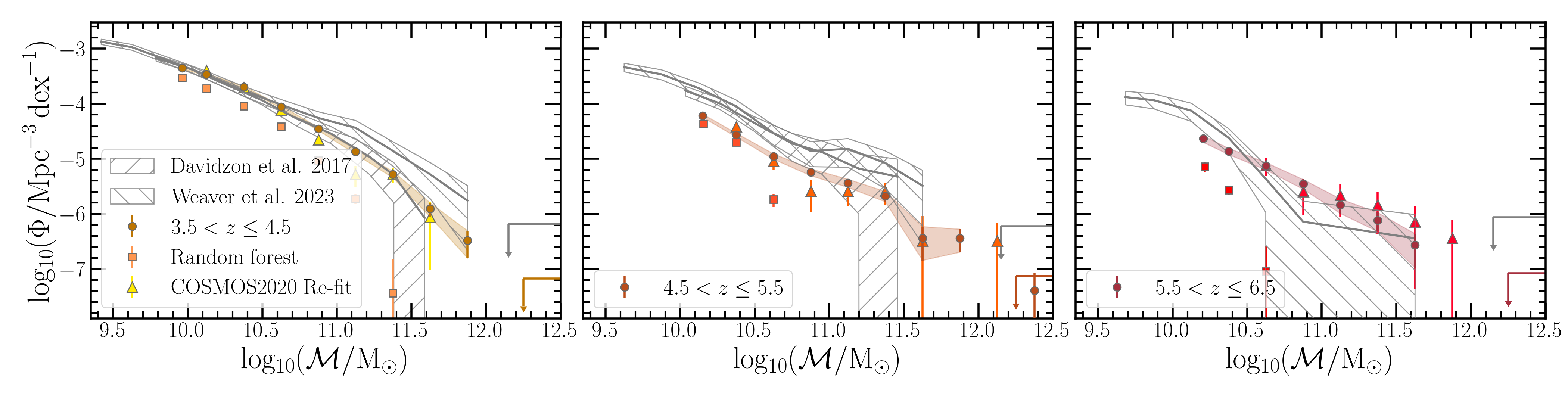}
    \caption{Same as Fig.~\ref{fig:smf_litcomp2}, but shows the additional data points from the random forest (squares) and the COSMOS2020 re-fit (triangles), as in Fig.~\ref{fig:smf_litcomp_extra1}.}
    \label{fig:smf_litcomp_extra2}
\end{figure*}

As a whole, the random-forest predicted quantities appear mostly reasonable, with the exception of the highest-redshift bins. Nonetheless, having calculated these values, it is possible to measure the evolution of the stellar mass function using the same stellar mass bins and redshift bins as above. The result is shown in Fig.~\ref{fig:smf_litcomp_extra1} for $z \leq 3.5$ and Fig.~\ref{fig:smf_litcomp_extra2} for $z>3.5$, alongside the values from re-fitting the modified COSMOS2020 catalogue with \lephare{}, the observed galaxy SMF from DAWN PL, and the results of \citetalias{Davidzon2017} and \citetalias{Weaver2023SMF}. 

At $z \leq 3.5$, there is strong agreement with DAWN PL, the result of re-fitting the COSMOS2020 catalogue, with \citetalias{Davidzon2017}. The agreement with \citetalias{Weaver2023SMF} is similar to what is achieved with DAWN PL. Overall, this exercise is another confirmation that there is sufficient information present in the DAWN PL catalogues to reliably infer the evolution of the galaxy stellar mass function. As noted previously, the random forest predicted values also successfully avoid the excess in massive galaxies seen at $1.5 < z \leq 2.5$. Referring back to Fig.~\ref{fig:rf_zvalid}, the problematic galaxies are likely placed at lower redshifts. At $z > 3.5$, the random forest appears poorly suited to measure the galaxy SMF. The stellar mass limit is high for DAWN PL (i.e., $\mathcal{M} 
\gtrsim10^{10}$ M$_{\odot}$ at $z\sim4$). Such galaxies are relatively rare in the training data (COSMOS2020) by absolute number. As such, lower stellar masses are preferred by the random forest, as are lower-redshift solutions, severely impacting the high-redshift galaxy SMF.

Given the improvement at low redshift, it may be advantageous to somehow utilise the random forest predicted galaxy properties alongside those obtained through SED fitting. However, it is not obvious how best to do so without introducing additional biases and systematic uncertainties to the analysis. A future work may explore this topic more fully.

\end{appendix}

\end{document}